\documentclass[a4paper,fleqn,usenatbib]{mnras}

\usepackage{newtxtext,newtxmath}

\usepackage[T1]{fontenc}
\usepackage{ae,aecompl}
\usepackage{relsize} 
\usepackage{color,soul}
\usepackage{subfig}

\usepackage{graphicx}
\usepackage{amsmath}
\usepackage{bm}
\usepackage{footnote}
\usepackage{bigints}

\defcitealias{Gie11}{GH11}
\defcitealias{Gil00}{GBG+00}
\defcitealias{Hamers17}{HT17}
\defcitealias{PaperI}{Paper I}

\title[Planet migration in globular clusters]{\vspace{-2mm}Forming short period sub-stellar companions in 47 Tucanae -- \\II.~Analytic expressions for the orbital evolution of planets in dense environments\vspace{-3mm}}

\author[A.~J.~Winter et al.]{Andrew~J.~Winter,$^{1,2,3}$\thanks{andrew.winter@uni-heidelberg.de} Cathie J.~Clarke$^{3}$, Giovanni Rosotti$^{2,4}$, Mirek Giersz$^{5}$\\
$^{1}$Zentrum f\"{u}r Astronomie, Heidelberg University, Albert Ueberle Str. 2, 69120 Heidelberg, Germany \\
$^{2}$School of Physics and Astronomy, University of Leicester, Leicester, LE1 7RH, UK \\
$^{3}$Institute of Astronomy, University of Cambridge, Madingley Road, Cambridge, CB3 0HA, UK \\
$^{4}$Leiden Observatory, Leiden University, P.O. Box 9513, NL-2300 RA Leiden, the Netherlands\\
$^{5}$Nicolaus Copernicus Astronomical Center, Polish Academy of Sciences, ul. Bartycka 18, Warsaw 00-716 Poland
\vspace{-3mm}
}

\date{Accepted X{\sevensize xxxx} XX. Received X{\sevensize xxxx} XX; in original form 2019 April XX}\vspace{-2mm}

\pubyear{2021}

\begin{document}
\label{firstpage}
\pagerange{\pageref{firstpage}--\pageref{lastpage}}
\maketitle

\begin{abstract}
{Short period, massive planets, known as hot Jupiters (HJs), have been discovered around $\sim 1$~percent of local field stars. The inward migration necessary to produce HJs may be `low eccentricity', due to torques in the primordial disc, or `high eccentricity' (HEM). The latter involves exciting high orbital eccentricity, allowing sufficiently close passages with the host star to raise circularising tides in the planet. We present an analytic framework for quantifying the role of dynamical encounters in high density environments during HEM. We show that encounters can enhance or suppress HEM, depending on the local stellar density and the initial semi-major axis $a_0$. For moderate densities, external perturbations can excite large eccentricities that allow a planet to circularise over the stellar lifetime. At extremely high densities, these perturbations can instead result in tidal disruption of the planet, thus yielding no HJ. This may explain the apparent excess  of HJs in M67 compared with their local field star abundance versus their apparent deficit in  47 Tuc. Applying our analytic framework, we demonstrate that for an initial massive planet population similar to the field, the expected HJ occurrence rate in 47 Tuc is $f_\mathrm{HJ}=2.2\times 10^{-3}$, which remains consistent with present constraints. Future large (sample sizes $\gtrsim 10^5$) or sensitive transit surveys of stars in globular clusters are required to refute the hypothesis that the initial planet population is similar to the solar neighbourhood average. Non-detection in such surveys would have broad consequences for planet formation theory, implying planet formation rates in globular clusters must be suppressed across a wide range of $a_0$. }
\end{abstract}

\begin{keywords} 
planets and satellites: formation,  dynamical evolution and stability, gaseous planets -- stars: kinematics and dynamics -- globular clusters: individual: 47 Tuc \vspace{-2mm}
\end{keywords}


\section{Introduction}

Hot Jupiters (HJs) are gas giant planets on short orbital periods ($\lesssim 10$~days) and are the class of planet to which the first discovered planet belongs \citep[51 Peg b --][]{Mayor95}. They remain over-represented among known exoplanets with respect to their occurrence rates, mainly due to their comparatively high detection efficiency among both transit and radial velocity surveys. A great deal of planet formation theory has been devoted to understanding their formation \citep[for recent reviews, see][]{dawson18, Fortney21}. Mass budget constraints imply the the majority of the mass of the HJ must be accreted outside of the current orbits, thus one of the main questions regarding their formation is how they migrated inwards. 

Broadly, the possible mechanisms for migration of HJs can be divided into two categories, labelled `low eccentricity migration' (LEM) and `high eccentricity migration' (HEM). In LEM, massive planets move inwards as a consequence of torques within the protoplanetary disk in which they form \citep[e.g.][]{Goldreich80, Lin86, Lin96}. In order to produce HJs, this requires efficient `type II' migration, operating once a gap in the gas surface density profile has opened. Current evidence suggests low viscosity in protoplanetary discs \citep[e.g.][]{Pinte16, Trapman20}, which would imply slow type II migration rates \citep[][]{Lega21}.  While this does not categorically rule out such a formation scenario, evidence for HJs around stars younger than $\sim 600$~Myr remains scarce \citep{Paulson06, Bailey18, Takarada20}, with younger candidates often the topic of debate -- e.g. around CI Tau \citep{Donati20} and V830 Tau \citep{Damasso20}. 

On the other hand, HEM represents a later mode of migration, relying on close passages with the host star and orbital circularisation due to tides raised at periastron \citep[e.g.][]{Goldreich66,Hut81, Adams06}. The necessary eccentricities may be excited by oscillations during interaction with an outer companion \citep[Kozai-Lidov --][]{Kozai62, Lidov62, Naoz16, Hamers17b, Fontanive19, Belokurov20} or by dynamical instability within a planetary system \citep{Rasio96, Carrera19}. Either of these scenarios may originate from the initial properties of the system (for example, see \citealt{Pu15} versus \citealt{Yee21}), or be due to perturbation by neighbouring stars in stellar clusters \citep{Bon01, Shara16, Brucalassi16, Li20, Winter20c, Rodet21, Wang22}. In favour of HEM as the origin of at least some HJs, \citet{Dong21} recently discovered a planet of mass $\sim 5\,M_\mathrm{J}$, TOI-3362b, that appears to be undergoing tidal circularisation. Statistically, the obliquity distribution of HJs appears consistent with late tidal damping after HEM \citep{Rice22}. Meanwhile, free-floating planets have recently been found to be abundant in the young Upper Scorpius association \citep{Miret-Roig22}, suggesting that a significant fraction of planetary systems have undergone chaotic dynamical decay. Finally, systems that have not undergone dynamical decay appear consistent with Hill radius limited growth within a protoplanetary disc, which suggests inefficient type II migration \citep[][]{Winter21}.

Disentangling the (dominant) formation pathway for HJs requires correlating their occurrence rates with the properties of their host star. This includes intrinsic properties of the host star, such as stellar mass and metallicity \citep[e.g.][]{Santos01, Boss02, Johnson10}, but also the external environment. This was the motivation of \citet[][hereafter \citetalias{Gil00}]{Gil00} in carrying out a \textit{Hubble Space Telescope} survey of the globular cluster 47 Tuc for short period massive planets. The non-detection of any transit signal among $34,091$ stars was initially thought to put upper limits on the HJ occurrence being $\lesssim 0.2$ percent, significantly fewer than the solar neighbourhood average \citep[$ 1.2\pm 0.4$~percent inferred from RV surveys --][]{Wright12}. However, \citet{Masuda17} applied an updated distribution of known HJ properties to demonstrate that the number of HJs in the \citetalias{Gil00} sample would be $2.2^{+1.6}_{-1.1}$ if the planet population is indistinguishable to those hosted by \textit{Kepler} stars of similar masses. The result may therefore be less significant than initially thought. This also applies to the wide field search for HJs by \citet{Weldrake05}. Whether or not HJs exist in 47 Tuc in comparable numbers to the field therefore remains an open question. 

If HJ formation is suppressed in 47 Tuc relative to the field, this could originate from lower formation rates due to the lower metallicity \citep{Santos01, Boss02, Ercolano10} or external irradiation of the planet forming disc by strong ultraviolet fields \citep{Johnstone98, Adams04, Facchini16, Winter18b, Haworth18}. The influence of external UV fields on giant planet occurrence remains uncertain, both empirically and theoretically. However, in terms of metallicity, \citet{Johnson10} estimated a scaling of occurrence rates of HJs as $10^{1.2\rm{[Fe/H]}}$, corresponding to approximately an order of magnitude for metallicity of 47 Tuc, with $\rm{[Fe/H]} \approx -0.7$, with respect to the \textit{Kepler} field with $\rm{[Fe/H]}\approx 0$. Due to the small number of stars with low metallicity in the \textit{Kepler} field, this dependence remains challenging to constrain with \textit{Kepler} data \citep[see dicussion by][]{Masuda17}. It is also unclear whether metallicity is a fundamental property that determines giant planet occurrence, or whether it is an extraneous property that correlates via formation conditions. 

Despite the above considerations, the apparent absence of HJs found in 47 Tuc may remain surprising within the paradigm of HEM. A naive expectation would be that if any planets at all exist in globular clusters then they should be more likely, not less likely, to undergo dynamical perturbation with subsequent circularisation. Such a trend has been hinted at by the marginally significant overabundance of HJs in the dense cluster M67 \citep{Brucalassi16}. Quantifying this expectation and  reconciling the tension between the findings in M67 and 47 Tuc partially motivates this work.

{In this, the second of a two paper mini-series, we consider the formation of HJs in 47 Tuc by HEM. We apply a Monte Carlo model for the dynamical evolution of 47 Tuc, introduced in \citetalias{PaperI} \citep{PaperI}, to follow the rate at which migrating planets undergo dynamical perturbation, interpreting this rate in terms of the efficiency of HJ production. The analysis presented in this work is complementary to that of \citet[][hereafter \citetalias{Hamers17}]{Hamers17}, who performed numerical simulations to show how the formation of HJs varies with stellar density due to encounters within some radius $R_\mathrm{enc}$. In this work, we offer a theoretical framework to interpret these results, allowing us to generalise the findings across a wide parameter space pertaining to both the properties of the star-planet system and external environment. Coupled with a dynamical model, our analytic prescription allows us to quantify the probabilities of various outcomes for planetary systems over the lifetime of 47 Tuc. }

The remainder of this manuscript is organised as follows.  We consider the theoretical rates of tidal circularisation and dynamical perturbations in Section~\ref{sec:Num_Method}. We apply our results in terms of the dynamical model for 47 Tuc in Section~\ref{sec:47Tuc}, wherein we also make predictions for future surveys. We summarise our conclusions in Section~\ref{sec:conclusions}.

\section{Orbital evolution theory}

\label{sec:Num_Method}
\subsection{Overview}
\label{sec:hem_pert}
\subsubsection{Motivation}
While the cause of the inward migration of HJs remains uncertain \citep[e.g.][for a recent review]{Fortney21}, we will here assume that gas giants on short orbital periods are produced by some dynamical perturbation of the initial formation configuration \citep[e.g][]{Rasio96,Ford08, Carrera19, Winter21, Miret-Roig22}, rather than from efficient migration within a stellar disc \citep[e.g.][]{Lin96, Baruteau14}. Such a dynamical perturbation can result in a sufficiently high eccentricity to yield close passages with the central star and tidal exchanges that shrink and circularise the orbit \citep[e.g.][]{Hut81,Eggleton98, Jackson08}. In this work, we consider how orbital perturbations due to stellar encounters influence a circularising planet.

\label{sec:post_process}

\subsubsection{Approach}

{In this section, we aim to produce an analytic estimate for various possible outcomes for a planet evolving in a high density environment. To this end, in Section~\ref{sec:tidal_circ_theory} we first discuss the theoretical tidal ciricularisation rate. Circularisation occurs due to close passages of the planet with the host star, during which the tides raised in the planet reduce its orbital energy and shrink the orbit while conserving the semi-latus rectum $l$.}

{We then quantify the changes of the orbital eccentricity due to encounters in a dense stellar environment. Such encounters can change how a planet circularises, possibly curtailing migration by reducing the eccentricity or inducing tidal disruption due to extremely close passages with the host star. We consider the encounter-driven evolution of eccentricity, rather than semi-major axis, for two reasons. Firstly, because the rate of circularisation for a planet on an highly eccentric orbit is strongly influenced by small changes in eccentricity. Secondly, because the change of angular momentum due to a stellar encounter scales as a power-law in closest approach distance $r_\mathrm{p}$ \citep{Heggie96}. On the other hand, changes in energy become exponentially smaller with increasing $r_\mathrm{p}$ \citep{Heggie75}. Thus the most common encounters, those occurring with large $r_\mathrm{p}$, predominantly alter eccentricity. }

{In Section~\ref{sec:pert_theory} we quantify the cross section for perturbations by a neighbouring star in terms of a small change in eccentricity $\epsilon$. We convert this to a rate of perturbation given a local stellar density and velocity dispersion in Section~\ref{sec:pert_rate}. We discuss the interpretation of these perturbation rates in Section~\ref{sec:cons_an}. We then apply the perturbation rates to quantify the statistical evolution of orbital eccentricity due to stellar encounters in Section~\ref{sec:stat_evol}. }

{Finally, we consider how dynamical perturbation influences tidal circularisation outcomes. In the first instance, we make arguments on the maximum possible $l$ along which a planet can circularise in Section~\ref{sec:tidal_acc}. This is set by the condition that the perturbation rate balances with the circularisation rate. We then compare these analytic predictions to a numerical experiment in Section~\ref{sec:circ_radii}. This allows us to interpret the fraction of tidally destroyed planets, which have a minimum pericentre distance that is too close to their host star to survive (Section~\ref{sec:HJ_surv}). We additionally consider the rate of ionisation of a planetary system in Section~\ref{sec:ionisation}. With these calculations, in Section~\ref{sec:fract_outcome} we establish the analytic framework for computing the relative outcome probabilities, with comparisons to the previous numerical experiments of \citetalias{Hamers17}. We apply this framework to 47 Tuc in Section~\ref{sec:47Tuc}. }

\subsection{Pseudo-synchronous tidal circularisation rate}
\label{sec:tidal_circ_theory}
In order to understand how eccentricity perturbations alter the evolution of a would-be HJ, we first need to estimate the circularisation rates. We will assume that the dissipation of orbital energy is dominated by tides raised in the planet, and that we are in the limit of low obliquity \citep[although see also][]{Alexander73}. In this case, the basic equations for the long term semi-major axis and eccentricity evolution of a planet circularising by successive close approaches with its host star are given by \citet{Hut81}:
\begin{multline}
\label{eq:adot}
    \dot{a}_\mathrm{tide} = -6 k_\mathrm{p} \tau_\mathrm{p} n^2  q^{-1}\left( \frac{R_\mathrm{p}}{a} \right)^{5} \frac{a}{(1-e^2)^{15/2}} \times \\ \times\left\{f_1(e^2) - (1-e^2)^{3/2} f_2(e^2)\frac{\Omega_\mathrm{p}}{n} \right\}
\end{multline}
\begin{multline}
\label{eq:edot}
    \dot{e}_\mathrm{tide} = -27 k_\mathrm{p} \tau_\mathrm{p} n^2  q^{-1} \left( \frac{R_\mathrm{p}}{a} \right)^{5} \frac{e}{(1-e^2)^{13/2}} \times \\ \times\left\{f_3(e^2) -\frac{11}{18} (1-e^2)^{3/2} f_4(e^2)\frac{\Omega_\mathrm{p}}{n} \right\}
\end{multline}
\begin{multline}
\label{eq:Omdot}
    \dot{\Omega}_\mathrm{p,tide} = 3 k_\mathrm{p} \tau_\mathrm{p} n \frac{q^{-2}}{r_\mathrm{g}^2} \left( \frac{R_\mathrm{p}}{a} \right)^{6} \frac{1}{(1-e^2)^{6}} \times \\ \times\left\{f_2(e^2) - (1-e^2)^{3/2} f_5(e^2)\frac{\Omega_\mathrm{p}}{n} \right\}
\end{multline}
where $\Omega_\mathrm{p}$ is the angular frequency of the rotating planet, $r_\mathrm{g}$ is the radius of gyration and
\begin{equation}
    n = \sqrt{\frac{G m_*(1+q)}{a^3}} ,
\end{equation} while $R_\mathrm{p}$, $k_\mathrm{p}$, and $\tau_\mathrm{p}$ are the planetary radius, apsidal motion constant, and tidal time lag. {We will generally follow \citetalias{Hamers17} in adopting $k_\mathrm{p}=0.25$ and $\tau_\mathrm{p} =0.66$~s, while we fix $R_\mathrm{p}=0.1\, R_\odot$.} The functions $f_i$ are defined:
\begin{equation}
    f_1(e^2) = 1 + \frac{31}{2}e^2 + \frac{55}{8} e^4 + \frac{185}{16}e^6 + \frac{25}{64}e^8
\end{equation}
\begin{equation}
    f_2(e^2) = 1 + \frac{15}{2}e^2 + \frac{45}{8} e^4 + \frac{5}{16}e^6 
\end{equation}
\begin{equation}
    f_3(e^2) = 1 + \frac{15}{4}e^2 + \frac{15}{8} e^4 + \frac{5}{64}e^6 
\end{equation}
\begin{equation}
    f_4(e^2) = 1 + \frac{3}{2}e^2 + \frac{1}{8} e^4 
\end{equation}
\begin{equation}
    f_5(e^2) = 1 + 3e^2 + \frac{3}{8} e^4 .
\end{equation} This set of equations dictates the tidal evolution of a low obliquity planet. 

In principle, one then must now choose an initial orbital frequency for the planet, as well as semi-major axis and eccentricity, to solve the system of equations~\ref{eq:adot}--\ref{eq:Omdot}. Indeed, in the parabolic limit $e\rightarrow 1$, this choice can dictate the outcome due to the {tidal force} on the planet orbit, where if the initial orbital frequency $\Omega_{\mathrm{p},0}$ exceeds a critical value then the planet will escape rather than circularise \citep{Hut82}. However, we expect that the rotational angular momentum of the planet is much smaller than its orbital angular momentum, which justifies the assumption that the orbitally averaged tidal torque is zero. This is equivalent to the pseudo-synchronisation condition $\dot{\Omega}_\mathrm{p} \approx 0$, or:
\begin{equation}
    \Omega_\mathrm{p} \approx n \frac{f_2(e^2)}{(1-e^2)^{3/2} f_5(e^2)}. 
\end{equation} This can be compared directly to equation 42 of \citet[][also the prescription of \citetalias{Hamers17}]{Hut81}. One can then rewrite equations~\ref{eq:adot} and~\ref{eq:edot}:
\begin{equation}
\label{eq:adot_ps}
    \dot{a}_\mathrm{tide} = -21 k_\mathrm{p} \tau_\mathrm{p} n^2  q^{-1} \left( \frac{R_\mathrm{p}}{a} \right)^{5} \frac{a e^2 f(e^2)}{(1-e^2)^{15/2}}
\end{equation}
\begin{equation}
\label{eq:edot_ps}
    \dot{e}_\mathrm{tide} = -\frac{21}{2} k_\mathrm{p} \tau_\mathrm{p} n^2  q^{-1} \left( \frac{R_\mathrm{p}}{a} \right)^{5} \frac{e f(e^2)}{(1-e^2)^{13/2}},
\end{equation}where
\begin{equation}
    f(e^2) = \frac{1 + \frac{45}{14}e^2 + 8 e^4 +\frac{685}{224}e^6 + \frac{255}{448}e^8 + \frac{25}{1792}e^{10}}{1 + 3 e^2 + \frac{3}{8} e^4}.
\end{equation} From equations~\ref{eq:adot_ps} and~\ref{eq:edot_ps} it is clear that a circularising planet that is slowly rotating always preserves the semi-latus rectum (SLR) $l = a(1-e^2)$, or equivalently the specific angular momentum $h\propto\sqrt{l}$.

\subsection{Perturbation cross section}

\label{sec:pert_theory}

We must now quantify the rate at which the eccentricity of an orbiting planet is altered by encounters with other stars. Any random (uncorrelated) dynamical encounter between stars can be expressed in terms of the effective cross section. This cross section is the effective area `seen' by neighbouring stars in a given environment, averaged over all possible orientations. In this case, we are interested in any perturbation that results in a significant change in the orbit of the planet. The distinction here compared to previous studies investigating perturbations of planetary systems by stellar flyby, is that we are not initially concerned by whether the planet is subsequently lost from the system \citep[cf.][for example]{Hill89,Davies01,Bon01,Fregeau06}. High eccentricity migration requires close passage {of the migrating planet within a few stellar radii of the host star over the entire circularisation time-scale $\tau_\mathrm{circ}$.} Thus, even slight external perturbations to the orbit may alter the migration rate. {The closest approach of an external perturber required in this case may therefore be far larger than for ionisation (or tidal capture -- cf. \citetalias{PaperI}).}

\begin{figure*}
    \centering
     \subfloat[\label{subfig:diffencrate}Differential planet perturbation rate with relative speed]{\includegraphics[width=\columnwidth]{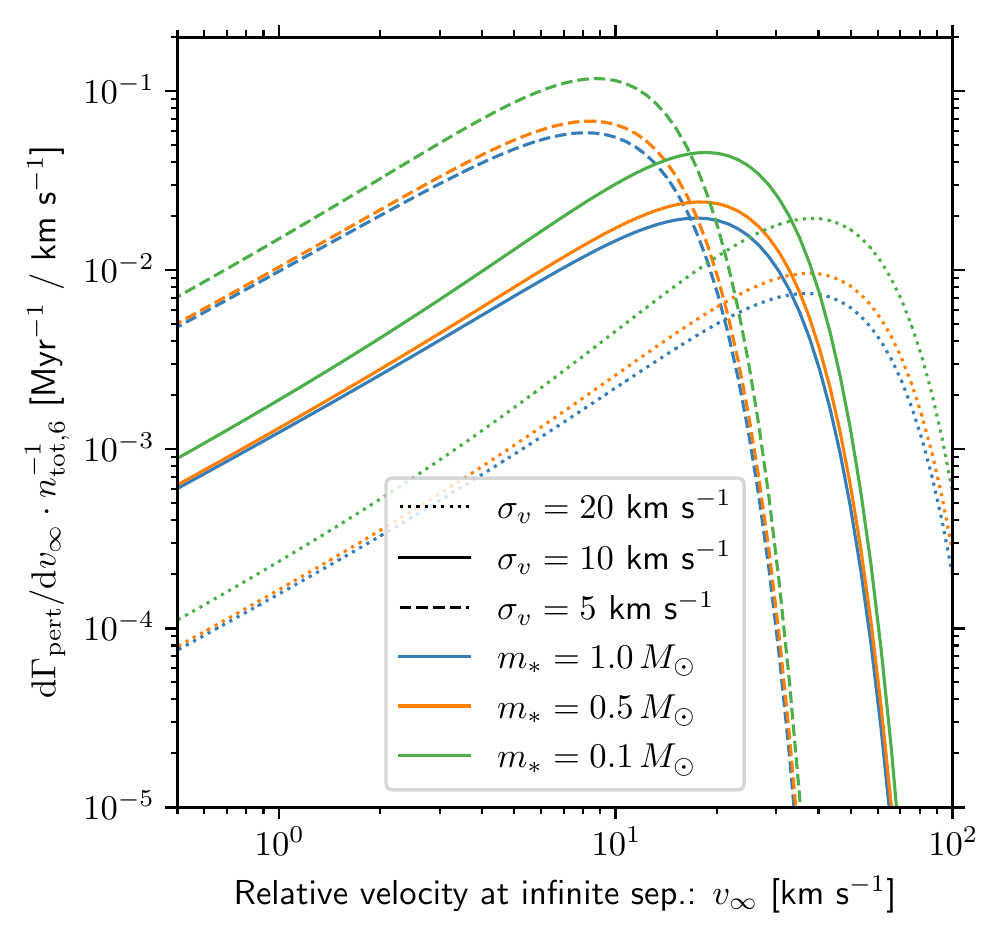}}
     \subfloat[\label{subfig:pertrate_sigv}Overall planet perturbation rate with velocity dispersion]{\includegraphics[width=\columnwidth]{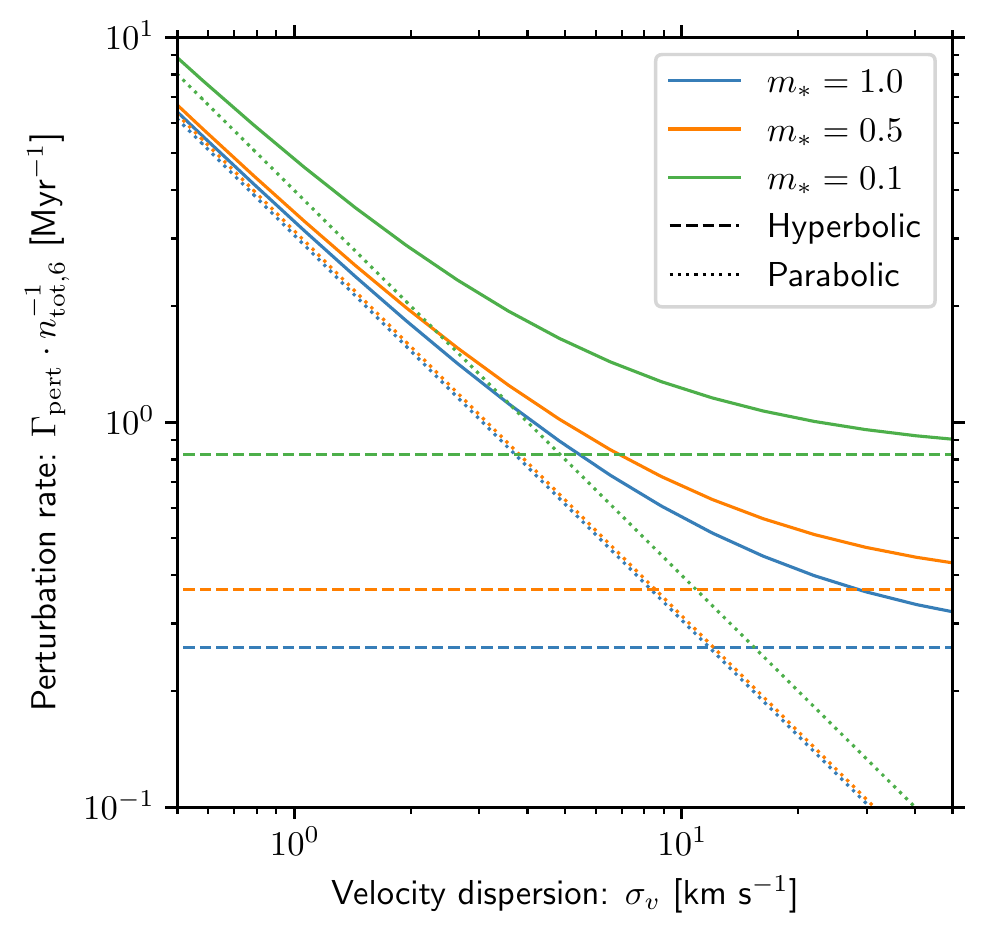}}
    \caption{The rate at which a planet is experiences an encounter that changes the eccentricity by $|\epsilon|>0.05$. In Figure~\ref{subfig:diffencrate} we show the differential perturbation rate of a planet orbiting with semi-major axis $a_\mathrm{0}=5$~au and eccentricity $e_0=0.9$ in a region of local density $n_{\rm{tot}}/10^6$~pc$^{-3} = n_{\rm{tot},6}=1$ and varying velocity dispersion $\sigma_v$, according to equation~\ref{eq:diffenc}. In each case, the differential rate is integrated over the perturbers adopting the initial mass function in our Monte Carlo model truncated above $5\, M_\odot$. Integrating these equations over relative velocities gives the overall encounter rates shown in Figure~\ref{subfig:pertrate_sigv}, divided into the hyperbolic (dashed) and parabolic (dotted) components, {while the solid line shows the sum of the two.}}
    \label{fig:diffencrate}
\end{figure*}

\subsubsection{Parabolic encounters}

 \citet{Heggie96} estimated the angle-averaged effective cross section for perturbation of the eccentricity $|\epsilon|$ {greater than some threshold eccentricity} $\epsilon_\mathrm{thr}$ of a binary with primary mass $m_*$, mass ratio $q$, initial semi-major axis $a_0$, eccentricity $e_0$ (their equation~19). This approximation is made by considering tidal and slow encounters, including only the quadrupole terms and taking the first order change in eccentricity:
\begin{equation}
    \epsilon \approx  \frac{\delta \bm{e} \cdot \bm{e}_0}{|\bm{e}_0|},
\end{equation}where $\bm{e}_0$ is the original eccentricity vector of the binary and $\delta \bm{e}$ is the change in this vector post-encounter. In the gravitationally focused limit, the cross section for interaction with a star of mass $m_\mathrm{pert}$ with relative velocity at infinite separation $v_\infty$ can be written \citep[see][]{Heggie96}:
\begin{align}
\nonumber
    \sigma_\mathrm{pert}^\mathrm{(foc)} &\approx 2\cdot \frac{9\sqrt{3}}{14 \pi}  \left( \frac{15\pi}{4} \right)^{2/3} \left[\Gamma\left(\frac{2}{3}\right)  \Gamma\left(\frac{5}{6}\right)\right]^2  \times
    \\ 
\nonumber & \quad \times \left(\frac{m_\mathrm{pert}^2}{(1+q)m_*m_\mathrm{tot} }\right)^{1/3} \frac{G m_\mathrm{tot}a_0}{v_\infty^2} e_0^{2/3} (1-e_0^2)^{1/3} \epsilon_\mathrm{thr}^{-2/3} \\
    &= C_\mathrm{foc} \cdot  \pi a_0^2 \cdot  \epsilon_\mathrm{thr}^{-2/3} y^{2/3} \left(\frac{v_\infty}{v_\mathrm{orb}}\right)^{-2} \left[1 +{q_\mathrm{pert}} \right],
\label{eq:sig_pert1}
\end{align} where $C_\mathrm{foc} \approx 2.73$ is a dimensionless constant and
\begin{equation}
  q_\mathrm{pert} =   \frac{m_\mathrm{pert}}{(1+q)m_*}.
\end{equation}We have defined:
\begin{equation}
   y \equiv e_0 \sqrt{1-e_0^2} \frac{m_\mathrm{pert}}{\sqrt{(1+q)m_* m_\mathrm{tot}}}, 
\end{equation}the planet orbital velocity is
\begin{equation}
    v_\mathrm{orb} = \sqrt{\frac{G m_\mathrm{*}(1+q)}{a_0}},
\end{equation}the total mass of the whole system is $m_\mathrm{tot}= m_*(1+q)+m_\mathrm{pert}$ and we include the first factor $2$ in the first expression of equation~\ref{eq:sig_pert1} because we initially do not care if $\delta  e$ is positive or negative.

\subsubsection{Hyperbolic encounters}

Equation~\ref{eq:sig_pert1} is derived assuming that the encounter is gravitationally focused (the eccentricity of the perturbing star is $e_\mathrm{pert}\approx 1$). However, in our context the three-dimensional velocity dispersion in the centre of 47 Tuc exceeds or is comparable to the typical orbital velocity $v_\mathrm{orb}\sim 10$~km~s$^{-1}$ (see \citetalias{PaperI}) and this may yield $e_\mathrm{pert}>1$ for the encounters of interest. The velocity dispersion is therefore too large to assume that focused encounters dominate the overall perturbation rate. The more general expression for $\epsilon$ due to a hyperbolic encounter is much less elegant, such that an equivalent of equation~\ref{eq:sig_pert1} must be derived numerically. We review the approach to this problem in Appendix~\ref{app:numeric_de}, where we show that the form of the hyperbolic equivalent of equation~\ref{eq:sig_pert1} is:
\begin{equation}
\label{eq:sigma_hyp}
      \sigma_\mathrm{pert}^\mathrm{(hyp)} =  C_\mathrm{hyp} \cdot \pi a_0^2 \cdot  y \epsilon_\mathrm{thr}^{-1}   \left(\frac{v_\infty}{v_\mathrm{orb}} \right)^{-1} \sqrt{1+ q_\mathrm{pert}},
\end{equation}where $C_\mathrm{hyp}$ is another dimensionless constant. The general perturbation cross section can then be approximated:
\begin{equation}
     \sigma_\mathrm{pert} \approx   \sigma_\mathrm{pert}^\mathrm{(foc)} \left[ 1+ \Delta_\mathrm{hf}  \frac{v_{\infty}}{v_{\mathrm{orb} }} \right]
\end{equation} for 
\begin{equation}
    \Delta_\mathrm{hf} = C_\mathrm{hf} {y}^{1/3} {\epsilon_\mathrm{thr}}^{-1/3} \left[1+ q_\mathrm{pert}\right]^{-1/2}.
\end{equation} We have introduced scaling parameter $C_\mathrm{hf}$, which we infer numerically to be $C_\mathrm{hf}\approx0.67$, or equivalently $C_\mathrm{hyp} = C_\mathrm{hf}\cdot C_\mathrm{foc} \approx 1.83$ (see Appendix~\ref{app:numeric_de}).

\subsection{Perturbation rate}
\label{sec:pert_rate}
From the cross sections derived in Section~\ref{sec:pert_theory}, we can estimate the perturbation rate $\Gamma_\mathrm{pert} = \tau_\mathrm{pert}^{-1}$ for a given local number density $n_\mathrm{tot}$ of (sub-)stellar objects and velocity dispersion $\sigma_v$. The  differential rate of perturbation is:
\begin{equation}
    \mathrm{d}\Gamma_\mathrm{pert} = v_\infty \, n_\mathrm{tot} \, \sigma_\mathrm{pert}(v_\infty) \, g(v_\infty; \sigma_v)  \xi (m_\mathrm{pert})  \, \mathrm{d} v_\infty \,  \mathrm{d}m_\mathrm{pert},
   \label{eq:diffenc}
\end{equation}where $n_\mathrm{tot}$ is the total local stellar density, $g$ is the $v_\infty$ distribution function that integrates to unity over all $v_\infty$ and $\xi$ is the mass function. {It is immediately clear by substituting equation~\ref{eq:sigma_hyp} into equation~\ref{eq:diffenc} that the perturbation rate for hyperbolic encounters is independent of the encounter velocity.} The overall differential encounter rate is shown in Figure~\ref{subfig:diffencrate} fixing $\epsilon_\mathrm{thr}=0.05$ {for illustration}. When computing perturbation rates, we hereafter adopt the initial mass function (IMF) similar to that used for the Monte Carlo simulation (see \citetalias{PaperI}):
\begin{equation}
\xi(m_*) \propto \begin{cases}  m_*^{-\alpha_1} \qquad & m_\mathrm{min} \leq m_* <m_\mathrm{br} \\
m_*^{-\alpha_2} \qquad &  m_\mathrm{br} \leq m_*\leq m_\mathrm{max}\\ 
0 \qquad \qquad & m_*>m_\mathrm{max} \, \mathrm{or} \, m_* < m_\mathrm{min}
\end{cases}    
\end{equation}for $\alpha_1=0.4$, $\alpha_2=2.8$, $m_\mathrm{br} = 0.8\,M_\odot$, $m_\mathrm{min}=0.08\, M_\odot$ and $m_\mathrm{max}=50\, M_\odot$ and normalisation constants such that $\xi$ is continuous and integrates to unity over all masses.  However, we modify the mass function such that $\xi$ is truncated above $m_\mathrm{max} = 5\, M_\odot$ to exclude short-lived OB stars (main sequence lifetimes shorter than $\sim 100$~Myr). {Note that these stars are still included and evolved for the sake of our Monte Carlo model in Section~\ref{sec:47Tuc}, thus contributing the local velocity dispersion.} The asymptotic relative velocity of two stars follows the Maxwell-Boltzmann distribution:
\begin{equation}
    g(v_\infty; \sigma_v) = \frac{v_\infty^2}{2\sqrt{\pi} \sigma_v^3} \exp \left(\frac{-v_\infty^2}{4\sigma_v^2} \right)
\end{equation}for three dimensional velocity dispersion $\sigma_v$.

The full encounter rate can be obtained by integrating equation~\ref{eq:diffenc}, as shown in Figure~\ref{subfig:pertrate_sigv} for varying velocity dispersion $\sigma_v$. The overall instantaneous perturbation rate can be written:
\begin{equation}
\label{eq:Gamma_pert}
    \Gamma_\mathrm{pert} = \Gamma_\mathrm{pert}^\mathrm{(foc)} +\Gamma_\mathrm{pert}^\mathrm{(hyp)}
\end{equation} where we have split the encounter cross section (and therefore the encounter rate) into a focused and hyperbolic component. The focused component is:
\begin{multline}
\label{eq:Gamma_pert_foc}
     \Gamma_\mathrm{pert}^\mathrm{(foc)} = 0.052\, (1+q)\epsilon_\mathrm{thr}^{-2/3} y'^{2/3}\mathcal{M}_{*}^\mathrm{(foc)} \frac{n_\mathrm{tot}}{10^6 \, \rm{pc}^{-3}} \frac{a_\mathrm{0}}{5\, \rm{au}} \times\\ \times \frac{ m_*}{1\, M_\odot} \frac{10 \, \rm{km s}^{-1}}{\sigma_v} \, \rm{Myr}^{-1}, 
\end{multline}
with
\begin{equation}
\mathcal{M}_{*}^\mathrm{(foc)}  = \int_0^\infty \, \mathrm{d}m_\mathrm{pert} \,{q_\mathrm{pert}^{2/3}} \left(1+ q_\mathrm{pert}\right)^{2/3}  \,\xi(m_\mathrm{pert}),
\end{equation} for
\begin{equation}
    y' = e_0 \sqrt{1-e_0^2}.
\end{equation} The hyperbolic component is
\begin{multline}
\label{eq:Gamma_pert_hyp}
     \Gamma_\mathrm{pert}^\mathrm{(hyp)} = 0.046 \, \sqrt{1+q}\epsilon_\mathrm{thr}^{-1} y' \mathcal{M}_{*}^\mathrm{(hyp)} \frac{n_\mathrm{tot}}{10^6 \, \rm{pc}^{-3}} \times \\ \times \,\left( \frac{m_* }{1\, M_\odot}\right)^{1/2} \,  \left(\frac{a_\mathrm{0}}{5\, \rm{au}}\right)^{3/2}  \, \rm{Myr}^{-1}, 
\end{multline} with
\begin{equation}
\mathcal{M}_{*}^\mathrm{(hyp)}  =  \int_0^\infty \, \mathrm{d}m_\mathrm{pert} \, q_\mathrm{pert}\, \xi(m_\mathrm{pert}) \approx 0.7 \left(\frac{m_*}{1\,M_\odot}\right)^{-1},
\end{equation}where the last approximation is true for our adopted IMF. 

The encounter rate described by equation~\ref{eq:Gamma_pert} is a shallow function of $\sigma_v$ and $m_*$ (cf. the tidal capture rate estimates in \citetalias{PaperI}). In the limit of large $\sigma_v$, $\Gamma_\mathrm{pert}$ is dominated by the hyperbolic component (equation~\ref{eq:Gamma_pert_hyp}), which is independent of $\sigma_v$ and only weakly dependent on $m_*$: $\Gamma_\mathrm{pert}^{\rm{(hyp)}}\propto m_*^{-1/2}$. The rate is somewhat dependent on the form of the mass function, but does not exhibit a precipitous decline for low host star masses as in the tidal capture case of \citetalias{PaperI}. 


\subsection{Consequences of the analytic perturbation rates}
\label{sec:cons_an}
The expressions for the focused and hyperbolic cross sections (encounter rates) are interesting for two reasons. First, there always exists some small change in eccentricity $\epsilon$ such that the growth of eccentricity is dominated by hyperbolic encounters for $\epsilon_\mathrm{thr} < \epsilon$. Dropping the `pert' subscript, we consider the perturbation rates $\Gamma^{\mathrm{(foc)}}(\epsilon)$ and $\Gamma^{\mathrm{(hyp)}}(\epsilon)$ as a function of a small change in eccentricity $\epsilon$. The relative contribution from the two types of encounters at $\epsilon$ can be written:
\begin{equation}
\label{eq:Gamma_rat}
   \left. \frac{\partial_\epsilon \Gamma^{\mathrm{(hyp)}}} {\partial_\epsilon \Gamma^{\mathrm{(foc)}}} \right. = \frac{3}{2} C_\mathrm{hf} y^{1/3} \left[1+q_\mathrm{pert}\right]^{-1/2}\frac{v_\infty}{v_\mathrm{orb}} \epsilon^{-1/3}.
\end{equation}Setting the LHS of equation~\ref{eq:Gamma_rat} equal to one, this gives a condition for the dominance of hyperbolic encounters:
\begin{equation}
\label{eq:epsilon}
    \epsilon < \epsilon_\mathrm{hyp} \equiv \left[\frac{3}{2} C_\mathrm{hf} \frac{v_\infty}{v_\mathrm{orb}}\right]^3 \left[1+q_\mathrm{pert}\right]^{-3/2} y.
\end{equation}
Given equation~\ref{eq:Gamma_pert_hyp}, we therefore obtain the important result that for a sufficiently large velocity dispersion the distribution of encounters in $\epsilon$ space is always the same. The total number of encounters scales linearly with local density, remaining independent of the local velocity dispersion.

While the condition in equation~\ref{eq:epsilon} is strongly dependent on $v_\infty$, for $v_\infty \gtrsim 2 v_\mathrm{orb}$ then $\epsilon_\mathrm{hyp} \gtrsim 1$ and eccentricity growth is always dominated by hyperbolic encounters (if ionising/resonant encounters are rare). In the context of 47 Tuc and a planet orbiting a solar mass star at semi-major axis $a = 5$~au, we are always in this regime. {In Figure~\ref{subfig:pertrate_sigv}, we show that the hyperbolic component dominates for $\sigma_v \gtrsim v_\mathrm{orb}\approx 10$~km/s for $\epsilon_\mathrm{thr}=0.05$. Here, $\sigma_v$ is the physical three dimensional dispersion. The one dimensional (radial) velocity dispersion towards the centre of 47 Tuc is $\sim 12$~km~s$^{-1}$ \citep{Gebhardt95}, thus the appropriate $\sigma_v$ is considerably higher ($\sigma_v \gtrsim 20$~km~s$^{-1}$ within $5$~pc of the centre -- see \citetalias{PaperI}). More generally, for sufficiently small $\epsilon$, hyperbolic encounters always dominate. When it comes to how a planet tidally circularises, we are interested in small changes in eccentricity $\epsilon\ll 0.05$.} These small changes in eccentricity will turn out to have a strong influence on circularisation. {Nonetheless, when $v_\mathrm{orb}$ exceeds $\sigma_v$ the hyperbolic encounter rate underestimates the rate of eccentricity change over long time-scales (see Section~\ref{sec:comp_numerical}).}

The second interesting consequence of the analytic perturbation rates is that the short-term eccentricity evolution is always dominated by the contribution of weak encounters (small $\epsilon$). As an illustration, we first assume that all encounters positively change the eccentricity. In this case, the rate of change of $e$ due to perturbations is:
\begin{equation}
  \dot{e} =  \int^\epsilon _{1-e} \tilde{\epsilon} \cdot  \partial_{\tilde{\epsilon}} \Gamma^{\mathrm{(hyp)}} \,\mathrm{d}\tilde{\epsilon}  \propto \ln\left(1-e\right) - \ln \epsilon.
\end{equation} This diverges as $\epsilon \rightarrow 0$, and we thus apparently have infinitely fast eccentricity growth as distant encounters are considered. Notice that this would not be the case if parabolic perturbations dominated. The solution to this apparent paradox is that we have both negative and positive changes to the orbital eccentricity. The evolution of the eccentricities can therefore be modelled as a random-walk (or diffusion) process, which strictly speaking must be treated in the continuum limit. This surprising result will be considered again in terms of our dynamical model in Section~\ref{sec:MC_mod}. In Section~\ref{sec:stat_evol} below, we consider the evolution of the probability distribution of the orbital eccentricity as a result of (infinite) distant encounters. 

\subsection{Statistical eccentricity evolution}

  \begin{figure*}
    \centering
    \subfloat[\label{subfig:ed_cont}IVP solution]{\includegraphics[width=\columnwidth]{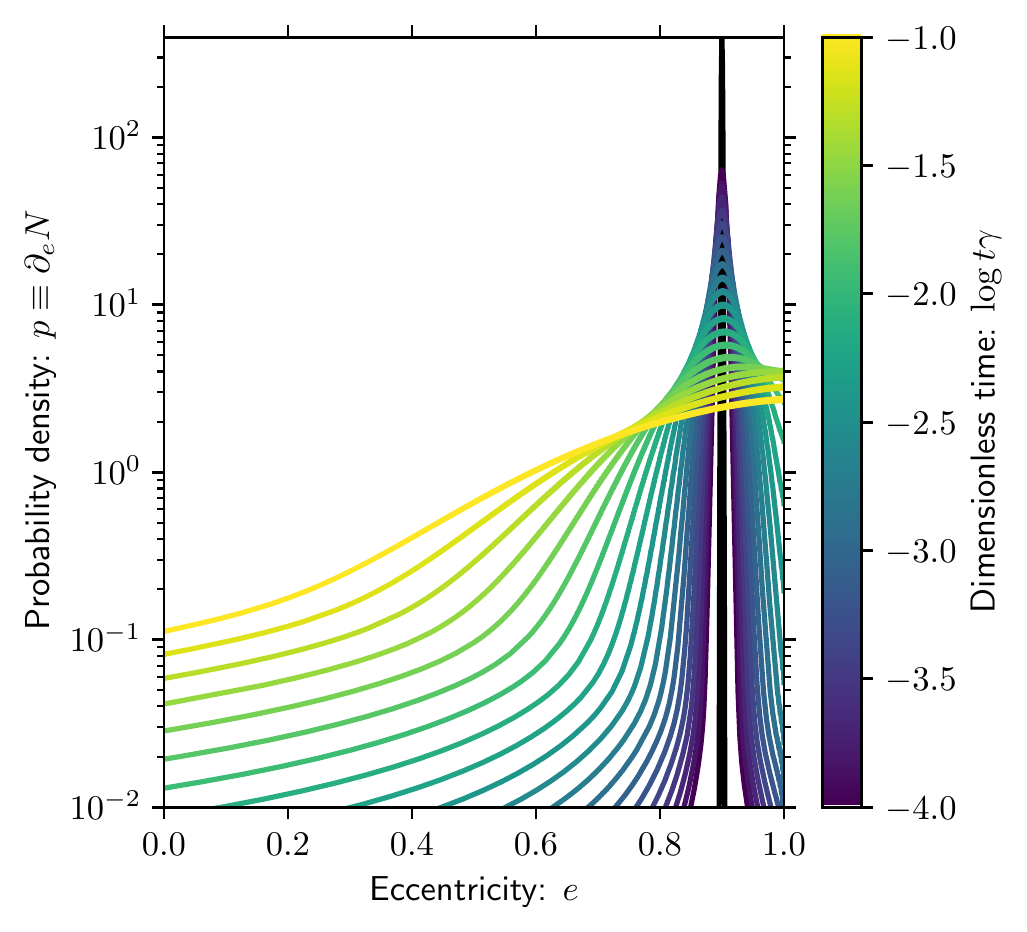}}
    \subfloat[\label{subfig:ed_rw}Random walk ($10^4$)]{\includegraphics[width=\columnwidth]{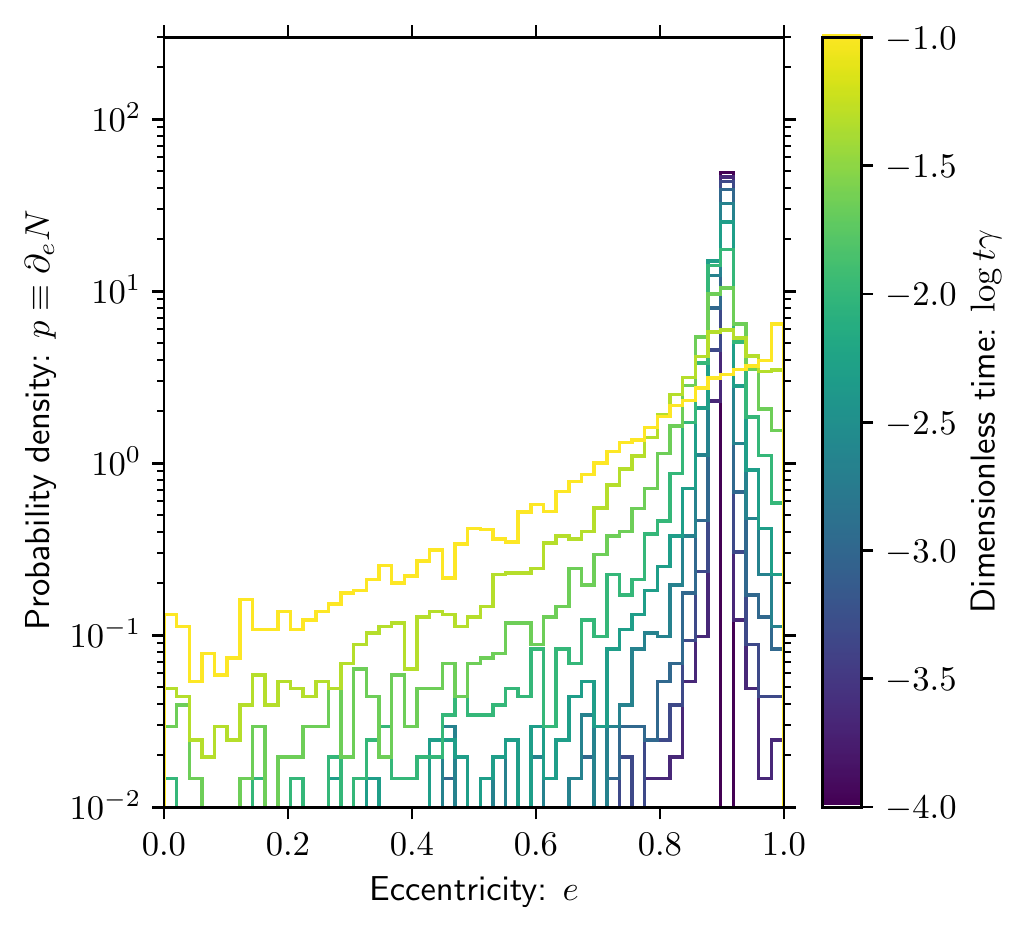}}
    \caption{Evolution of the probability density function $p(e)$ of a planets orbital eccentricity $e$ due to (hyperbolic) dynamical perturbations at fixed semi-major. The time $t$ is normalised by the encounter time-scale $\gamma^{-1}$ (equation~\ref{eq:gamma}), such that the results shown are independent of the absolute encounter rate. Figure~\ref{subfig:ed_cont} shows the result of numerically solving the initial value problem (IVP; equation~\ref{eq:partial_de}) for an initially Gaussian distribution with mean eccentricity $e_0=0.9$ and dispersion $\sigma_{e,0}= 10^{-3}$. Figure~\ref{subfig:ed_rw} is the same but for $10^4$ stochastic `random walk' experiments, where all eccentricities are initially $0.9$. Histogram bin sizes are $2\cdot 10^{-2}$ in eccentricity space. See text for details. }
    \label{fig:eccdiff_long}
\end{figure*}

\label{sec:stat_evol}
\subsubsection{Governing equations}
\label{sec:dpdt_e}
 {Given the encounter rate computed in Section~\ref{sec:pert_rate}, we can estimate the statistical evolution of the eccentricity of a star-planet system. It is straight forward to show from equations~\ref{eq:delta_e_general} and~\ref{eq:def_theta} that in the approximate prescription for the encounter cross section we have applied, the change in eccentricity $\epsilon$ due to an encounter has an equal probability of being positive or negative. In Appendix~\ref{app:stat_e_evol}, we show that if a planet initially has some known eccentricity $e_0$, then at time $t\lesssim \gamma^{-1}$ the probability density $p(e,t) \equiv \mathrm{d} N/\mathrm{d}e$ for $N$ the relative number of planets, follows something similar to the diffusion equation with an extra term:}
 \begin{multline}
 \label{eq:partial_de}
    \frac 2 {\gamma} \partial_t {p}(e, t) \approx  \, \partial_{e} \left[ e\sqrt{1-e^2}\partial_e p(e,t)\right] + \Delta (e, t),
 \end{multline} where we have defined the diffusion factor
 \begin{equation}
 \label{eq:gamma}
     \gamma \equiv  0.046 \, \sqrt{1+q} \mathcal{M}_{*}^\mathrm{(hyp)} \frac{n_\mathrm{tot}}{10^6 \, \rm{pc}^{-3}}  \left( \frac{m_* }{1\, M_\odot}\right)^{1/2} \,  \left(\frac{a_\mathrm{0}}{5\, \rm{au}}\right)^{3/2}  \, \rm{Myr}^{-1}, 
 \end{equation}and assumed we are always in the hyperbolic limit for perturbing encounters. 
 
 The last term in equation~\ref{eq:partial_de} is a non-local term:
 \begin{multline}
 \label{eq:Delta_I}
\Delta  \equiv       \lim_{\epsilon\rightarrow 0} \left\{\int_{e+\epsilon}^1 {\partial_{\tilde{e}} p(\tilde{e},t)} \frac{\tilde e\sqrt{1-\tilde e^2}}{\tilde{e}-e} \, \mathrm{d} \tilde e  \right. \\\left.- \int_0^{e-\epsilon} {\partial_{\tilde{e}} p(\tilde{e},t)} \frac{\tilde e\sqrt{1-\tilde e^2}}{e-\tilde{e}} \, \mathrm{d} \tilde e \right\} .
 \end{multline}This accounts for the non-zero possibility of large, instantaneous changes in eccentricity due to dynamical encounters, which is in contrast to the usual diffusion scenario, where only local properties of $p$ are important. Although the individual terms on the RHS of equation~\ref{eq:Delta_I} diverge as $\epsilon\rightarrow 0$, we show in Appendix~\ref{app:stat_e_evol} that the difference between them remains finite such that $\Delta$ is well-defined for all $e$ at which $p(e,t)$ is twice continuously differentiable in $e$. For a given initial distribution $p_0(e) = p(e,0)$, one can therefore solve equation~\ref{eq:partial_de} numerically for the time evolution of $p$. 
 
 \subsubsection{Eccentricity evolution and random walk comparison}
 \label{sec:IVP_vs_rw}

The long term evolution of the orbital eccentricity of a planet with an initial eccentricity $e_0=0.9$ is shown in Figure~\ref{fig:eccdiff_long}. We show results for the solution to the initial value problem (IVP) from equation~\ref{eq:partial_de} 
in Figure~\ref{subfig:ed_cont}. We solve the IVP over $1000$ grid-points between $0$ and $1$ in eccentricity space with the fourth order Runge-Kutta method of the \texttt{integrate.solve\_ivp} module from \textsc{Scipy} \citep{Virtanen20}. Boundaries are reflective at $e=0,1$ and we adopt the initial distribution:
\begin{equation}
\label{eq:p0}
    p_0(e) = p(e, 0) = \frac{1}{\sqrt{2\pi}\sigma_{e,0}} \exp\left(-\frac{(e-e_0)^2}{2\sigma_{e,0}^2} \right)
\end{equation} with $\sigma_{e,0}=10^{-3}$. We see that at early times, the non-local term is negligible and we approximately recover a solution similar to that of the standard diffusion with diffusion coefficient $D\approx \gamma/2 \cdot e\sqrt{1-e^2}$. This solution is just equation~\ref{eq:p0} with the substitution: 
\begin{equation}
\label{eq:diffusion_coeff}
    \sigma_{e,0}^2 \rightarrow 2Dt = \gamma t e\sqrt{1-e^2}. 
\end{equation}

When considering the influence of perturbations on tidal circularisation it will be convenient to apply a stochastic approach, {such that individual planet evolution scenarios subject to circularisation tides and external perturbation can be tracked.} While it is not possible to model arbitrarily small perturbations from the initial eccentricity, we can perform such an experiment if we are interested in changes of eccentricity greater than some $\epsilon_\mathrm{min}$. 

In order to compare the stochastic eccentricity evolution of a planet to the solution to the IVP, we perform a Monte Carlo experiment wherein we draw a set of random values $u_i \in (0,1)$ where $i=1,\dots, N_\mathrm{res}$. We then compare these with corresponding $\xi_i$ which are the probabilities of a perturbation in time-step of size $\Delta t$ of size $\epsilon_i$:
\begin{equation}
    \xi_i = \gamma e\sqrt{1-e^2}/2 \cdot  \Delta t \cdot \Delta \epsilon_i \cdot |\epsilon_i|^{-2},
\end{equation} where $\Delta \epsilon_i$ is the size of the $i^\mathrm{th}$ eccentricity perturbation bin. For $u_i<\xi_i$, an encounter perturbing the planet by an amount $\epsilon_i$ is assumed to have occured within the time-step of size $\Delta t$, and is added to the eccentricity for the following time-step. We repeat this for negative and positive $\epsilon_i$ -- i.e. $\epsilon_i\in (-e, -\epsilon_\mathrm{min})$ and $\epsilon_i\in (\epsilon_\mathrm{min}, 1-e)$ respectively. In this instance, the grid in perturbation size $\epsilon$ is uniform in $\epsilon$ and we adopt $\epsilon_\mathrm{min}=10^{-2}$ and $N_\mathrm{res}=10^4$. The time-step $\Delta t$ is chosen such as $ \xi_i <0.1$ everywhere in the grid (i.e. for $\epsilon_\mathrm{min}$). 

The results of $10^4$ iterations of this experiment are shown in Figure~\ref{subfig:ed_rw}. For eccentricity bin sizes $2\cdot 10^{-2}$ (larger than $\epsilon_\mathrm{min}$) we find good agreement with the direct integration of the IVP in Figure~\ref{subfig:ed_cont}. We conclude that this approach is valid for sufficiently large changes in eccentricity, subject to the choice of $\epsilon_\mathrm{min}$. This offers an efficient method for computing the evolution of a planet undergoing both perturbations and evolution due to tides, which we will apply again in Section~\ref{sec:rw_sims}.

\subsection{Tidal acceleration}
\label{sec:tidal_acc}
\subsubsection{Maximum circularisation radius}
We are interested in comparing the rate at which eccentricity is changed by external encounters to the rate at which tidal circularisation occurs. Since a planet undergoing purely pseudo-synchronous circularisation conserves the SLR, $l$, it is useful to consider the change in $l$ over the time $\tau_\mathrm{circ}$ that a planet takes to circularise as a result of dynamical perturbation. Given our analytic calculation of the eccentricity evolution as a result of hyperbolic encounters (Section~\ref{sec:stat_evol}), this change can be approximated:
\begin{equation}
    (\Delta l)^2 \approx \int_0^{\tau_\mathrm{circ}} \mathrm{d}t \, {\gamma  } e \sqrt{1-e^2}\cdot 4 a^2 e^2 , 
\end{equation} {where the squared term on the LHS comes from the diffusion coefficient in the short term solution to the IVP equation (see equation~\ref{eq:diffusion_coeff}).} We can rewrite this in terms of an integral over the semi-major axis $a$:
\begin{equation}
\label{eq:Delta_l}
    (\Delta l)^2 \approx \int_{a_0}^{l} \mathrm{d}a \, \frac{\gamma(a,e)}{\dot{a}} e \sqrt{1-e^2}\cdot 4 a^2e^2,
\end{equation}where $a_0$ is the initial semi-major axis, and eccentricity $e$ is a function of instantaneous semi-major axis $a$. Again, the SLR $l$ is conserved along the path integral. Now, we have $\dot{a}=\dot{a}_\mathrm{tide}$ and for convenience, we can rewrite equation~\ref{eq:adot_ps}:
\begin{equation}
    \dot{a} =\dot{a}_\mathrm{tide} = \alpha_\mathrm{tide}(e) a^{-7} e^2 (1-e^2)^{-15/2},
\end{equation} where 
\begin{equation}
\label{eq:alpha_tide}
    \alpha_\mathrm{tide} = -21 k_\mathrm{p} \tau_\mathrm{p} G m_*^2 M_\mathrm{p}^{-1} R_\mathrm{p}^5 f(e^2)
\end{equation} is  independent of  $a$. Unfortunately, it remains somewhat dependent on eccentricity via the function $f(e^2)$. However, we are primarily interested in large eccentricities, such that for $e\rightarrow 1$ we can estimate $f(e^2) \approx 4059/1120$ to simplify the integral. To make the semi-major axis dependence explicit in the integrand of equation~\ref{eq:Delta_l}, we also rewrite:
\begin{equation}
\tilde{\gamma} \equiv \gamma a^{-3/2}. 
\end{equation} With this we have 
\begin{equation}
\label{eq:Delta_l2}
    (\Delta l)^2\approx \int_{a_0}^{l} \mathrm{d}a \, \frac{4\tilde{\gamma}  a^{5/2} l^8}{ \alpha_\mathrm{tide} } = \frac{8 \tilde{\gamma} l^8}{7\alpha_\mathrm{tide}} \left( a_0^{7/2} - l^{7/2}\right),
\end{equation}where in the limit of large initial eccentricity $e_0$ we also have $a_0 \gg l$. 

Finally, our condition for circularisation comes from noting that when the change in eccentricity $\Delta l \rightarrow l$, then a planet becomes unlikely to circularise with semi-major axis $\sim l$. Hence, writing $\Delta l/l < 1$, we have an expression for the maximum allowed SLR:
\begin{equation}
    l_\mathrm{max} =   \mathcal{C}_\mathrm{\Delta} \left( \frac{7\alpha_\mathrm{tide} }{8 \tilde{\gamma} a_0^{7/2}} \right)^{1/6},
\end{equation} or more helpfully:
\begin{multline}
   \left( \frac{l_\mathrm{max}}{0.02\,\rm{au}} \right)^6 =  1.9 \, \mathcal{C}_\mathrm{\Delta}
    \, \frac{k_\mathrm{p}}{0.25} \frac{\tau_\mathrm{p}}{0.66 \,\rm{s}} \left(\frac{m_*}{1\,M_\odot}\right)^{2}\left(\frac{M_\mathrm{p}}{1\,M_\mathrm{J}}\right)^{-1}   \times \\ \times  \left(\frac{R_\mathrm{p}}{0.1\,R_\odot}\right)^5 
   \left( \frac{a_0}{5 \,\rm{au}}\right)^{-2}
   \left( \frac{ \gamma_0}{10^{-4} \,\rm{Myr}^{-1}}\right)^{-1},
   \label{eq:lmax}
\end{multline}where $\gamma_0 = \gamma(a_0)$. {We have introduced a correction term $\mathcal{C}_\mathrm{\Delta}$ that originates from the modification of the encounter rate $\gamma_0$ to incorporate the non-local term in the PDE that governs the probability density function for $e$ (i.e. $\Delta$ in equation~\ref{eq:partial_de}). This term becomes important for the largest values of the minimum eccentricity required to circularise:
\begin{equation}
\label{eq:emin}
   e_\mathrm{min} = \sqrt{1-\frac{l_\mathrm{max}}{a_0}}.
\end{equation}
As $e_\mathrm{min} \rightarrow 1$, the diffusive term in equation~\ref{eq:partial_de} at $l_\mathrm{max}$ becomes small. In this case, encounters that produce $l\lesssim l_\mathrm{max}$ may become dominated by the non-local term. Because $l_\mathrm{max}$ scales only with $\mathcal{C}_\Delta^{1/6}$ this correction factor only has a moderate (order unity) influence on $l_\mathrm{max}$ across any reasonable range of $\gamma_0$. We will initially adopt $\mathcal{C}_\Delta=1$, and revisit this value in Section~\ref{sec:circ_radii} and Appendix~\ref{app:rw_exp}. 
}

Substituting in $\gamma_0=\gamma$ from equation~\ref{eq:gamma} yields a version of equation~\ref{eq:lmax} that is similar (although not identical) to the estimate of the final semi-major axis given by equation 37 of \citetalias{Hamers17}, which is derived in a different way. The expression highlights that the maximum angular momentum (or SLR) at which a planet can circularise is very weakly dependent on the encounter rate parameter $\gamma$. {This is because the rate of tidal circularisation is very strongly dependent on $l$.} Nonetheless, the typical $l$ for HJs coincides with a sensible density range for dense stellar clusters. HJs have $l \sim 0.02{-}0.1$~au, with only a factor of a few in dynamical range; the shortest period HJ discovered has a semi-major axis of $0.0143$~au \citep{McCormac20}. For sensible encounter rates, $10^{-7}\,\rm{Myr}^{-1} \lesssim \gamma \lesssim  10^{-1} \,\rm{Myr}^{-1}$ ($1 \,\rm{pc}^{-3} \lesssim n_\mathrm{tot} \lesssim 10^6\,\rm{pc}^{-3}$), we are in the interesting regime, for which the typical HJ $l\sim l_\mathrm{max}$, given reasonable initial orbital, stellar and planetary properties. In general, density can vary by several orders of magnitude even within the same cluster environment, which may therefore result in a change of a factor several in the typical circularisation radii $l$.

\subsubsection{Critical stellar density}
We can recast equation~\ref{eq:lmax} into a critical number density, comparable to that inferred by \citetalias{Hamers17}. From equation~\ref{eq:gamma} we have $\gamma\propto n_\mathrm{tot}$, the local density. By assuming that HJ formation requires $l_\mathrm{max}\gtrsim 0.02$~au, we can rewrite equation~\ref{eq:lmax} to yield:
\begin{multline}
\label{eq:n_crit}
 n_\mathrm{crit} \sim 5\cdot 10^3\, \rm{pc}^{-3} \, \frac 1 {\mathcal{M}_{*} ^{\mathrm{(hyp)}}} \frac{k_{\rm{p}}}{0.25} \frac{\tau_\mathrm{p}}{0.66 \,\rm{s}} \left(\frac{m_*}{1\,M_\odot}\right)^{3/2} \left(\frac{M_\mathrm{p}}{1\,M_\mathrm{J}}\right)^{-1}   \times \\ \times  \left(\frac{R_\mathrm{p}}{0.1\,R_\odot}\right)^5
   \left( \frac{a_0}{5 \,\rm{au}}\right)^{-7/2},
\end{multline}where we have assumed $q\ll 1$. This density is comparable to that inferred by \citetalias[][see their equation 40]{Hamers17}. Unlike the expression in that work, our expression is not dependent on the local velocity dispersion or encounter radius because we have used the general expression for eccentricity evolution as a function of many distant, hyperbolic encounters. This has similarly allowed us to eliminate the initial pericentre distance, or equivalently eccentricity, as a free parameter. However, while the scaling in equation~\ref{eq:n_crit} is accurate, the normalisation is somewhat arbitrary due to the strong dependence on the choice of $l_\mathrm{max}$ threshold (equivalently, the initial pericentre distance in the derivation by \citetalias{Hamers17}). We will therefore consider a more useful metric, which is the destruction fraction itself; we quantify this fraction by first considering the distribution of circularisation radii due to encounters (Section~\ref{sec:circ_radii}) and then the conditions for HJ survival (Section~\ref{sec:HJ_surv}).

\subsubsection{Circularisation time-scale and the low density limit}

One can also interpret the threshold we have derived by considering the rate of acceleration of tidal circularisation. The change in $l$ from equation~\ref{eq:Delta_l} is dominated by encounters for maximal $a$ -- i.e. $a=a_0$. We therefore obtain a similar requirement for $l$ by considering the initial, instantaneous rate of change of the tidal circularisation time-scale:
\begin{equation}
\label{eq:tau_circ}
    \tau_\mathrm{circ} =  \left(\frac{\dot{a}_\mathrm{tide}^2}{a^2}  + \dot{e}_\mathrm{tide}^2\right)^{-1/2}
\end{equation} when $a\gg l$ and large $e$. If 
\begin{equation}
\label{eq:tidal_acc}
    \dot{\tau}_\mathrm{circ} = \frac{\gamma}{2} e \sqrt{1-e^2} \partial_e \tau_\mathrm{circ} \lesssim 1, 
\end{equation}{then the rate of change of the circularisation rate is slower than the rate of circularisation itself. This means that over the time required for a planet to circularise, $\tau_\mathrm{circ}$ (or equivalently the SLR, $l$) remains approximately constant.} Computing equation~\ref{eq:tidal_acc} yields a condition for $l$ similar to expression~\ref{eq:lmax}. With this interpretation, it is understood that HJ circularisation radii are limited by the initial rate of tidal acceleration relative to circularisation time-scale in a sufficiently dense cluster environment.

{With the definition of the circularisation time-scale, there exists a further constraint on the possible values of the SLR $l$ for circularisation to occur. In the limit of low $e$ (or large $a$), the time-scale for circularisation $\tau_\mathrm{circ}$ will exceed the age of the system. From equations~\ref{eq:adot_ps},~\ref{eq:edot_ps} and~\ref{eq:tau_circ}, we have:
\begin{equation}
\label{eq:tau_Hubble}
    \frac{\tau_\mathrm{age}}{\tau_\mathrm{circ}} = \tau_\mathrm{age} \cdot |\alpha_\mathrm{tide}|  \frac{e}{a^8 (1- e^2)^{13/2}} \sqrt{ \frac 1 4  + \frac{e^2}{(1-e^2)^2}}. 
    \end{equation}By requiring that the LHS of equation~\ref{eq:tau_Hubble} is greater than one, we have the requirement for circularisation:
    \begin{equation}
        l< l_\mathrm{age},
    \end{equation}where 
    \begin{multline}
    \label{eq:lage}
     \left( \frac{ l_\mathrm{age}}{0.02 \, \rm{au}}\right)^{15/2} = 2400 \frac{\tau_\mathrm{age}}{10 \, \rm{Gyr}} \cdot f(e^2) \frac{k_\mathrm{p}}{0.25} \frac{\tau_\mathrm{p}}{0.66\,\rm{s}} \left(\frac{m_*}{1\,M_\odot}\right)^2 \times \\ \times \left(\frac{M_\mathrm{p}}{1\,M_\mathrm{J}}\right)^{-1}  \left(\frac{R_\mathrm{p}}{0.1\,R_\odot}\right)^{5}  \left(\frac{a}{5\, \rm{au}} \right)^{-1/2} {e} \sqrt{ \frac {{(1-e^2)^2}} 4  + {e^2}},
    \end{multline} and we have substituted $\alpha_\mathrm{tide}$ from equation~\ref{eq:alpha_tide}:
    \begin{multline}
        \alpha_\mathrm{tide} \cdot \left({0.02 \, \rm{au}}\right)^{-15/2} 
        \cdot \left({5\,\rm{au}} \right)^{-1/2}= -240 \cdot f(e^2) \frac{k_\mathrm{p}}{0.25} \frac{\tau_\mathrm{p}}{0.66\,\rm{s}} \times \\ \times \left(\frac{m_*}{1\,M_\odot}\right)^2 \left(\frac{M_\mathrm{p}}{1\,M_\mathrm{J}}\right)^{-1}  \left(\frac{R_\mathrm{p}}{0.1\,R_\odot}\right)^{5} \, \rm{Gyr}^{-1}.
    \end{multline}
    Equation~\ref{eq:lage} is simplified by adopting $e\approx 1$ and thus $f = 4059/1120$ as before. In this way, we obtain the second maximal constraint $l_\mathrm{age}$ as a function of semi-major axis $a$. }

\subsection{Circularisation radii}
\label{sec:circ_radii}
\subsubsection{Random walk with circularisation}

\label{sec:rw_sims}

\begin{figure*}
    \centering
    \subfloat[\label{subfig:te_m4}$\gamma = 10^{-4}$~Myr$^{-1}$ at $a=5$~au]{\includegraphics[width=0.5\textwidth]{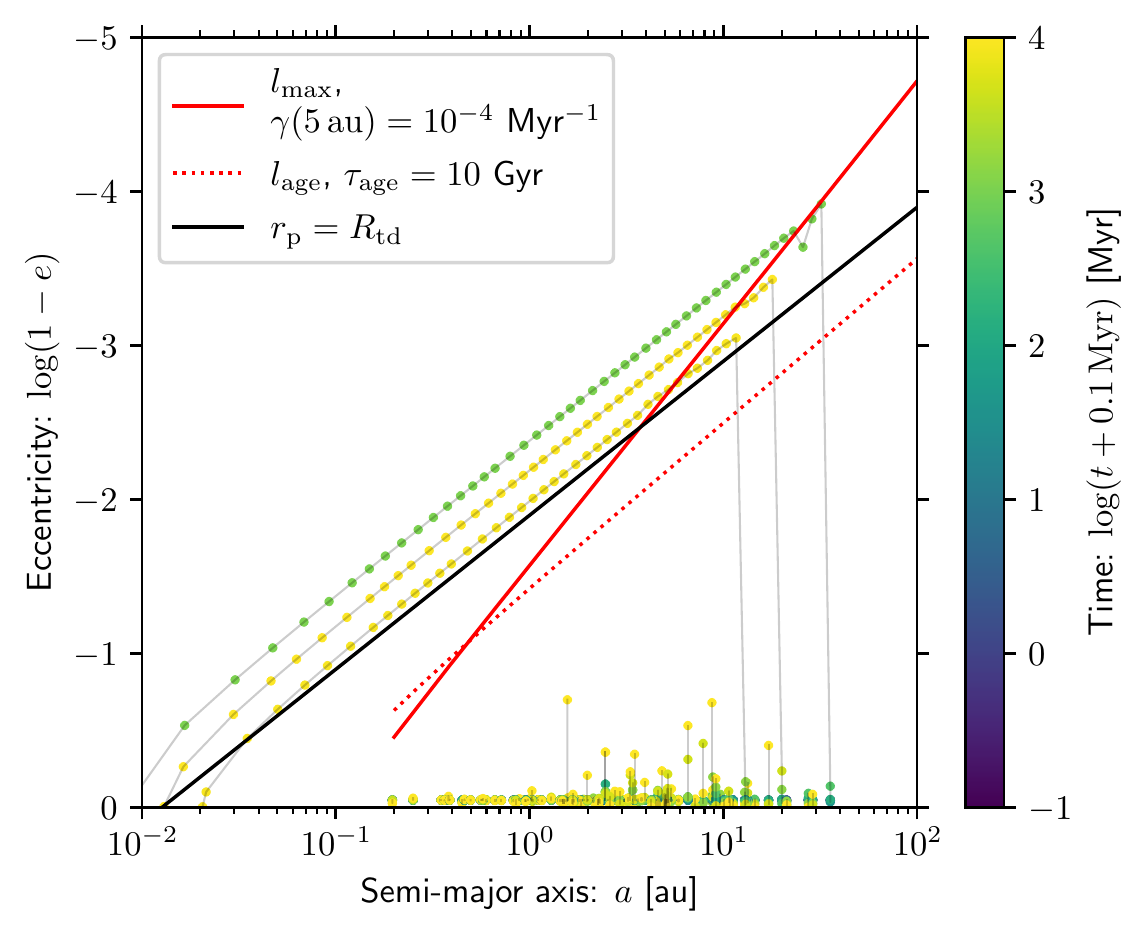}}
    \subfloat[\label{subfig:te_m3}$\gamma = 10^{-3}$~Myr$^{-1}$ at $a=5$~au]{\includegraphics[width=0.5\textwidth]{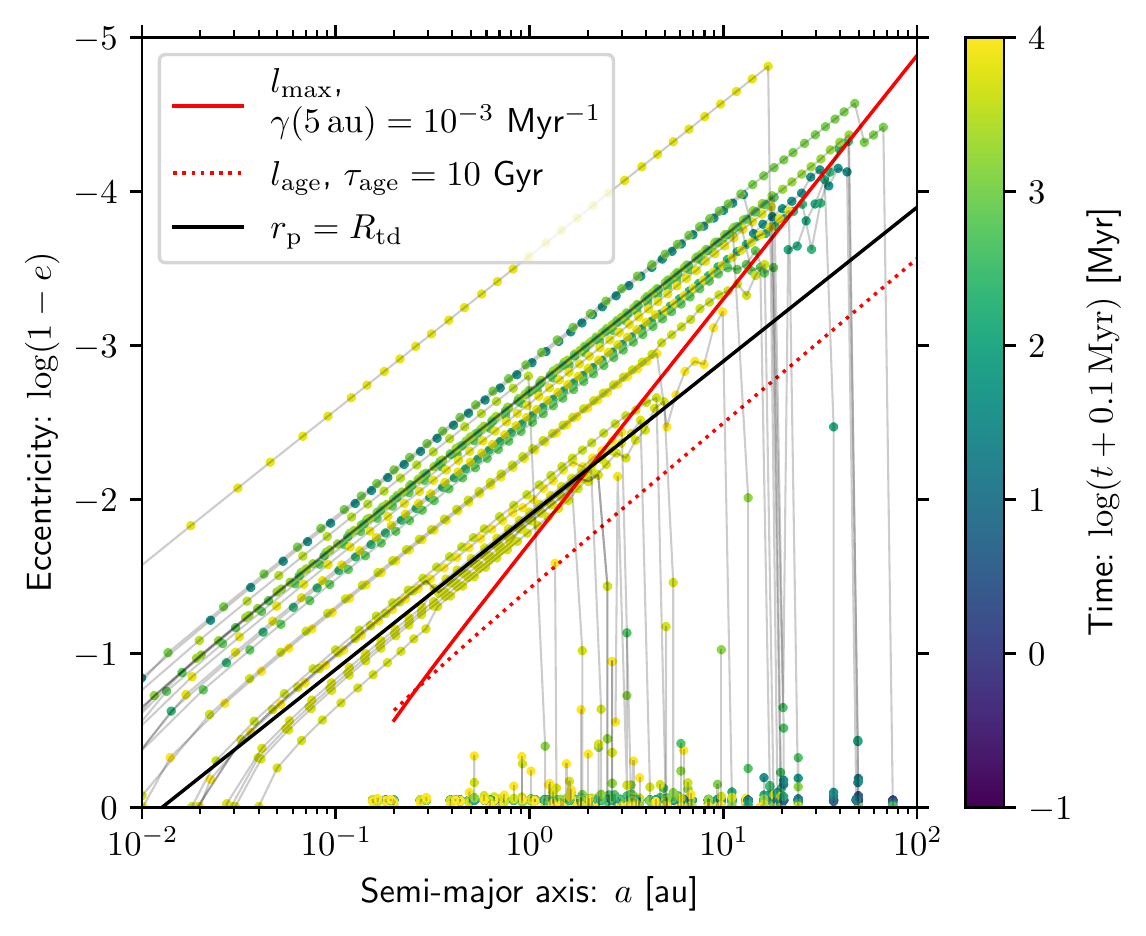}}\\
    \subfloat[\label{subfig:te_m2}$\gamma = 10^{-2}$~Myr$^{-1}$ at $a=5$~au]{\includegraphics[width=0.5\textwidth]{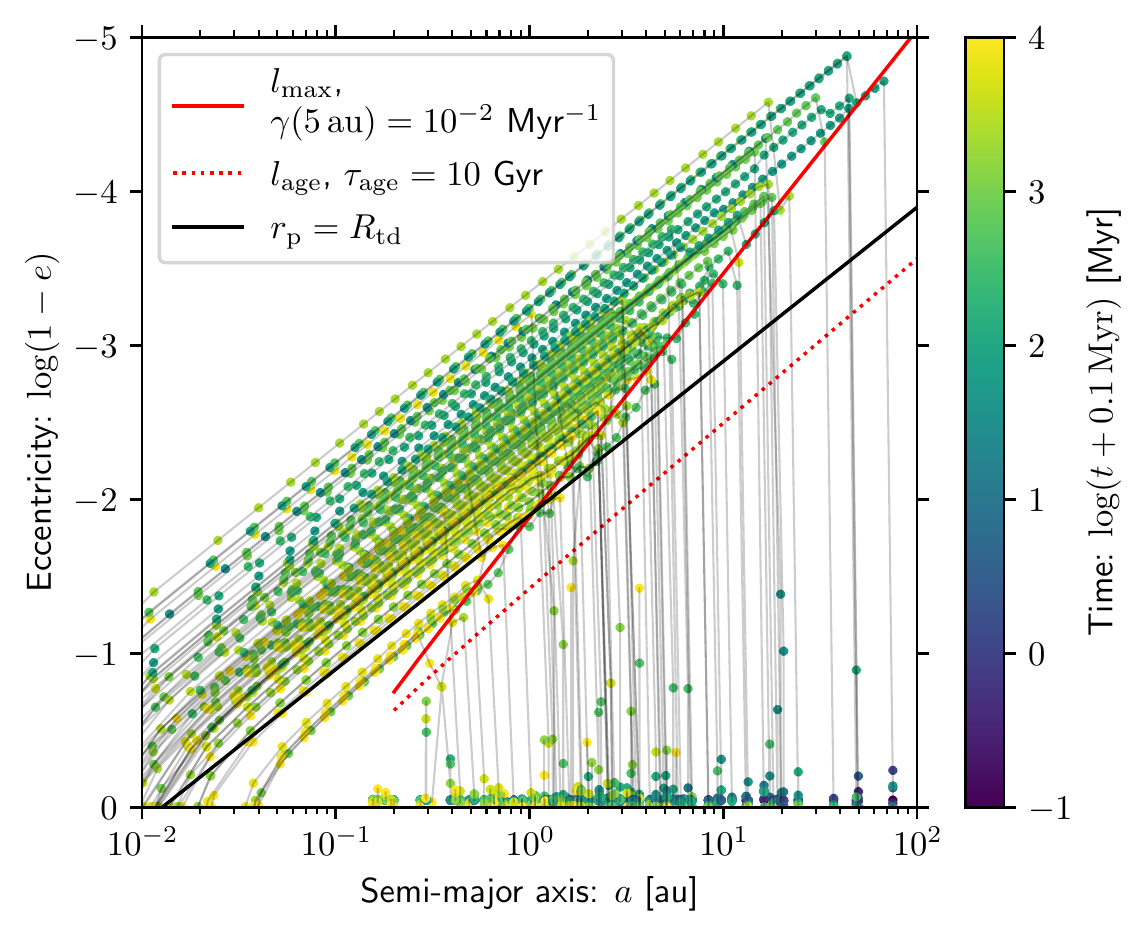}}
    \subfloat[\label{subfig:te_m1} $\gamma = 10^{-1}$~Myr$^{-1}$ at $a=5$~au]{\includegraphics[width=0.5\textwidth]{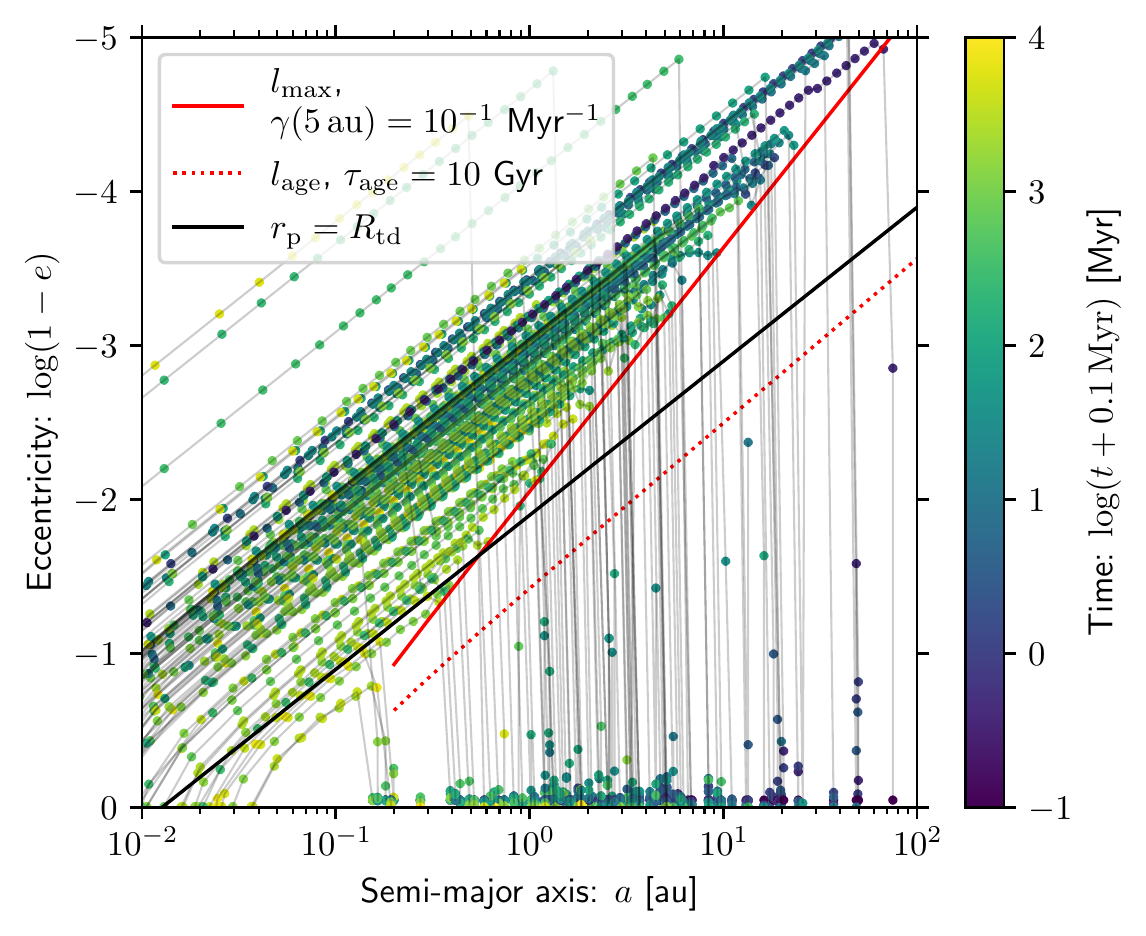}}
    \caption{The eccentricity $e$ and semi-major axis $a$ evolution resulting from one hundred numerical experiments for the stochastic perturbations to the orbital eccentricity with tidal circularisation. {Planets are initialised with a log-normal semi-major axis (mean $\overline{\log a/1\,\rm{au}} = 0.5$, 0.5~dex dispersion) and {initial eccentricity $e_0=0.1$}. They evolve due to perturbations, undergoing a random walk in the $y$-axis (eccentricity space). The planets are always subject to tidal forces from the host star, which try to circularise the planet while conserving SLR, $l=a(1-e^2)$. However, tides can only circularise the planet efficiently when $l< l_\mathrm{max}$, which we estimate analytically using equation~\ref{eq:lmax}, here with $\mathcal{C}_\Delta=1$, shown as a solid red line. The numerical experiment shows good agreement with the analytic expression. The dotted red line shows the maximum SLR, $l_\mathrm{age}$, for which a planet can circularise over the age of the system ($\tau_\mathrm{age}=10$~Gyr), using equation~\ref{eq:lage}.} Each panel shows a different value of the local encounter rate, parameterised by $\gamma$ at $5$~au. The time intervals are shown as coloured points by the time indicated in the colour bar, with faint connecting lines for each realisation. The black line shows the tidal destruction radius $R_\mathrm{td}$, discussed in Section~\ref{sec:HJ_surv}.  {Thus, most of the circularised HJs for these experiments would physically be tidally destroyed rather than circularise; circularisation requires slower encounter rates. However, we continue to evolve planets when pericentre distances $r_\mathrm{p}< R_\mathrm{td}$ because here we are only interested in the final circularisation radii due to the tidal forces and external eccentricity perturbations. } }
    \label{fig:ea_evol}
\end{figure*}

{In order for a planet to migrate via HEM, it must reach a sufficiently high eccentricity such that tidal forces act more quickly to shrink the orbit than external perturbations act to alter the eccentricity. This is achieved at $l_\mathrm{max}$, the SLR that yields a balance between tidal forces and perturbative encounter rates. If such an eccentricity is reached due to external perturbations, we have shown that the short term eccentricity evolution is a diffusive process. Thus we expect the SLR $l$ of the planet to fluctuate close to $l_\mathrm{max}$ before finally circularising. Thus, if the SLR is conserved under tidal circularisation, the typical circularisation radii for HJs approximately coincide with $l_\mathrm{max}$.}

We therefore proceed on the assumption that encounters play an important role in determining the circularisation radii of HJs. If initially very extreme eccentricities that yield {orbits that can circularise ($l > l_\mathrm{max}$) are rare}, then the distribution of circularisation radii $l$ {achieved by the diffusive eccentricity evolution} should only depend on the value of $l_\mathrm{max}$ (we use the SLR $l$ interchangeably with circularisation radius, since $a = l$ for $e=0$). Working on this principle, we perform the following numerical experiment to quantify the distribution of circularisation radii.

We perform a similar random walk calculation as described in Section~\ref{sec:IVP_vs_rw}, but this time include tidal circularisation. In order to ensure that all potentially important encounters are included, we this time adopt a minimum change in eccentricity $\epsilon_\mathrm{min} = 5\cdot 10^{-6}$ from a single encounter. {Here we define our grid of  $10^4$ $\epsilon$ values to be logarithmic spaced in $1-e$ between $\epsilon_\mathrm{min}$ and $1-\epsilon_\mathrm{min}$, such that changes that yield large eccentricities -- i.e. those in which we are primarily interested -- are well-resolved.}   Circularisation is treated with a $4^\mathrm{th}$ order Runge-Kutta scheme, following equations~\ref{eq:adot_ps} and~\ref{eq:edot_ps}, while the time-step is determined by the most stringent of several conditions. The first is simply the Courant–Friedrichs–Lewy condition for the rate of circularisation. Secondly, to ensure that the fractional change of semi-major axis is small so as to allow accurate computation of the encounter rate in a given time-step, we also ensure that the semi-major axis $|\Delta a| < a/10$. Finally, we must avoid rapid changes to the circularisation radius (rate) due to encounters within successive time-steps.  We achieve this by ensuring that encounters that give large change in the eccentricity $\Delta e_\mathrm{pert}$ due to all perturbations on a given time-step do not frequently (on successive time-steps) exceed certain values. Perturbations that give $\Delta e_\mathrm{pert} > 1-e_\mathrm{min}$ are considered too large. In addition, we ensure a dimensionless acceleration parameter remains small:
\begin{equation}
    C  = C_a+C_e < C_\mathrm{max}
\end{equation} where
\begin{equation}
\label{eq:C_cond}
    C_a = \Delta t \left| \frac{\dot{a}_{\mathrm{tide}}(a, e+\Delta e_\mathrm{pert}) - \dot{a}_{\mathrm{tide}}(e)}{\dot{a}_{\mathrm{tide}}(a,e)} \cdot \frac{\dot{a}_{\mathrm{tide}}(a,e+\Delta e_\mathrm{pert})}{{a}} \right|, 
\end{equation}and similar for $C_e$ but substituting $e, \dot{e}_\mathrm{tide}$ instead of $a,\dot{a}_{\mathrm{tide}}$. The value of $C$ is large if the acceleration of the circularisation rate and the resultant change of semi-major axis per time-step are large. By ensuring $\Delta t$ remains small enough to keep $C$ small, we ensure that we do not allow the planet to `walk' into and out of a region in $a{-}e$ space in which it should go through rapid circularisation within a single time-step. We must still allow for occasional large instantaneous changes in eccentricity, thus these conditions may be violated on a single time-step. However, such events should be infrequent -- i.e. not occurring on successive time-steps. We reduce $\Delta t$ by an order of magnitude if $C > C_\mathrm{max}=0.1$ or $\Delta e_\mathrm{pert} > 1-e_\mathrm{min}$. Testing with different thresholds yields similar results. 

For initial conditions, we draw semi-major axis from a lognormal distribution with a mean at $\log(a/1\,\mathrm{au}) = 0.5$ with $0.5$~dex scatter, and fix the initial eccentricity $e_0=0.1$. {We choose this to better reflect a sensible initial eccentricity of a planet \citep[e.g.][]{Dunhill13}, rather than one that has already been excited to high eccentricity by dynamical interactions.} We evolve each realisation for $12$~Gyr. The results of this experiment are shown in Figure~\ref{fig:ea_evol} for three different values of $\gamma(a)$ at $a=5$~au. The red line in each plot traces the value of $l_\mathrm{max}$ in $a{-}e$ space, assuming $\mathcal{C}_\Delta=1$ and adopting $a_0=a$. As $\gamma$ increases, $l_\mathrm{max}$ decreases. In each case, $l_\mathrm{max}$ traces the maximal possible circularisation radius. This suggests that the tidal acceleration condition discussed in Section~\ref{sec:tidal_acc} is appropriate.

One counter-intuitive consequence of the tidal acceleration condition is that it becomes difficult to circularise planets on wide orbits, despite higher perturbation rates. This is because $l_\mathrm{max}$ decreases with increasing $a$. {For systems that start at large $a$, or occupy dense regions and are therefore subject to more frequent perturbations, tidal inspiral only dominates over perturbations for values of $l$ that are so low that it becomes hard to avoid tidal destruction of the planet (see Section~\ref{sec:HJ_surv}).} Thus HEM migration may be efficiently suppressed at sufficiently high densities. 

\subsubsection{Analytic circularisation radius distribution}

\begin{figure}
    \centering
    \includegraphics[width=\columnwidth]{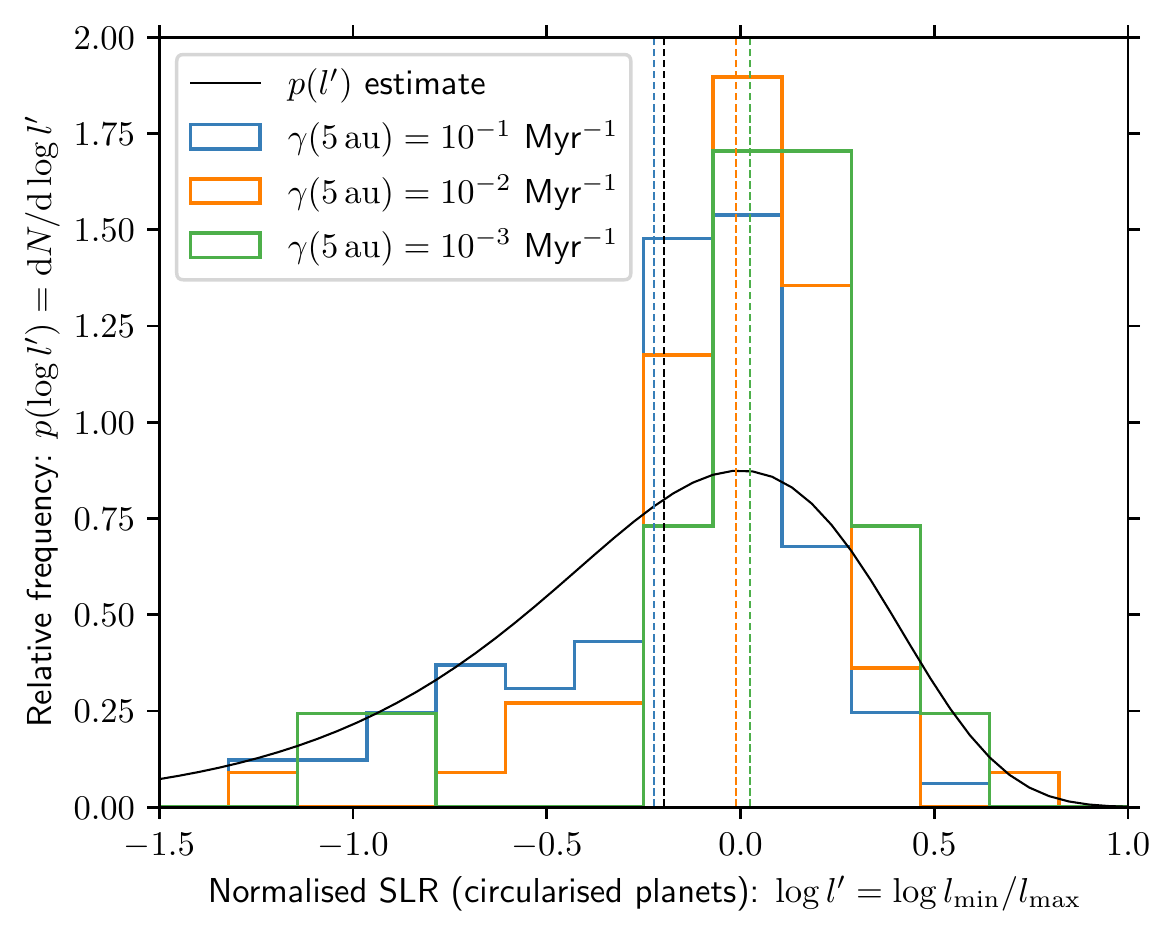}
    \caption{Histogram of the minimum semi-latus rectum (SLR, $l_\mathrm{min}$) distribution for the planets that eventually circularise (defined to have final $e<10^{-3}$, and $a<0.1$~au) in the suite of random walk simulations. The minimum SLR is normalised by the maximum value $l_\mathrm{max}$ predicted theoretically using equation~\ref{eq:lmax}. Our heuristic estimate for the distribution of $l' = l_\mathrm{min}/l_\mathrm{max}$, as described by equation~\ref{eq:prob_l}, is shown as a black line.  The logarithmic mean of each distribution is shown as a vertical  in the appropriate colour.  }
    \label{fig:l_dist}
\end{figure}

We are interested in how tidal acceleration influences the fraction of planets that do not survive circularisation. In order to do this, we must first relate the maximum value to the overall distribution of circularisation radii (or $l$ values). To this end, we first define the normalised SLR:
\begin{equation}
    l' \equiv l_\mathrm{min} /{l_\mathrm{max}}.
\end{equation} In this expression we have introduced $l_\mathrm{min}$, which is the lowest value of $l$ reached by a circularising planet. This value is of interest because if the majority of dynamical perturbations occur early during circularisation, then this quantity relates directly to the smallest pericentre distance reached. As discussed in Section~\ref{sec:HJ_surv}, this distance determines whether the would-be HJ survives. {In defining $l_\mathrm{max}$, we here adopt equation~\ref{eq:lmax} with the correction factor $\mathcal{C}_\Delta$ estimated as in Appendix~\ref{app:rw_exp}. }

We now search for sensible probability density function $p(l')$ that describes the distribution of $l'$ at any given perturbation rate $\gamma$. {In the absence of greater constraints, the functional form should have support in the range $(0, \infty)$ and yield $p(l')$ that drops to zero as $l'$ exceeds unity. We choose an exponential distribution, which is a maximal entropy solution that satisfies this condition:
\begin{equation}
\label{eq:prob_l}
    p(l') =\exp (-l') .
\end{equation}We compare the probability distribution given by our heuristic equation~\ref{eq:prob_l} to the distribution of $l'$ for circularised planets obtained from the numerical random walk experiment described in Section~\ref{sec:rw_sims}.} {We find that the median of the analytic distribution is always within one standard deviation of the random walk medians, and by eye we have reasonable agreement between the overall shape} across two orders of magnitude in the encounter rate (we exclude $\gamma(5\,\rm{au}) = 10^{-4}$ due to the low number of circularised planets in our simulations). {Given that we have achieved reasonable agreement with a simple and maximal entropy distribution without any degrees of freedom, we do not perform a formal statistical comparison with alternative models and fitting parameters.} We hence adopt equation~\ref{eq:prob_l} to simplify computation of the survival rates of HJs as follows.

\subsection{Hot Jupiter survival}
\label{sec:HJ_surv}

For sufficiently small pericentre distance $r_\mathrm{p}$ (or $l$), a would-be HJ may not survive the circularisation process. Of course, we immediately have $r_\mathrm{p}>R_*+R_\mathrm{p}$ to avoid collision. However, more stringently \citet{Guillochon11} estimate that to avoid destruction or ejection {(due to asymmetric mass removal)} they must have radius greater than:
\begin{align}
\nonumber
    r_\mathrm{p} > R_\mathrm{td} &= \eta R_\mathrm{Roche} \\ &= 6.8 \times 10^{-3} \cdot \eta \cdot \frac{R_\mathrm{p}}{0.1~\rm{au}} \left( \frac{m_*}{1\,M_\odot} \right)^{1/3} \left( \frac{M_\mathrm{p}}{1\,M_\mathrm{J}}\right)^{-1/3}\, \mathrm{au},
\end{align}where $R_\mathrm{Roche}$ is the Roche radius and $\eta \approx 1.87$ to give $R_\mathrm{td}\approx 0.013$~au for a solar mass star and a Jupiter-like planet. We show this tidal destruction contour in $a{-}e$ space as a black line in Figure~\ref{fig:ea_evol}. {However, at this stage we have allowed the planet orbits to evolve in the absence of tidal destruction, since we are interested in the final distribution of circulisation radii due to tidal forces and external eccentricity perturbations.}

We now wish to relate the survival condition to the distribution of circularisation radii. The pericentre of the circularising planet is related to the maximum SLR $l_\mathrm{max}$ of a planet with initial eccentricity $e_0$, semi-major axis $a_0$ by the expression:
\begin{equation}
    r_\mathrm{p, max} = l_\mathrm{max}/(1+e_0) = \frac{l_\mathrm{max}}{1+\sqrt{1-l_\mathrm{max}/a_0}}. 
\end{equation} For large $a_0\gg l_\mathrm{max}$ (requiring large $e_0$ for circularisation), we have  $r_\mathrm{p, max} \approx l_\mathrm{max}/2$. We then define:
\begin{equation}
    x \equiv \frac{r_\mathrm{p}}{R_\mathrm{td}} \approx \frac l {2 R_\mathrm{td}}.
\end{equation}We can write the maximum value of this ratio:
\begin{align}
\nonumber
    x_\mathrm{max} &= \frac{r_\mathrm{p, max}}{R_\mathrm{td}} \\ &\approx \frac{3.67}{\eta} \frac{l_\mathrm{max}}{0.05\,\rm{au}} \left(\frac{R_\mathrm{p}}{0.1\,\rm{au}}\right)^{-1} \left(\frac{m_*}{1\,M_\odot}\right)^{-1/3} \left(\frac{ M_\mathrm{p}}{1\,M_\mathrm{J}} \right)^{1/3} ,
\end{align} where $l_\mathrm{max}$ is evaluated via equation~\ref{eq:lmax} and the prefactor $3.67/\eta \approx 2$. Since $x\propto l$ and $x_\mathrm{max} \propto l_\mathrm{max}$, we can also define $x' = x/x_\mathrm{max}$ to give $p(x')\propto p(l')$. Now the fraction of circularising planets that are tidally destroyed is:
\begin{equation}
\label{eq:f_td}
    f_\mathrm{td} = \int_0^{\frac 1 {x_\mathrm{max}} }p(x') \mathrm{d}x' =1-\exp(-1/x_\mathrm{max}),
\end{equation} where the integral is evaluated using the form of $p(x')$ from equation~\ref{eq:prob_l}.

\subsection{Ionisation rate}
\label{sec:ionisation}
In the close encounter limit, the orbital energy as well as angular momentum is changed, which can lead to ionisation or exchange of the planet. The corresponding scattering cross-section $\sigma_\mathrm{ion}$ for equal mass components following \citet{Hut83} is:
\begin{equation}
\label{eq:Hut_83}
   \sigma_\mathrm{ion} \approx \frac{20}{9} \pi  q_\mathrm{pert}^{1/3} a_{0}^2 \left(\frac{v_{\rm{c}}}{v_\infty}\right)^2,
\end{equation}where $q_\mathrm{pert} = m_\mathrm{pert}/m_*$ and
\begin{equation}
    v_\mathrm{c}^2 =   \frac{G m_\mathrm{tot} }{a_0},
\end{equation} for total mass $m_\mathrm{tot}$. \citet{Fregeau04} find that the scattering cross-section is almost independent of the binary mass-ratio $q$. Finally, the scaling with $q_\mathrm{pert}^{1/3}$ is an approximation based on the scaling in the test particle limit for a gravitationally focused encounter \citep{Ostriker94,Breslau14, Winter18}. 

The cross section from equation~\ref{eq:Hut_83} can be converted to an encounter rate in the same way as in Section~\ref{sec:pert_rate} for the perturbation rate. That is:
\begin{equation}
    \mathrm{d}\Gamma_\mathrm{ion} = v_\infty n_\mathrm{tot} \sigma_\mathrm{ion} g(v_\infty;\sigma_v) \xi(m_\mathrm{pert}) \,\mathrm{d}v_\infty \mathrm{d}m_\mathrm{pert} 
\end{equation}to give:
\begin{multline}
\label{eq:Gamma_ion}
    \Gamma_\mathrm{ion} = 0.028 \mathcal{M}^{\rm{(ion)}}_* \left(\frac{ \sigma_v}{10 \,\mathrm{km\,s}^{-1}}\right)^{-1}\frac{m_*}{1\, M_\odot} \times \\ \times \frac{ a_0}{5 \, \rm{au}} \frac{n_\mathrm{tot}}{10^6 \, \rm{pc}^{-3}}  \, \mathrm{Myr}^{-1}.
\end{multline}Here we have defined:
\begin{equation}
    \mathcal{M}^{\rm{(ion)}}_*  = \int_0^\infty \mathrm{d}m_\mathrm{pert}\,  (1+q_\mathrm{pert}+q) q_\mathrm{pert}^{1/3}  \xi(m_\mathrm{pert}).
\end{equation}{Comparing to equation~\ref{eq:Gamma_pert_hyp}, we see that the ionisation rate is comparable to the rate at which extreme eccentricities are excited in a single encounter ($\epsilon_\mathrm{thr}\sim 1$).} We also see that the dominant type of encounter depends on the local velocity dispersion $\sigma_v$.

\subsection{Fractional outcomes}
\label{sec:fract_outcome}
\subsubsection{Analytic expressions}
\label{sec:analytic_Pdot}
We now consider the fraction of planets that are ionised compared to those that are circularised or tidally destroyed in an environment with fixed properties ($n_\mathrm{tot}$, $\sigma_v$). We first simply make the distinction between ionised planets and those influenced by tides, regardless of whether a planet is circularised or destroyed. The relevant probabilities satisfy:
\begin{equation}
    P_\mathrm{tide} + P_\mathrm{ion} = 1 - \exp\left[-(\Gamma_\mathrm{ion} + \Gamma_\mathrm{tide}) t \right],
\end{equation}where $P_\mathrm{tide}$, $P_\mathrm{ion}$ is the cumulative probability of a planet being tidally influenced or ionised respectively. We have also used the tidal rate:
\begin{equation}
\label{eq:Gamma_tide_main}
    \Gamma_\mathrm{tide} \approx \frac{\gamma_0 e_0\sqrt{1-e_0 }}{2 (e_\mathrm{tide}-e_0)}
\end{equation} for

\begin{equation}
    e_\mathrm{tide} = \mathrm{min} \left\{ \max\{e_\mathrm{min}, e_\mathrm{age} \}, e_\mathrm{td}\right\}
\end{equation}where
\begin{equation}
    e_\mathrm{min} = \sqrt{1-\frac{l_\mathrm{max}}{a_0}}
\end{equation} is the minimum eccentricity required for tidal forces to dominate over perturbative encounters and 
\begin{equation}
    e_\mathrm{age}= \sqrt{1-\frac{l_\mathrm{age}}{a_0}}
\end{equation} is the minimum eccentricity that allows circularisation over the age of the system. Finally

\begin{equation}
    e_\mathrm{td} = {1-\frac{R_\mathrm{td}}{a_0}}
\end{equation}is the eccentricity above which the planet will be tidally disrupted at periastron. {Equation~\ref{eq:Gamma_tide_main} is the rate at which individual encounters perturb a planet with initial eccentricity $e_0$ to eccentricity $>e_\mathrm{tide}$. This estimate is justified in Appendix~\ref{app:rw_exp}, in which we compare with the circularisation times obtained from our random walk experiments.}

With these rates, we can write the differential equations that govern each probability:
\begin{align}
    &\dot{ P}_\mathrm{tide}  = (1-P_\mathrm{tide}- P_\mathrm{ion} )\Gamma_\mathrm{tide}\\
    & \dot{ P}_\mathrm{ion}  = (1-P_\mathrm{tide}- P_\mathrm{ion} )\Gamma_\mathrm{ion}.
\end{align} The solutions to these equations are simply:
\begin{align}
\label{eq:P_tide}
   &P_\mathrm{tide}  = \frac{\Gamma_\mathrm{tide}}{\Gamma_\mathrm{tide} + \Gamma_\mathrm{ion}} \left\{ 1 - \exp\left[-(\Gamma_\mathrm{ion} + \Gamma_\mathrm{tide})t \right]  \right\}\\
\label{eq:P_ion}
  & P_\mathrm{ion}  = \frac{\Gamma_\mathrm{ion}}{\Gamma_\mathrm{tide} + \Gamma_\mathrm{ion}} \left\{ 1 - \exp\left[-(\Gamma_\mathrm{ion} + \Gamma_\mathrm{tide})t \right]  \right\}.
\end{align}Finally, we have:
\begin{equation}
\label{eq:P_HJ_td}
    P_\mathrm{HJ} = (1- f_\mathrm{td}) P_\mathrm{tide} \qquad  P_\mathrm{td} = f_\mathrm{td} P_\mathrm{tide} ,
\end{equation}where $P_\mathrm{HJ}$, $P_\mathrm{td}$ are the probabilities of HJ formation or tidal destruction respectively, and $f_\mathrm{td}$ is defined by equation~\ref{eq:f_td}. Equations~\ref{eq:P_tide},~\ref{eq:P_ion} and~\ref{eq:P_HJ_td} are an analytic approximation for the relative fractions of the three outcomes as a function of time in any dense stellar environment. 
\begin{figure}
    \centering
    \includegraphics[width=\columnwidth]{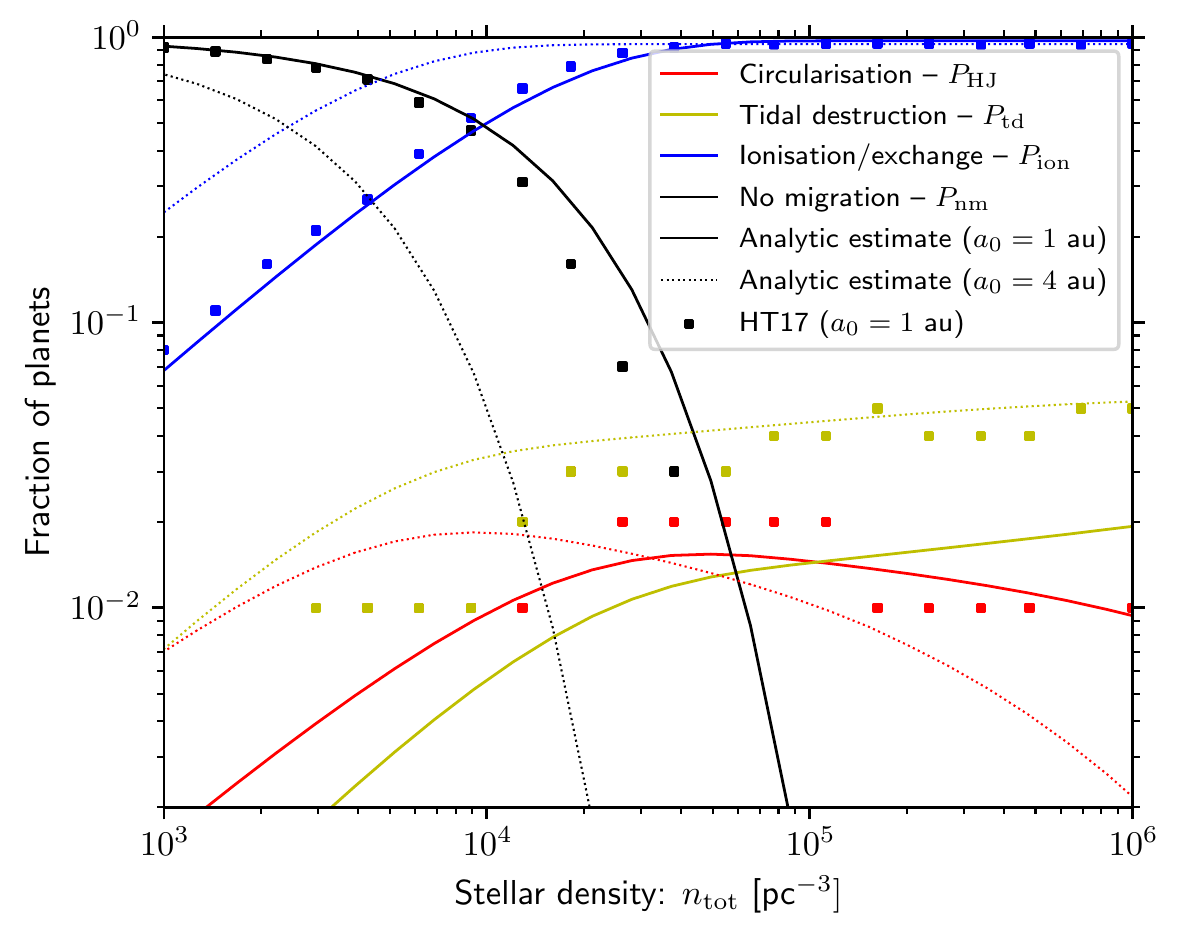}
    \caption{Direct comparison between the outcome probabilities for planets in a dense environment computed with the analytic approximation presented in this work (solid lines) versus those obtained by the numerical experiments of \citetalias[][square points]{Hamers17}. The results for are shown for initial semi-major axis $a_0=1$~au and $e_0=0.3$, stellar mass $m_*=1\,M_\odot$,  velocity dispersion $\sigma_v=6$~km~s$^{-1}$, planetary radius $R_\mathrm{p}=1\,R_\mathrm{J}$ and  mass $M_\mathrm{p}= 1\, M_\mathrm{J}$. The dotted lines are the theoretical results but for $a_0=4$~au, which show good agreement with the simulation results shown in Figure 5 of \citetalias{Hamers17}. For the simulations, the authors adopt an encounter radius $R_\mathrm{enc}=100$~au. Results are computed over a time interval of $10$~Gyr at each stellar density $n_\mathrm{tot}$, while the assumed mass function of perturbers is fixed to be appropriate at $5$~Gyr. Ionisation and transfer outcomes are shown as blue lines, red lines show planets that undergo circularisation (hot or warm Jupiters), yellow-green lines are for tidal destruction, while black lines are for the planets that undergo no significant migration. See text for details. }
    \label{fig:HTcomp}
\end{figure}

\subsubsection{Comparison to simulation results}
\label{sec:comp_numerical}

\citetalias{Hamers17} performed simulations tracking the evolution of massive planets under the influence of external perturbations and stellar tides for $10$~Gyr at constant local stellar density $n_\mathrm{tot}$ and velocity dispersion $\sigma_v$, including only encounters within a certain radius $R_\mathrm{enc}$. We here compare our analytic estimates to the outcomes of those experiements, particularly in Figure 5 of that work. The authors use a mass function that is initially that of \citet{Sal55}, but evolved for $5$~Gyr. We approximately reproduce this mass function by truncating the mass function above $m_\mathrm{pert} = 1.3\,M_\odot$. We adopt the same $\sigma_v=6$~km~s$^{-1}$ and $m_*=1\,M_\odot$. {The authors also adopt a Rayleigh distribution in $e_0$, with a rms of $0.33$ and truncated above $e_0=0.6$. We adopt $e_0 =0.3$ as a typical initial eccentricity.  }

The results of our analytic calculations are shown in Figure~\ref{fig:HTcomp} for initial semi-major axes $a_0=1$~au and $a_0=4$~au. We compare the results of the simulations performed by  experiment for the $a_0=1$~au case with $R_\mathrm{enc}=100$~au, as tabulated by \citetalias{Hamers17}. The $a_0=4$~au outcomes are not tabulated in that study, but can be visually compared with Figure 5 in that work. 

For the $a_0=1$~au case, we find that our analytic expressions qualitatively reproduce the numerical experiment outcomes. HJ production peaks around $n_\mathrm{crit}\sim 3\cdot10^4$~pc$^{-3}$, while the tidal destruction fraction continues to increase slowly with density. The only quantitative disagreement is that the analytic results under-predict the frequency of tidal destruction outcomes by a factor $\sim 2$. This may be due to the regime of the experiments performed by \citetalias{Hamers17}. In particular, the $a_0=1$~au case has large $v_\mathrm{orb} \gg \sigma_v$ in this case. Our treatment addresses encounters with $v_\mathrm{orb} \lesssim \sigma_v$, and we ignore parabolic encounters such that we may underestimate the number of perturbations that result in large eccentricity changes (see  Section~\ref{sec:cons_an}). {Comparison with the $a_0=4$~au results (with smaller $v_\mathrm{orb}$) again demonstrate good qualitative  agreement with all the outcomes. Quantitatively, we slightly overestimate the tidal outcomes (destruction and circularisation) by a factor of order unity.  This may be due to an overestimate of $\Gamma_\mathrm{tide}$ (see Appendix~\ref{app:rw_exp}), or the fact that we adopt a single initial eccentricity $e_0=0.3$ rather than the distribution of \citetalias{Hamers17}. Nonetheless, our results remain a good estimate and within the Poisson error ($0.02$ for the $2000$~realisations) of the Monte Carlo experiments of \citetalias{Hamers17}. We conclude that our analytic estimate is appropriate for sufficiently large $\sigma_v$ and/or $a_0$.}

\subsection{Summary}

In Section~\ref{sec:Num_Method} we have reduced the question of the statistical evolution of planetary systems in dense environments to an evaluation of the local conditions during their evolution. We therefore need only track these conditions within a given dynamical model to obtain fractional expected outcomes. We apply these results to a dynamical model for 47 Tuc in Section~\ref{sec:hj_ion}.

\section{Application to 47 Tuc}
\label{sec:47Tuc}
 \label{sec:hj_ion}
 \subsection{Summary of Section~\ref{sec:Num_Method}}
 {In Section~\ref{sec:Num_Method} we developed an analytic framework for computing the statistical likelihood of four possible outcomes for a planet orbiting a star in a dense stellar environment. The possible outcomes are:
 \begin{enumerate}
    \item \textit{No migration:} The orbital energy of the planet is not greatly altered from the initial energy. 
     \item \textit{Circularisation:} The orbital eccentricity is sufficiently excited by stellar encounters to yield a periastron distance of a few stellar radii. In this case, the planet feels strong tidal forces at periastron that allow it to migrate to a short-period orbit over its lifetime.
     \item \textit{Tidal disruption:} The rate of dynamical perturbation is so rapid that the planet cannot circularise in the usual way without already having undergoing further changes to its eccentricity. As the encounter rate increases, the periastron distance required for circularisation to act more quickly than dynamical perturbations decreases. For sufficiently fast encounter rates, the planet becomes tidally disrupted at closest approach with the host star.
     \item \textit{Ionisation/exchange:} The star-planet system is disrupted due to a close encounter that imparts kinetic energy and unbinds the planet. 
 \end{enumerate}  }
 
 {Having analytically quantified the relative frequencies of each of these outcomes, given system and environmental properties, we benchmarked our results against numerical experiments. Given the agreement between the analytic expectation and the simulation results, we here apply our analytic framework to a dynamical model of 47 Tuc. For this globular cluster there exist observational constraints on the fraction of HJs.}
 
 \subsection{Summary of observational constraints}
\label{sec:obs_comp}

In this section we consider the expected efficiency of HJ production with respect to the observational constraints. In terms of the latter, the transit survey by \citetalias{Gil00} is discussed in some detail in \citetalias{PaperI}. In brief, $34,091$ stars were observed with visual magnitudes in the range $17.1 <V < 21.1$, bounded by sensitivity constraints and the requirement that all target stars remain on the main sequence. This corresponds to stellar masses $0.52\, M_\odot < m_* < 0.88 \, M_\odot$ \citep{Bergbusch92} and the targets were typically separated from the cluster centre by a (projected) distance of $\sim 1$~pc. The authors assumed that the field occurrence rate of HJs is $1$~percent \citep{Wright12}, then adopted a planet radius $R_\mathrm{p} = 1.3 \, R_{\rm{J}}$ and a typical period of $3.5$~days, with a ten percent chance of geometric transit. This yielded an expected number of detected HJs within the sample to be $17$. Given the non-detection of any HJ in their sample, the authors concluded that the HJ fraction in 47 Tuc is lower than in the field with high statistical significance. 

However, a number of considerations have since been highlighted that reduce the significance of this finding. In the first instance, HJ incidence inferred from \textit{Kepler} planets was estimated by \citet{Howard12} to be $0.5\pm 0.1$~percent. This fraction is lower than the $\sim 1.2$~percent inferred from an RV surveys by \citet{Marcy05} and \citet{Wright12}. A slightly lower fraction of $0.89\pm 0.36$~percent was inferred by \citet{Mayor11}, where HJs were defined to have masses $M_\mathrm{p}>50\, M_\oplus$ and periods $P_\mathrm{orb}< 11$~days. This is approximately consistent with the value inferred by \citet{Howard12}. The \textit{Kepler} detection rate may additionally be somewhat reduced by pipeline considerations such as the harmonic filter applied to remove periodic stellar activity, affecting transit recovery for planets with $P_\mathrm{orb}\lesssim 3$~days \citep{Jenkins10, Christiensen15}. Meanwhile, the statistics on the distribution of stellar properties and planet radii are challenging to recover from the RV data, for which many detections do not transit. 

\citet{Masuda17} revisited the finding by \citetalias{Gil00} using updated statistics for HJs around \textit{Kepler} planets, including the distribution of radii and orbital periods. This is the most direct comparison that has been made between the surveyed 47 Tuc sample and the \textit{Kepler} planets, finding an expected number of detections of $ N_\mathrm{det}= 2.2_{-1.1}^{+1.6}$. This corresponds to an occurrence rate $f_\mathrm{HJ}$ of $0.43^{+0.07}_{-0.06}$~percent among \textit{Kepler} targets with similar masses as the \citetalias{Gil00} sample. Based on this estimate, we can therefore write the expected number of detections in 47 Tuc:
\begin{equation}
   \langle  N_\mathrm{det} \rangle \approx 2.2  \cdot  \frac{f_\mathrm{HJ}}{4.3\times 10^{-3}}  \cdot  \frac{N_\mathrm{samp}}{3.4 \cdot 10^{4}},
\end{equation} where $N_\mathrm{samp}$ is the sample size and the stellar properties and sensitivity are similar to those of \citetalias{Gil00}. The corresponding probability of obtaining no detection is:
\begin{equation}
    P_\mathrm{nd} = { N_\mathrm{\mathrm{samp}} \choose 0}  \left( 1- f_\mathrm{det}\right)^{N_\mathrm{samp}}
\end{equation}where 
\begin{equation}
    f_\mathrm{det} = \frac{\langle  N_\mathrm{det} \rangle }{N_\mathrm{samp}}.
\end{equation}This yields $ P_\mathrm{nd} \approx 0.11$ for $N_\mathrm{samp} = 3.4 \cdot 10^4$. Including the $21 \,920$ stars surveyed in 47 Tuc by \citet{Weldrake05}, assuming comparable sensitivity, yields $P_\mathrm{nd} \approx 0.027$ -- i.e. a marginally significant suppression with respect to the field population. 

We can define the maximum $f_\mathrm{HJ}$ that is consistent with the constraints:
\begin{equation}
    f_\mathrm{HJ, max} \approx 2 \cdot 10^{-3}  \cdot {3.4 \cdot 10^{4}}\left(1 - P_\mathrm{nd, max}^{1/N_\mathrm{samp}}  \right).
\end{equation}We will generally adopt $P_\mathrm{nd, max}=0.05$ ($2\,\sigma$ significance) to give:
\begin{equation}
\label{eq:fHJ_max}
     f_\mathrm{HJ, max} \approx 2 \cdot 10^{-3} \frac{10^{5}}{N_\mathrm{samp}}, 
\end{equation}or $f_\mathrm{HJ, max} \approx 3.6\times 10^{-3}$ for $N_\mathrm{samp} = 5.6 \cdot 10^4$. We will adopt the approximate expression equation~\ref{eq:fHJ_max} for the remainder of this work, with the caveat that future surveys must consider the sensitivity and stellar properties of their sample when computing significance of non-detections \citep[as in][]{Masuda17}.

{When comparing the observational constraints to the formation efficiency of HJs from our simulation,  we will first adopt the extreme assumption that {the number of massive planets per star $N_\mathrm{mp} =1$ for every $a_0$ -- i.e. that} 100~percent of stars host planets with a given semi-major axis $a_0$. Our results can then be generalised by multiplying the fraction of planets that circularise in our simulations by the expected initial occurrence rate within some range {$\delta a_0$ around $a_0$ -- i.e. $\delta a_0 \cdot \mathrm{d}N_\mathrm{mp}/\mathrm{d}  a_0$}. We reconsider our findings in terms of inferred occurrence rates for field stars in Section~\ref{sec:occ_rates}.   }
 
 \subsection{Dynamical model}
\label{sec:MC_mod}

We model the dynamical evolution of 47 Tuc using the Monte Carlo code \textsc{Mocca} \citep{Hyp13,Giersz13} with parameters motivated by the findings of \citet{Gie11}. The model, including the density and velocity dispersion evolution, is discussed in \citetalias{PaperI}. In brief, the model is initiated with $2\cdot 10^6$ stars and an equal number of brown dwarfs and is evolved for 12~Gyr to yield density and velocity distributions that consistent with the present day 47 Tuc.

In this model, we include a population of $2\cdot 10^4$ `migrating planets', {which we use for benchmarking only, and not to compute the tidal outcomes for which we apply the analytic expressions derived in Section~\ref{sec:Num_Method}.} The planets in the simulation are initiated with eccentricity $e_0=0.9$ and semi-major axes $a_0 = 5$~au, paired with stars drawn from the same IMF as single stars. While the stellar evolution is calculated using the code by \citet{Hurley00, Hurley02} in \textsc{Mocca}, we do not include tidal forces in the orbital evolution of the binaries or planetary systems. 

This population cannot be used to directly compute the eccentricity evolution and/or circularisation rates over the lifetime of 47 Tuc. The reasons for this are discussed in detail in Appendix~\ref{app:analytic_vs_MC}. In brief, the sampling framework in \textsc{Mocca} does not allow arbitrarily weak encounters in a given time-step. One therefore loses the encounters that yield small absolute changes in eccentricity $|\epsilon| \lesssim 0.05$, which are those in which we are most interested in this context. To capture such encounters accurately would require a much smaller time-step that would make the Monte Carlo simulations impracticable, essentially becoming an N-body simulation similar to that of \citetalias{Hamers17}. We therefore simply show in Appendix~\ref{app:analytic_vs_MC} that the relative number of encounters resulting in a change of eccentricity of magnitude $|\epsilon|$ scales with $|\epsilon|^{-2}$ for sufficiently large $|\epsilon|$. This is as expected from the theoretical hyperbolic cross sections, as described in Section~\ref{sec:pert_theory}. {The power-law index of $-2$ rather than $-1$ comes from the derivative of the cross section with respect to $\epsilon$, since the cross section pertains to encounters of \textit{at least} $\epsilon$ (or $\epsilon_\mathrm{thr}$).} In conjunction with the comparison to the simulation results of \citetalias{Hamers17} presented in Section~\ref{sec:comp_numerical}, this validates our analytic treatment such that we can apply it to our dynamical model. 

\subsection{Mass segregation}

\begin{figure*}
\subfloat[$100$~Myr\label{subfig:mrf_100Myr}]{\includegraphics[width=0.33\textwidth]{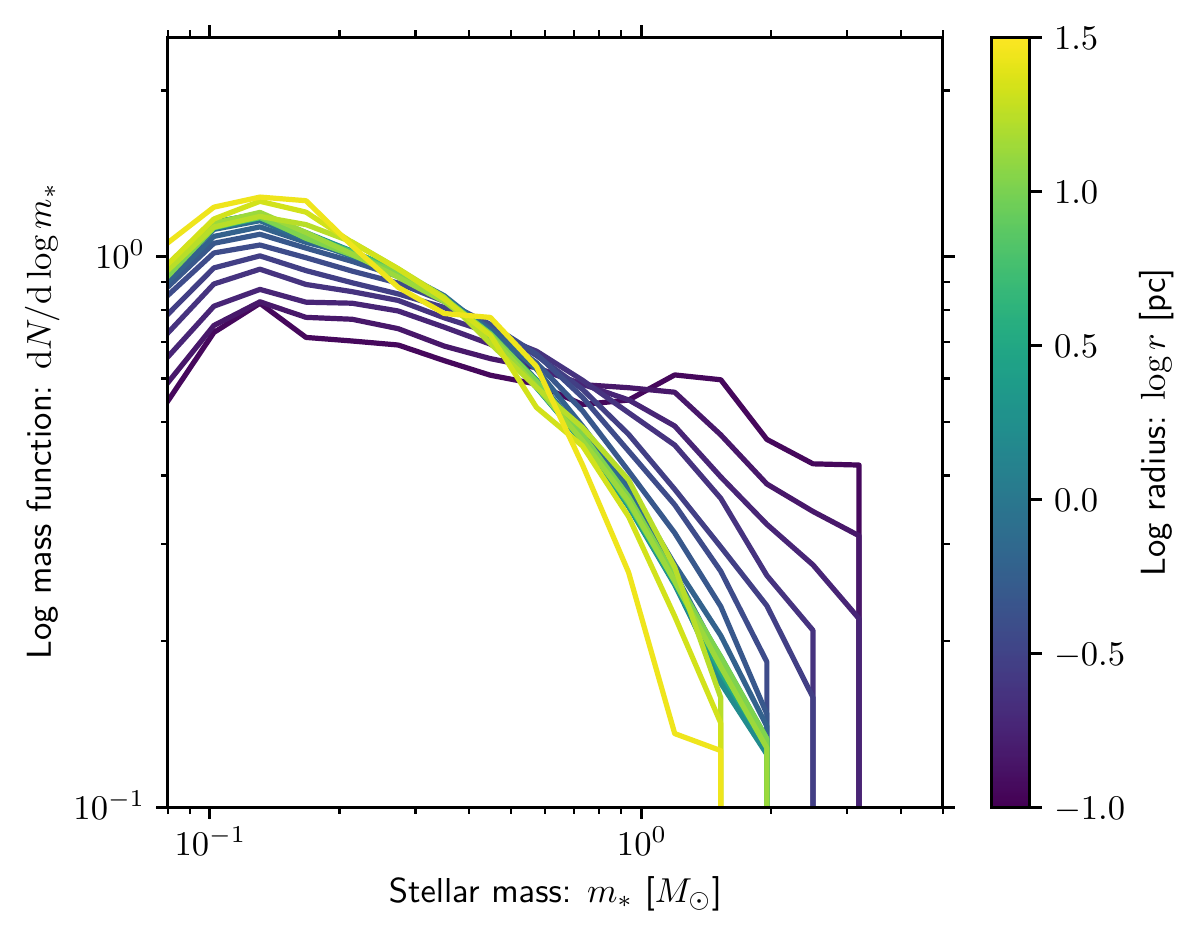}}
\subfloat[$1$~Gyr\label{subfig:mrf_1Gyr}]{\includegraphics[width=0.33\textwidth]{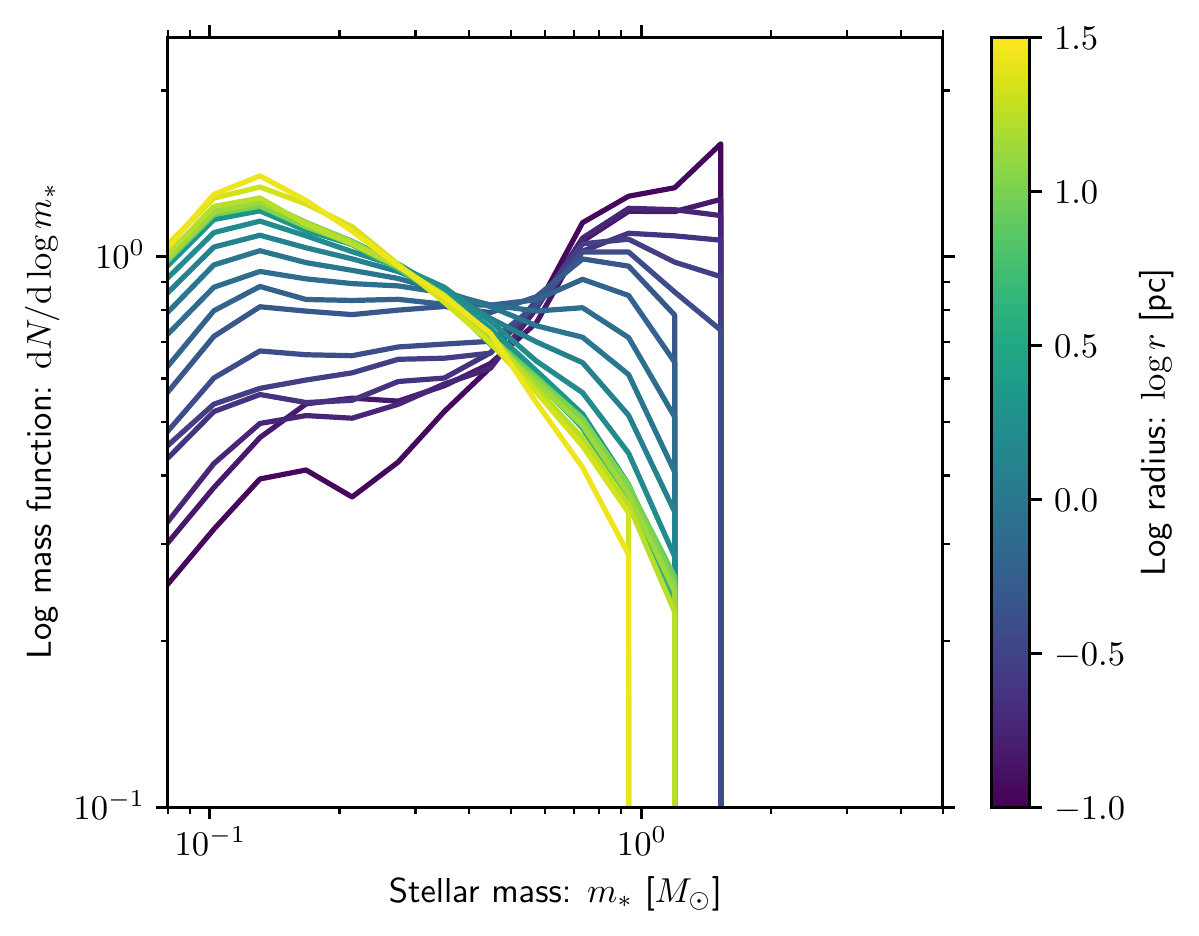}}
\subfloat[$10$~Gyr\label{subfig:mrf_10Gyr}]{\includegraphics[width=0.33\textwidth]{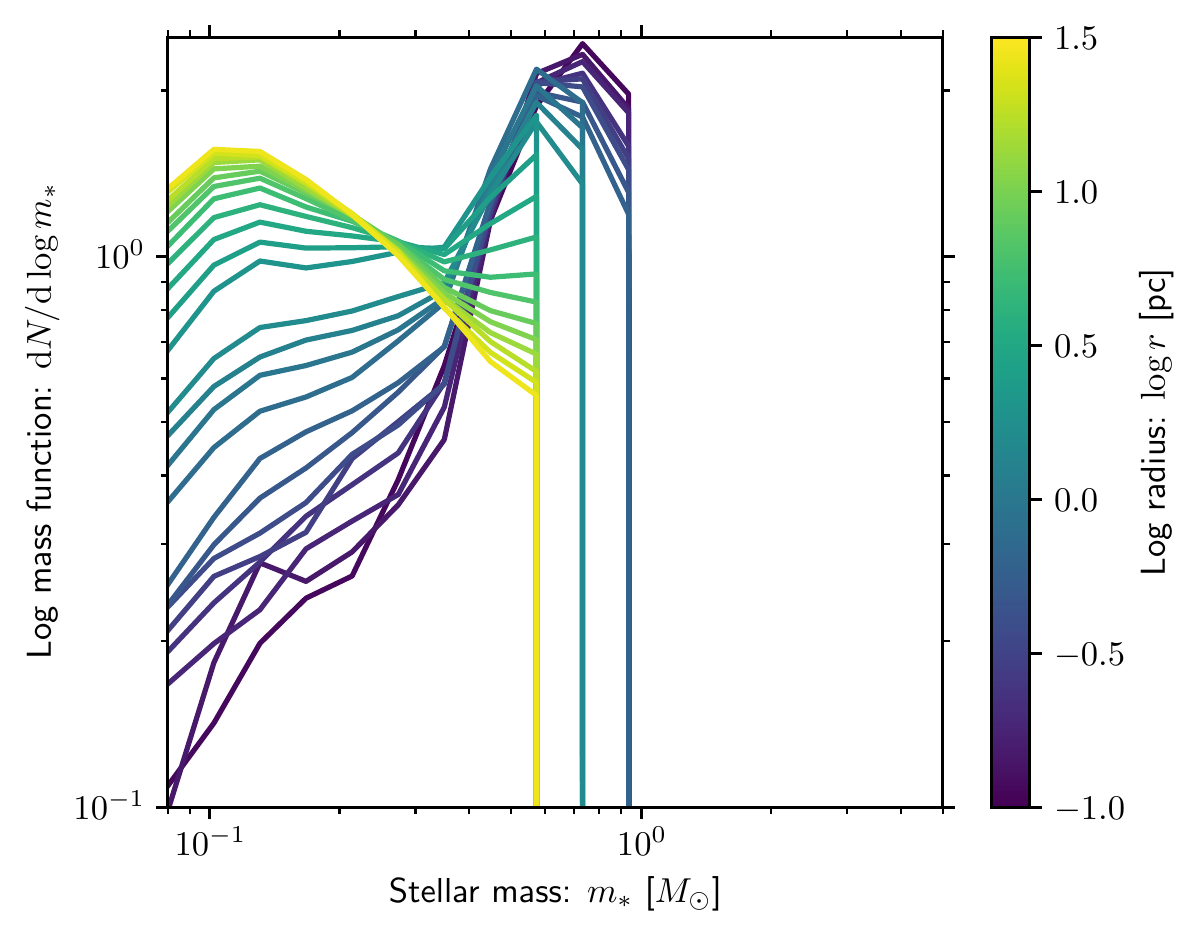}}
\caption{\label{fig:massfunc} Stellar mass function binned by three dimensional radius within the cluster, with bin size $0.1$~dex, as shown by the colour bar. The results are shown for the stellar population in our dynamical model at $100$~Myr (Figure~\ref{subfig:mrf_100Myr}), $1$~Gyr (Figure~\ref{subfig:mrf_1Gyr}) and $10$~Gyr (Figure~\ref{subfig:mrf_10Gyr}).}
\end{figure*}

The rate of ionisation, circularisation and tidal destruction of a planet is dependent not only on the local velocity dispersion and density, but also the local mass function. The local mass function varies both temporally and spatially, which should be accounted for in computing the orbital evolution of the planets. We therefore define the local mass function numerically from the dynamical model. 

At a given snapshot, we compute the percentiles in the mass distribution of stars within thirty logarithmically spaced radial bins between $0.1$~pc and $10^{1.5}$~pc (i.e. with width 0.1~dex). The percentiles we compute are at intervals of five percent, except for the extreme upper and lower end where we include $1^\mathrm{st}$, $2^\mathrm{nd}$, $98^\mathrm{th}$ and $99^\mathrm{th}$ percentiles. We then numerically determine the derivative of the cumulative distribution function within the $1^\mathrm{st}-99^\mathrm{th}$ percentiles, interpolating to estimate the local mass function. We repeat this procedure at $100$~Myr intervals up to the end of the simulation ($12$~Gyr). 

The mass functions we compute are shown at $100$~Myr, $1$~Gyr and $10$~Gyr in Figure~\ref{fig:massfunc}. Initially the mass segregation only strongly influences high stellar masses and the inner regions (Figure~\ref{subfig:mrf_100Myr}). Progressively more low mass stars are cleared from the centre of the cluster, eventually resulting in radically different mass functions in the inner and outer regions (Figure~\ref{subfig:mrf_10Gyr}). 

When we compute the local encounter rates in our dynamical model, we adopt the mass function first at the closest snapshot and then the closest radial position. These mass functions are then numerically integrated over when calculating the relevant encounter rates for a given star. 

\subsection{Orbital integration}

\begin{figure*}
\subfloat[$a_0=1$~au\label{subfig:a01_Poc}]{\includegraphics[width=0.33\textwidth]{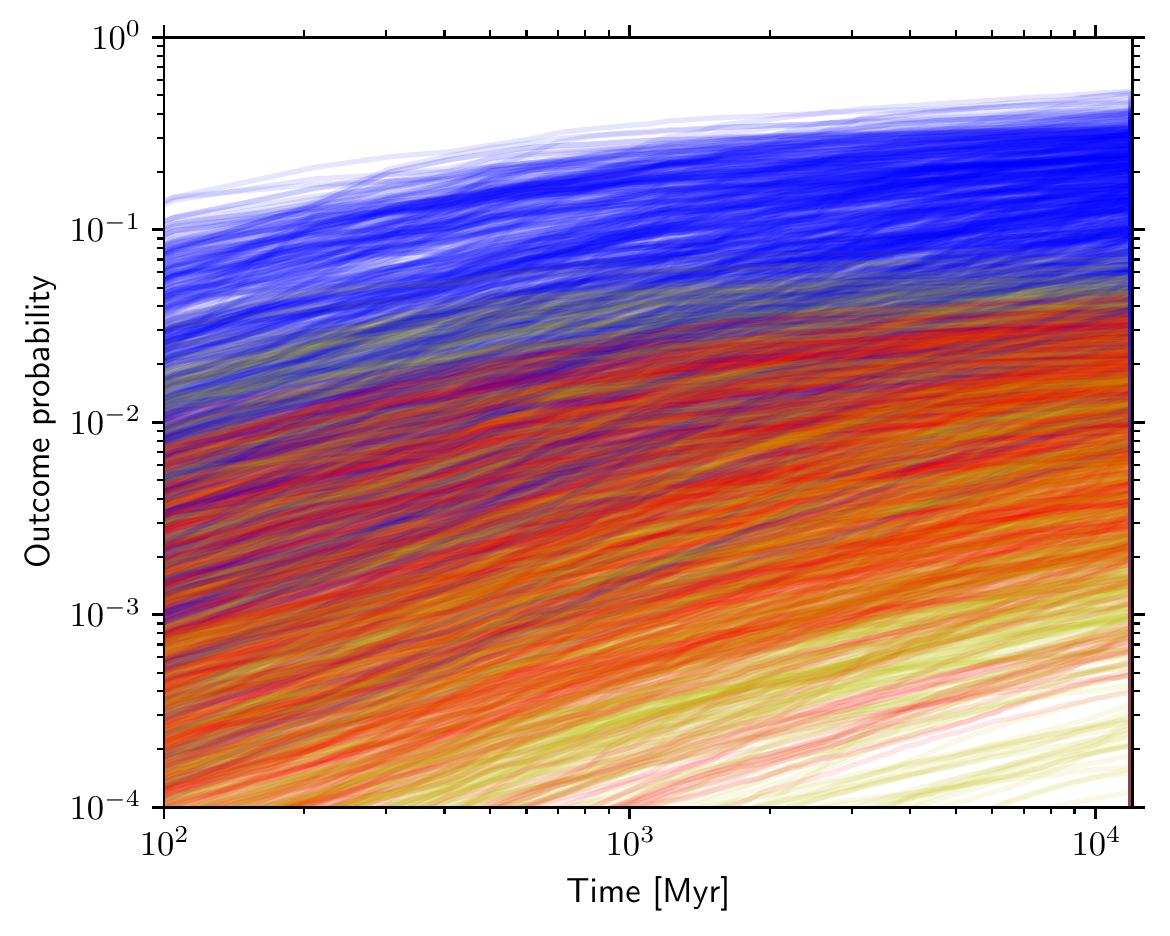}}
\subfloat[$a_0=5$~au\label{subfig:a05_Poc}]{\includegraphics[width=0.33\textwidth]{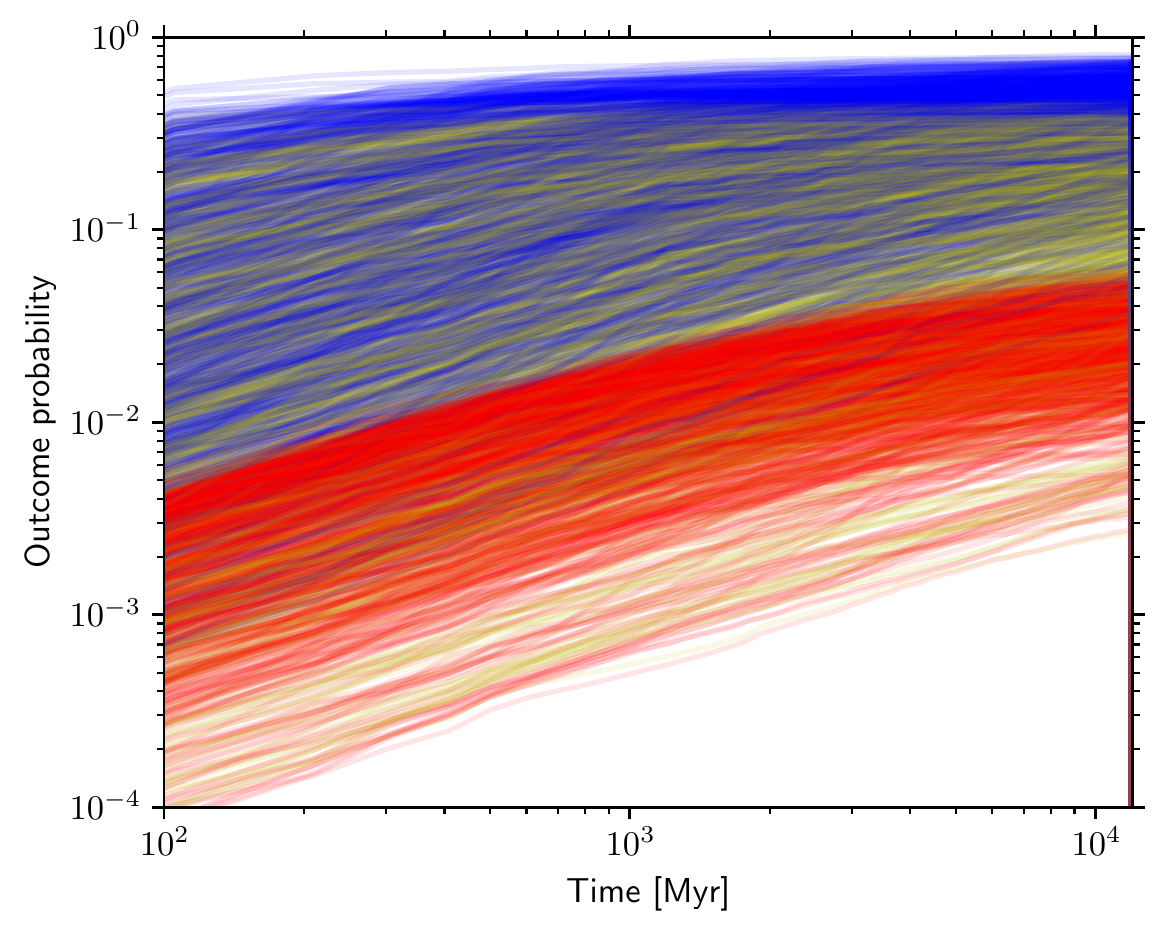}}
\subfloat[$a_0=25$~au\label{subfig:a025_Poc}]{\includegraphics[width=0.33\textwidth]{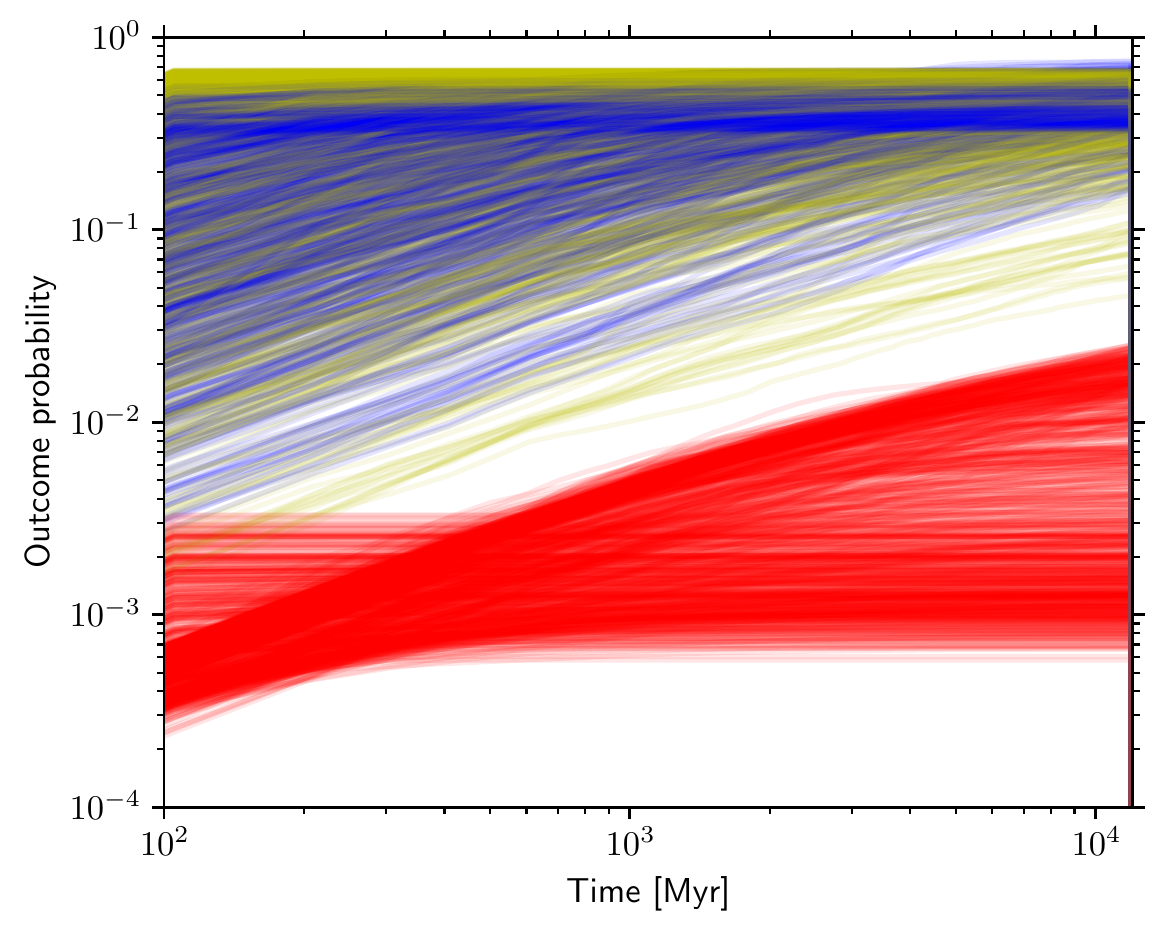}}
\caption{\label{fig:ptime_}Evolution of the probability of the different outcomes for a planet evolving in 47 Tuc. Each line corresponds to a single star in a subset of 845 stars with masses $m_*<0.88\, M_\odot$ in our dynamical model. Probabilities for ionisation or exchange (blue lines), circularisation (red lines) or tidal destruction (yellow-green lines). Results are shown for initial planet semi-major axes $a_0=1$~au (Figure~\ref{subfig:a01_Poc}), $5$~au (Figure~\ref{subfig:a05_Poc}) and $a_0=25$~au (Figure~\ref{subfig:a025_Poc}). {When the lines become practically horizontal (constant in time), the probabilities are `locked in', in that there is very little chance of a planet having escaped all of the three possible outcomes.}}
\end{figure*}

To compute the time-dependent evolution of the statistical outcomes for planets evolving in our model for 47 Tuc, we first draw a random subset of $1000$~stars. Of these stars, $845$ have masses $m_*<0.88\,M_\odot$, which is the maximum mass that remains on the main sequence up to the 12~Gyr age of the cluster (see Section~\ref{sec:obs_comp}). For each of the stars in the sample we obtain the radial position and azimuthal and radial velocity, updated at $100$~Myr intervals. We then obtain the local density and velocity dispersion averaged over an epicycle at each time-step. To do this, we fit an approximate analytic double power-law density profile:
\begin{equation}
\label{eq:rho_pot}
    \rho_* = \frac{M_\mathrm{s}}{4\pi a_\mathrm{s}^3} (r/a_\mathrm{s})^{-\alpha}(1-r/a_\mathrm{s})^{\alpha-\beta}
\end{equation} to the stellar mass density of the cluster, where $M_\mathrm{s}$, $a_\mathrm{s}$, $\alpha$ and $\beta$ are fitting constants. With these parameters, we construct a spherically symmetric potential using the \texttt{TwoPowerSphericalPotential} class of \textsc{Galpy}\footnote{\url{http://github.com/jobovy/galpy}} \citep{Bovy15}. {A number of alternative spherically symmetric profiles with fewer fitting parameters are possible, and allow faster integration of orbits. However, we adopt this density profile because it reliably reproduces the physical density profile in our Monte Carlo model (see \citetalias{PaperI}).} Due to the spherical symmetry, we are only interested in the radial oscillations in the stellar position. We therefore average $\sigma_v$ and $n_\mathrm{tot}$ for a single epicycle. 

With the averaged environmental properties we can then adopt the expressions in Section~\ref{sec:analytic_Pdot} to compute the evolution of a given outcome `oc', $P_\mathrm{oc}$, by writing:
\begin{equation}
\label{eq:Poc_deltat}
    P_\mathrm{oc}(t+\Delta t) = P_\mathrm{oc}(t) + \Delta t \cdot \dot{P}_\mathrm{oc}(t).
\end{equation}Note that we can choose a time-step that is smaller than that which we update the orbital solutions in our dynamical model ($100$~Myr). For each star, we compute the probability over $3000$ equal time-steps up to $12$~Gyr. {For computing the tidal rates, we will assume the initial planet eccentricity is $e_0=0.1$, which is a typical value for the eccentricity of planets forming in hydrodynamic simulations of protoplanetary discs \citep[e.g.][]{Bitsch10,Dunhill13, Ragusa18}.}

\subsection{Time evolution of outcome probabilities}

The results of computing the outcome probabilities according to equation~\ref{eq:Poc_deltat} are shown in Figure~\ref{fig:ptime_}, for ionisation (blue), circularisation (red) and tidal destruction (yellow-green). We find that for initial semi-major axis $a_0=1$~au HJ production is efficient, producing circularised planet at a yield of $P_\mathrm{circ}\sim 0.02$. {If 100~percent of systems hosted planets at these separations, such high numbers of HJs would be in tension with the observed absence of short period companions, discussed in Section~\ref{sec:obs_comp}. However, as discussed in Section~\ref{sec:occ_rates}, this is not expected given field star occurrence rates.} In addition, as the initial semi-major axis of the planet increases, this efficiency decreases. This is due to increasing ionisation and tidal destruction rates. We explore this further in terms of the final outcomes as follows.

\subsection{Final outcomes}

\subsubsection{Projection averaging}

We are interested in quantifying the observable dependence on the outcomes as a function of projected separation $d$ from the centre of the cluster. In two dimensions, we can geometrically average the outcome probabilities $P_\mathrm{oc}$:
\begin{equation}
\label{eq:tau_mean}
  \langle P_\mathrm{oc}\rangle_\mathrm{2D}(d) =\frac{1}{n_{d}}\sum_{r_i>d}\frac{P_{\mathrm{oc},i}}{r_i}\frac{d}{\sqrt{r_i^2-d^2}}
\end{equation} where
\begin{equation}
    n_d=  \sum_{r_i>d} \frac{d}{r_i\sqrt{r_i^2-d^2}}.
\end{equation} and $P_{\mathrm{oc},i}$ 
is the outcome probability for each star $i$ at three dimensional radius $r_i$.

\subsubsection{Dependence on final projected radial position}

\begin{figure}
    \centering
   \includegraphics[width=0.48\textwidth]{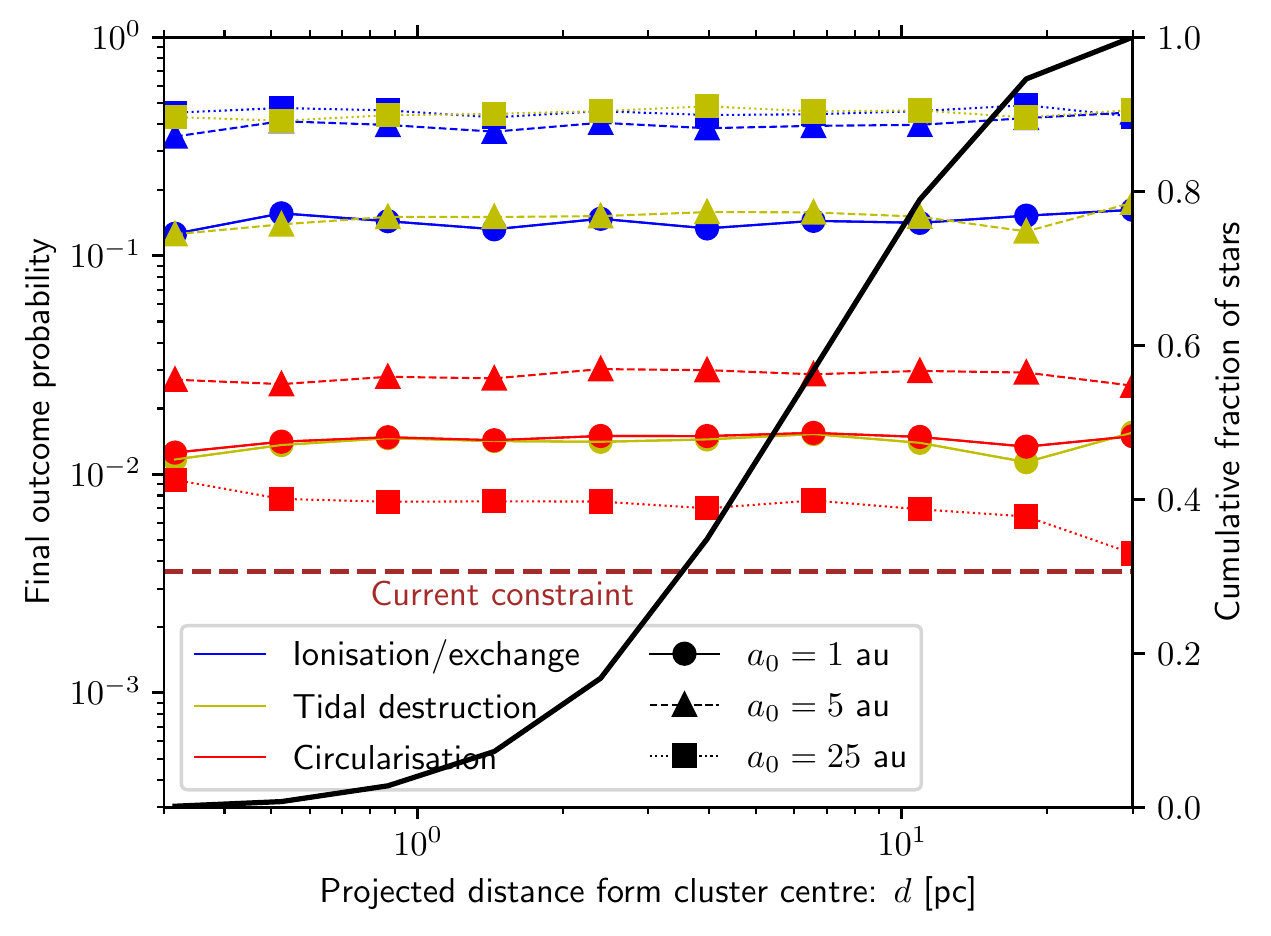}
    \caption{Statistical outcomes for planets evolving in 47 Tuc as a function of the final projected separation from the cluster centre after $12$~Gyr of evolution. Results are shown for a planet initially at $a_0= 1$~au, $5$~au and $25$~au. Ionisation/exchange probabilities are shown in blue, tidal destruction in yellow-green, circularisation in red. The dashed brown line is the estimated $2\sigma$ constraint on the total fraction of HJs in the aggregate samples of \citetalias{Gil00} and \citet{Weldrake05} (see discussion in Section~\ref{sec:obs_comp}). {This constraint can be compared with the yields per planet (red lines) by multiplying by the expected number of planets per star.} The solid black line shows the cumulative fraction of the sample within each projected separation. }
    \label{fig:circ_prob}
\end{figure}

We show the results of computing equation~\ref{eq:tau_mean} for varying projected separation $d$ in our simulation at $t=12$~Gyr in Figure~\ref{fig:circ_prob}. We also show the approximate upper limit for the fraction of HJs inferred from the aggregated sample of \citetalias{Gil00} and \citet{Weldrake05} -- this should be understood as the upper limit if the planet occurrence was $100$~percent. Thus across all $a_0$~au the rate of HJ production is close to this upper limit if occurrence rates are significantly samller than this (e.g. $10$~percent). We explore the dependence on $a_0$ in greater detail in Section~\ref{sec:sma_dep}. 

It is clear from Figure~\ref{fig:circ_prob} that the final outcomes are practically independent of the projected separation from the cluster centre. This is somewhat suprising due to the strong dependence on the rates of ionisation, circularisation and tidal destruction on local density and velocity dispersion. In fact, this finding also applies to the final position in three dimensions, and is not a result of our projected separation averaging. This suggests that the origin of this finding is that most outcomes are `locked-in' early during the dynamical evolution, and dynamical mixing subsequently washes out any trends. We confirm this hypothesis as follows. 

\subsubsection{Dependence on initial radial position}

\begin{figure*}
\subfloat[$a_0=1$~au\label{subfig:ra01_Poc}]{\includegraphics[width=0.33\textwidth]{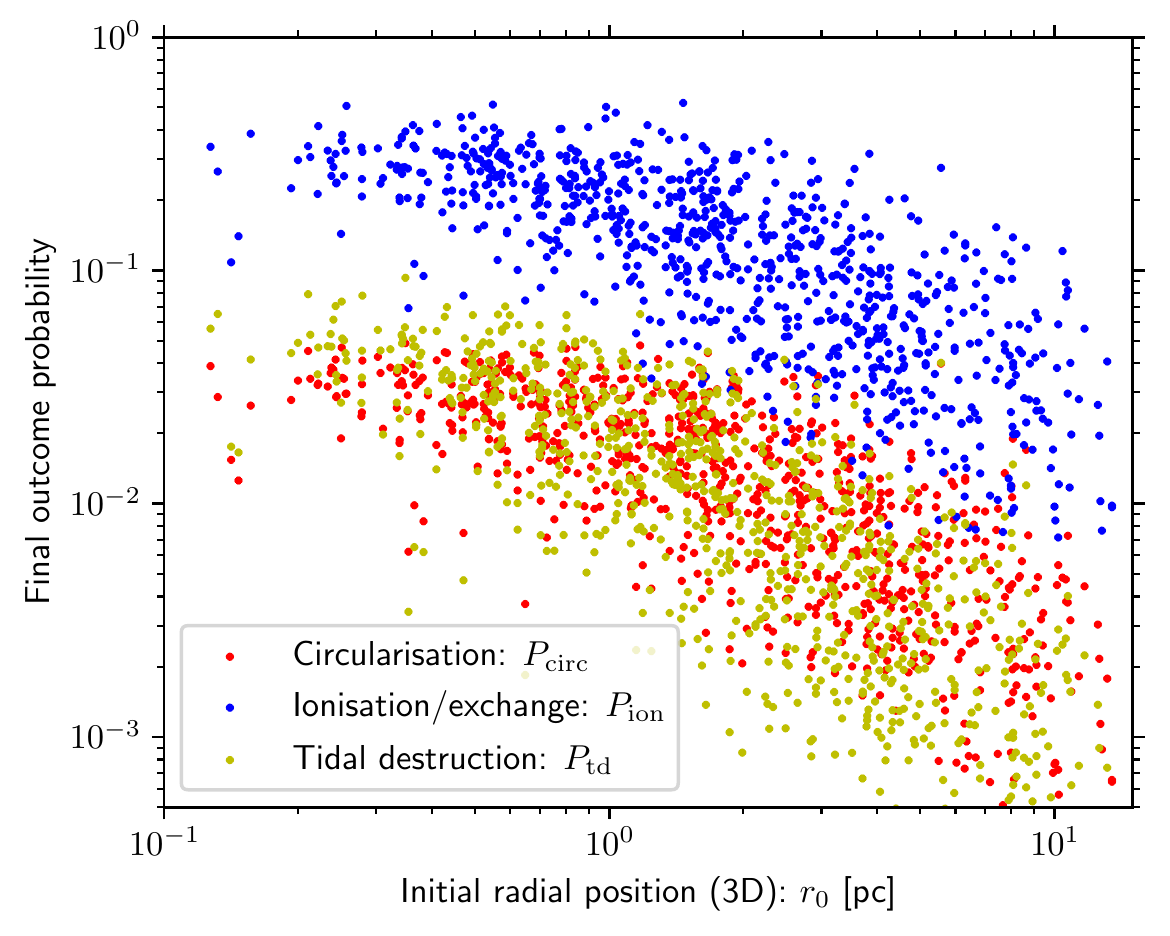}}
\subfloat[$a_0=5$~au\label{subfig:ra05_Poc}]{\includegraphics[width=0.33\textwidth]{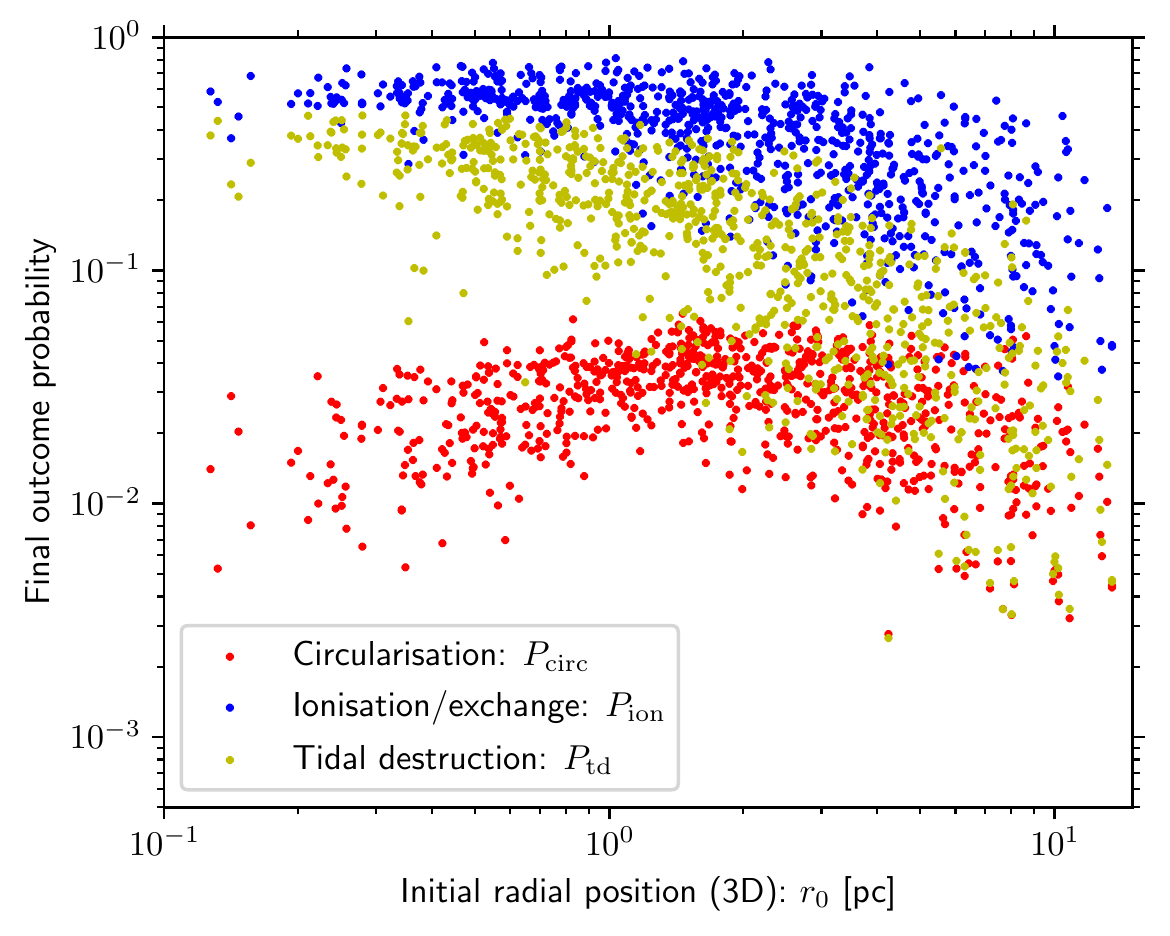}}
\subfloat[$a_0=25$~au\label{subfig:ra025_Poc}]{\includegraphics[width=0.33\textwidth]{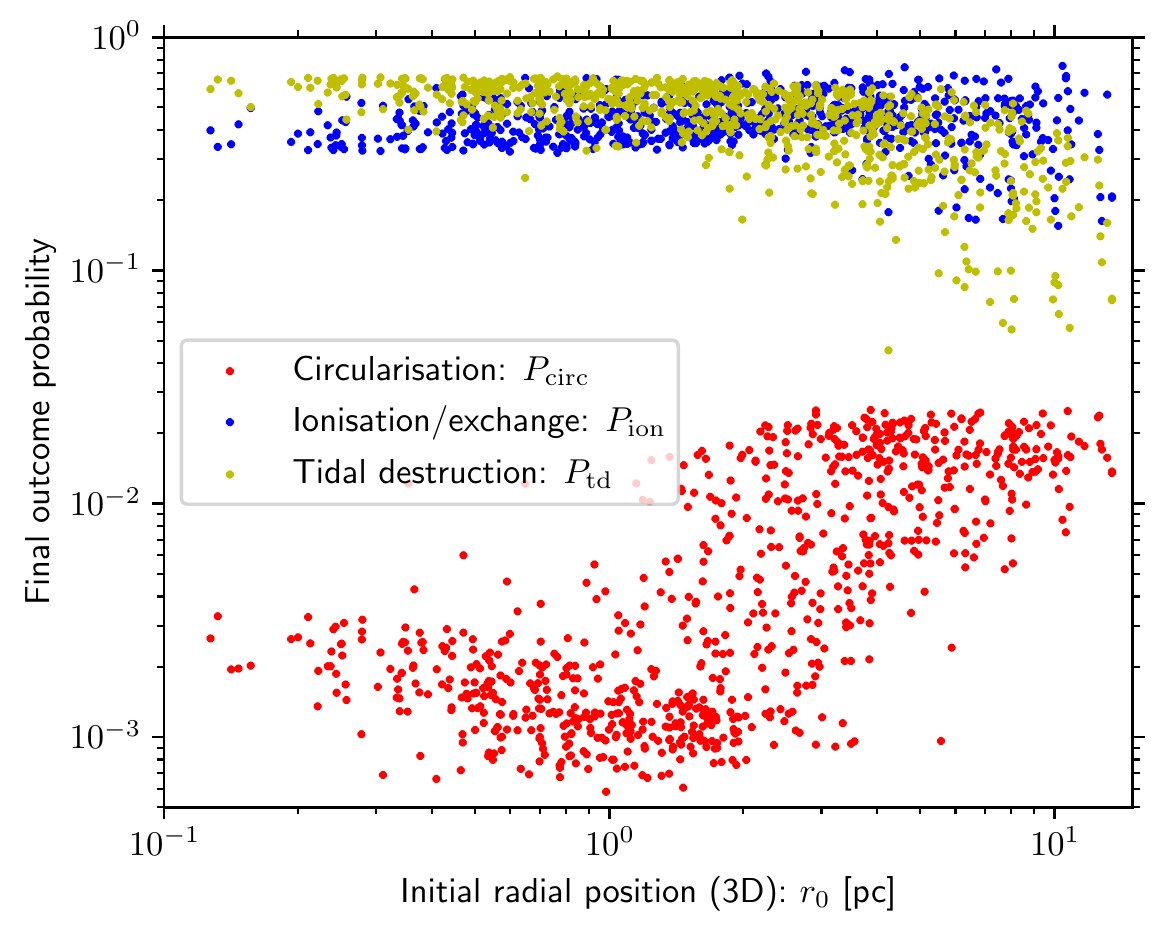}}
\caption{\label{fig:p_radius}Dependence of the final outcome probability on the initial three dimensional radial position. We show results for initial semi-major axes $a_0= 1$~au (Figure~\ref{subfig:ra01_Poc}), $5$~au (Figure~\ref{subfig:ra05_Poc}) and $25$~au (Figure~\ref{subfig:ra025_Poc}). The blue points show the probability that a planet has undergone ionisation, yellow-green for tidal destruction and red for circularisation.  }
\end{figure*}

We wish to examine whether the initial radial location of a star-planet system in the cluster is a better predictor of the outcome for planets than the final position. We therefore consider the final outcome probabilities versus the initial radius $r_0$ in Figure~\ref{fig:p_radius}. We find that the final outcome is indeed a strong function of the initial position in the cluster. The previous result that there is no strong dependence of outcomes on the final position in the cluster is therefore a result of dynamical mixing. 

In particular, for all $a_0$ the sum of all outcomes -- i.e. the probability that the environment significantly influences the star-planet -- decreases with increasing radius. This is expected because the stellar density is greater in the inner regions, and therefore stellar encounters are more frequent at smaller $r_0$. 

There are also some qualitative changes in the radial dependence of the fractional outcomes with the initial semi-major axis. For small $a_0=1$~au (Figure~
\ref{subfig:a01_Poc}), all outcomes behave similarly, with probabilities declining with $r_0$. However, as $a_0$ increases, tidal destruction begins to dominate over circularisation in the central regions. This results in declining HJ formation rates with decreasing $r_0$, seen clearly in Figure~\ref{subfig:a025_Poc}. This is due to the {large perturbation rate (or  $\gamma$ value, equation~\ref{eq:gamma})}, which results in small $l_\mathrm{max}$ and large $f_\mathrm{td}$ (see Section~\ref{sec:tidal_acc} and~\ref{sec:HJ_surv}) -- i.e. a planet cannot circularise without already undergoing tidal disruption. HJ formation is therefore inefficient for planets at large $a_0$ and stars born in the inner regions of globular clusters. 

\subsubsection{Semi-major axis dependence}

\label{sec:sma_dep}
\begin{figure}
    \centering
    \includegraphics[width=\columnwidth]{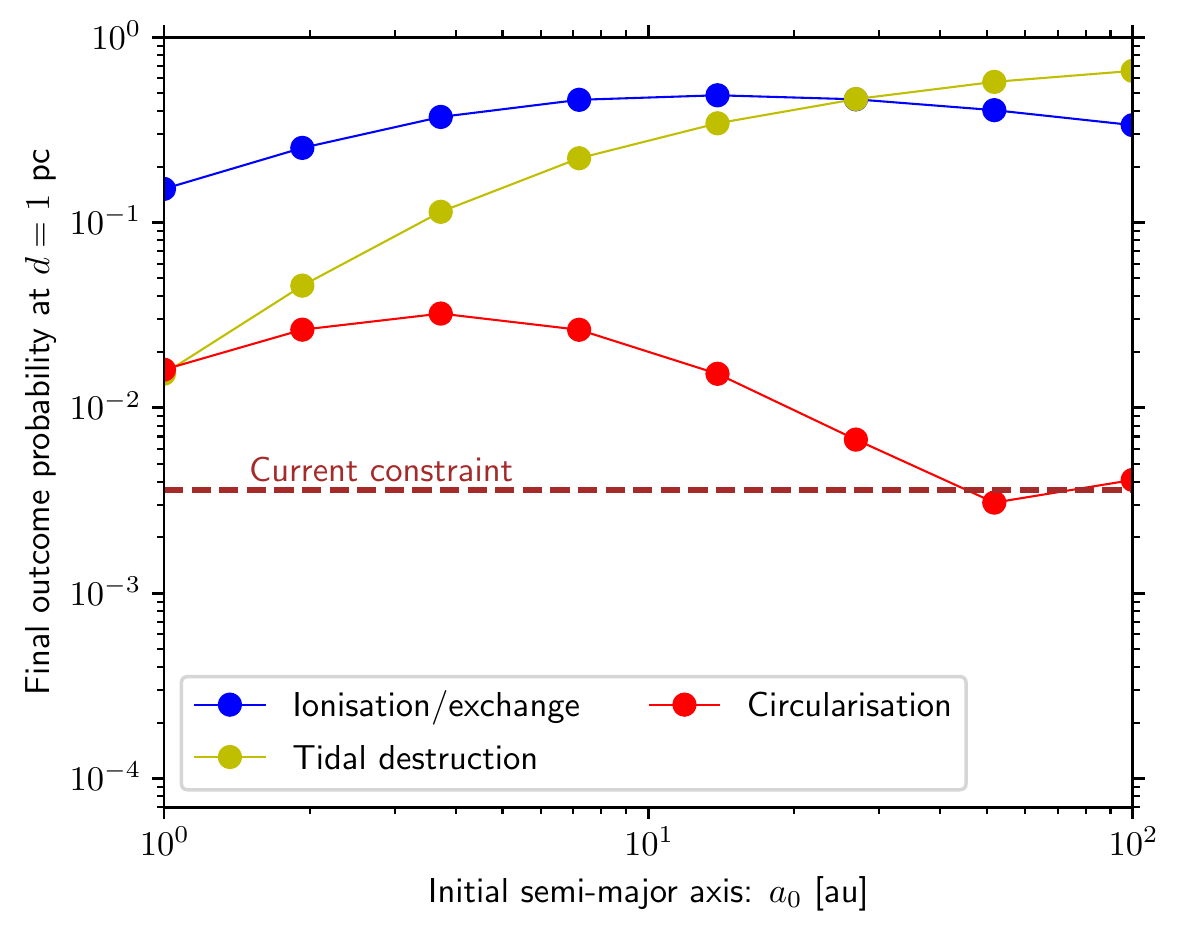}
    \caption{Probability of outcomes for a planetary system evolving within our dynamical model of 47 Tuc as a function of the initial semi-major axis $a_0$. A projected average probability at $d=1$~pc is adopted, although the results are not strongly dependent on this choice (Figure~\ref{fig:circ_prob}). The points show the discrete values of $a_0$ that are adopted. Blue lines are for ionisation/exchange, yellow-green for tidal destruction and red for circularisation. The dashed brown line shows the $2\sigma$ constraint on the HJ fraction from the aggregated \citetalias{Gil00} and \citet{Weldrake05} samples, {which can be compared with final outcome probabilities by multiplying by the per star occurrence rates}.   }
    \label{fig:Poc_a0}
\end{figure}

In Figure~\ref{fig:Poc_a0} we show the variation of the final outcome probabilities as a function of initial semi-major axis $a_0$. We choose the values averaged at projected separation $d=1$~pc, although in practice this choice makes little difference, as shown in Figure~\ref{fig:circ_prob}. We find that the fraction of planets that experience tidal destruction increases with $a_0$, and is the most likely outcome for $a_0\gtrsim 30$~au. For $a_0\gtrsim 50$~au, the frequency of HJ production decreases below the upper limit constraint for $f_\mathrm{HJ}$ aggregated across the surveys of \citetalias{Gil00} and \citet{Weldrake05}. {Thus even a 100~percent occurrence rate of planets at these semi-major axes would not be expected to yield any HJs in these samples. The slight increase in the fraction of circularised planets at large $a_0$ is due to the decrease in the ionisation rate compared with the perturbation rate ($\Gamma_\mathrm{ion} \propto a_0$ while $\Gamma_\mathrm{per}^\mathrm{(hyp)} \propto a_0^{3/2}$), which is related to the inefficiency of ionisation in high velocity dispersion environments.}

\subsubsection{Stellar mass dependence}

\begin{figure}
    \centering
   \includegraphics[width=0.48\textwidth]{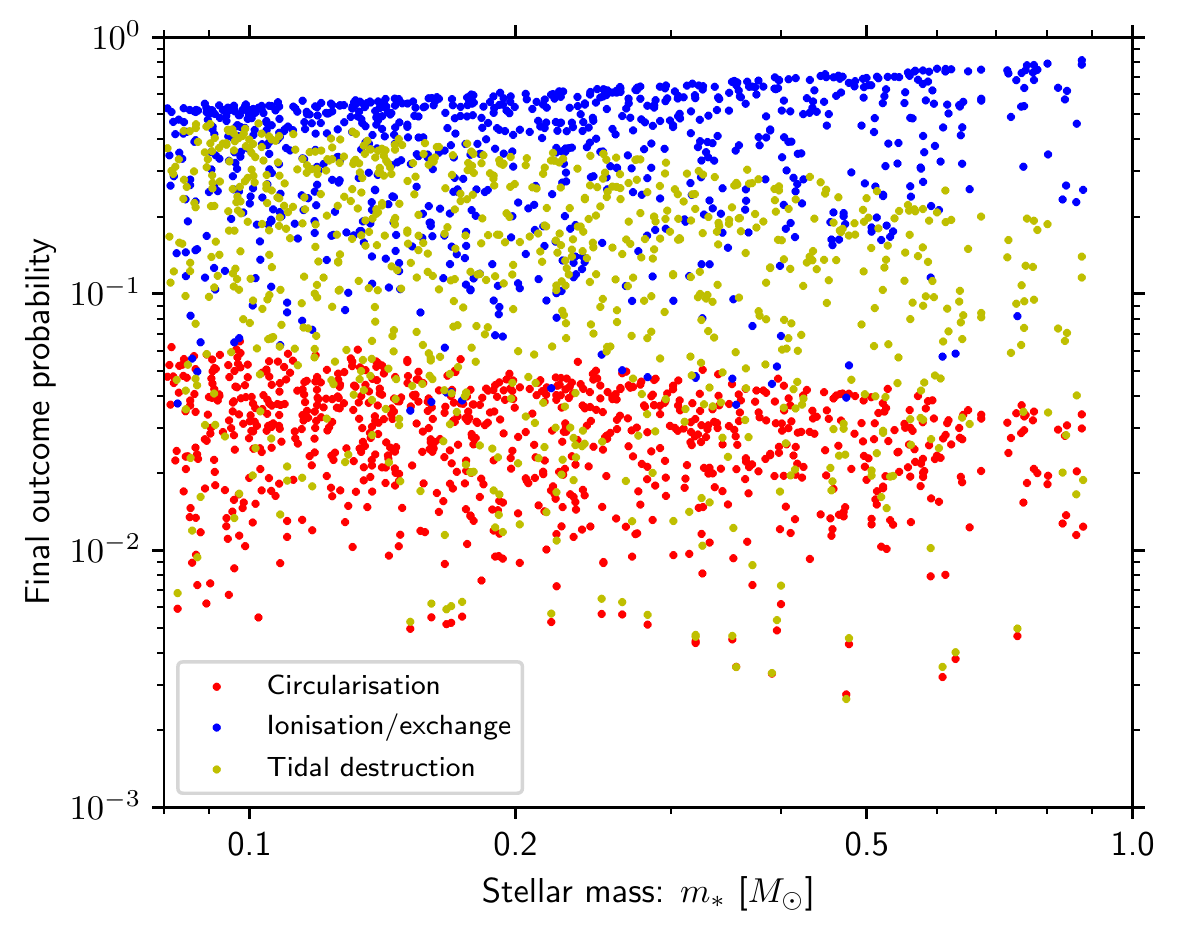}
    \caption{The distribution of outcome probabilities for planets around 845 randomly selected stars of mass $m_*<0.88 \, M_\odot$ in our dynamical model as a function of $m_*$ after $12$~Gyr. The initial semi-major axis in this case is $a_0=5$~au. The outcomes considered are ionisation/exchange of the planet (blue points), circularisation (red points), and tidal destruction (yellow-green points).  }
    \label{fig:mst_probs}
\end{figure}

When considering future surveys of globular clusters, we may be interested in whether the outcome probabilities are dependent on the stellar mass. We show the probabilities for semi-major axis $a_0=5$~au in Figure~\ref{fig:mst_probs} (the results are similar for varying $a_0$). We find that there is only a very weak dependence of the different outcomes across the relevant range of stellar masses. This is because the differences between ionisation and perturbation cross sections is only moderately dependent on the stellar mass via the ratio with the local stellar mass function. Mass segregation further suppresses this difference by yielding encounters that preferentially have order unity mass ratio. In conclusion, we do not expect the mass function of surveyed stars to strongly influence the occurrence rate of HJs. However, this does not apply to detectability \citep[see discussion by][]{Masuda17}.


\subsection{Initial planet population}

\label{sec:occ_rates}

\subsubsection{Massive planet occurrence}

In order to interpret our findings, we need to convert the efficiency at which planets are converted into HJs into an occurrence rate of HJs \textit{per star}. {To do this we must appeal to field star planet occurrence rates as a function of $a_0$,  ${\mathrm{d} N_\mathrm{mp}}/{\mathrm{d}\log a_0}$. We have previously implicitly assumed that one hundred percent of stars host a planet at each initial semi-major axes $a_0$, or ${\mathrm{d} N_\mathrm{mp}}/{\mathrm{d}\log a_0} = \delta(a_0')$ where $\delta$ is the Dirac delta distribution and $a_0 '$ is any value of $a_0$ we have adopted. We now consider empirical constraints on the true occurrence rates of giant planets.} 

Particularly useful in this context are the results of the \textit{Gemini Planet Imager Exoplanet Survey} (GPIES) presented by \citet{Nielsen19}. This survey is sensitive to planets/brown dwarfs with semi-major axes $3\,\rm{au}\lesssim a \lesssim 100$~au  and masses $3 \, M_\mathrm{J} \lesssim M_\mathrm{p} \lesssim 100 \, M_\mathrm{J}$. The authors extrapolate their findings to estimate the frequency of planets with $0.03\,\rm{au}\lesssim a \lesssim 100$~au and $1 \, M_\mathrm{J} \lesssim M_\mathrm{p} \lesssim 13 \, M_\mathrm{J}$ around solar mass stars, combining with the previous estimates of \citet{Cumming08} and \citet{Fernandes19} for closer-in massive planets (see Figure 18 of \citealt{Nielsen19}). The resultant total per star occurrence rate for massive planets in this range is $N_\mathrm{mp} \sim 0.1$. 

{In order to convert our results to an expected occurrence rate of HJs in 47 Tuc, we must make some assumptions about how the occurrence rate varies with semi-major axis ($\mathrm{d}N_\mathrm{mp}/\mathrm{d} \log a$). Interpreting the occurrence rates of massive planets in the field is not straight forward in this context. This is not only because definitions of the occurrence rates depend on the mass range considered (and physical/observational covariance with $a$), but also because we work on the premise that some fraction of planets migrate by HEM. In this case, the occurrence rates as a function of semi-major axis $a$, is by definition not the same as that of the initial semi-major axis $a_0$ \citep[see][for example]{Winter21}.}

{For the above reasons, we make a simplified estimate for the form of $\mathrm{d}N_\mathrm{mp}/\mathrm{d} \log a$, with the caveat that this form remains uncertain. Based on the findings of \citet{Nielsen19}, $\mathrm{d}N_\mathrm{mp}/\mathrm{d} \log a$ is not well constrained at large separations. In the approximately solar mass stellar sample, the detected companion with the greatest semi-major axis has $a\sim 30$~au (projected separation $23.65\pm0.08$~au). The results therefore appear broadly consistent with an occurrence rate that is uniform in $\log a$, truncated outside of $30$~au. We impose an inner truncation radius of $1$~au, motivated by the fact that \citet{Fernandes19} find very few planets with $M_\mathrm{p}>1\, M_\mathrm{J}$ inside this separation (see their Figure 1). We then normalise over this range to give an overall occurrence rate of $N_\mathrm{mp} =0.1$:}
\begin{equation}
\label{eq:dNdloga}
    \frac{\mathrm{d} N_\mathrm{mp}}{\mathrm{d}\log a_0} = \begin{cases} 0.0677 &\qquad 1\,\rm{au} <a_0 < 30\,\rm{au} \\
    0 &\qquad \rm{otherwise}
    \end{cases}
\end{equation}
{Visually considering such a mass function on the top panels of Figure 18 of \citet{Nielsen19}, we see that equation~\ref{eq:dNdloga} would remain broadly consistent with the posterior distributions inferred by \citet{Cumming08}, \citet{Fernandes19} and \citet{Nielsen19} for $1$~au$< a< 30$~au. }

\subsubsection{Application to 47 Tuc}

{We now compute the expected number of HJs forming in 47 Tuc:
\begin{equation}
    f_\mathrm{HJ} = \int \! P_\mathrm{HJ}(a_0) \cdot  \frac{\mathrm{d} N_\mathrm{mp}}{\mathrm{d}\log a_0} \,  \mathrm{d} \log a_0  .
\end{equation}To compute this occurrence rate, we use the numerical results for the probability of a given planet with initial semi-major axis $a_0$ becoming a HJ, $P_\mathrm{HJ}(a_0)$, shown as the red line in Figure~\ref{fig:Poc_a0}. Adopting the occurrence rate of massive planets in semi-major axis space as defined by equation~\ref{eq:dNdloga}, we obtain $f_\mathrm{HJ} = 2.2 \times 10^{-3}$. This remains consistent with the constraint $f_\mathrm{HJ, max} \approx 3.6\times 10^{-3}$ inferred in Section~\ref{sec:obs_comp} for the combined samples of \citetalias{Gil00} and \citet{Weldrake05}.  }

\textit{Our findings therefore suggest that in order to determine whether the planets in globular clusters are significantly different to the field, transit surveys with sample size $\gtrsim 10^5$ (based on equation~\ref{eq:fHJ_max}) at the survey sensitivity of \citet{Gil00} are required.} Alternatively, smaller sample sizes with greater survey sensitivities could in principle rule out a more abundant lower mass planet population. {This latter possibility has the caveat that the tidal distortion (or $Q$-factor) is dependent on the planet properties. While the rate of tidal circularisation is dependent on the apsidal motion constant $k_\mathrm{p}$ and tidal lag time $\tau_\mathrm{p}$ (e.g. equation~\ref{eq:adot_ps}), the maximum SLR $l_\mathrm{max}$ (equation~\ref{eq:lmax}) is only weakly dependent on these assumed constants. Thus our results should be a reasonable estimate of the outcome probabilities of (massive) planets in general. Nonetheless, future application to lower mass planets should consider variations in these constants, as well as the closest approach distance required for tidal disruption (Section~\ref{sec:HJ_surv}).   }

\subsection{Future survey ramifications}

{We can ask what the consequences of detection or non-detection of HJs in future surveys of 47 Tuc would be, both for the local planet population and more generally for the formation pathways of HJs in the field. In this section, we summarise the consequences of future detection (Section~\ref{sec:detection}) or non-detection (Section~\ref{sec:nondetection}) of HJs in globular clusters. We consider in greater detail each possible isolated formation mechanism in Sections~\ref{sec:planet_planet},~\ref{sec:outer_comps} and~\ref{sec:LEM}.}

\subsubsection{Detection in a large/sensitive survey}
\label{sec:detection}
{Detecting the first HJ in a globular cluster is a tantalizing prospect. Such an object would in itself represent the outcome of planet formation in an environment completely different from local star forming regions. However, from a statistical perspective, a single discovery constraining the population of HJs to an occurrence rate $f_\mathrm{HJ}\sim 10^{-3}$ would not distinguish between a range of interpretations. It could mean that normal HJ formation is via a mechanism that is unaffected by dynamical encounters (such as disc-induced migration, see Section~\ref{sec:LEM}), or it could mean that these HJs were created by scattering of a population of wider planets that are similar to that of the solar neighbourhood. Future follow-up on such a discovery may offer ways to distinguish between these possibilities.  }

\subsubsection{Non-detection in a large/sensitive survey}
\label{sec:nondetection}
{Statistically, the most interesting constraints would originate from non-detection of any HJs in a large and sensitive future transit survey, constraining the massive ($M_\mathrm{p}\gtrsim 1\,M_\mathrm{J}$) HJ  occurrence rate to $f_\mathrm{HJ, max}<2.2 \times 10^{-3}$ (or larger if greater sensitivity than the survey of \citetalias{Gil00}).\textit{ Non-detection in such a sample would suggest both a paucity of massive planets at wide separations relative to the solar neighbourhood and that any formation mechanism producing HJs in low density environments does not operate.} }

{In the case of non-detection, the general pathway for HJ formation must be one which can be viably disrupted in a dense environment. The formation pathways that should most obviously be disrupted would be those of HEM (see Sections~\ref{sec:planet_planet} and~\ref{sec:outer_comps}). In this case, non-detection in a large/sensitive transit sample could indicate that HEM is the origin of HJs in the field. However, LEM migration mechanisms may also be suppressed in a low metallicity environment (see Section~\ref{sec:LEM}). Thus non-detection could also imply that planet formation is completely suppressed across a large dynamical range in semi-major axis. We discuss each of the isolated HJ formation mechanisms in further detail as follows.  }


\subsubsection{Planet-planet scattering}
\label{sec:planet_planet}
{Planet-planet scattering within isolated planetary systems has been suggested as a possible origin of the eccentricity excitation required to produce HJs via HEM \citep{Rasio96, Carrera19}. Whether or not this formation channel produces enough HJs to explain the observed occurrence rates depends sensitively on the initial stability of planetary systems. For example, it remains unclear whether \textit{Kepler} multiple systems statistically `pile-up' close to the stability limit \citep{Pu15, Yee21}, which is a necessary measurement to quantify the frequency of chaotic dynamical decay. 

{To gain an intuition as to how HJ formation via planet-planet scattering may proceed in the context of globular clusters we consider a simple thought experiment. We consider a planet excited by internal scattering to a SLR $l<l_\mathrm{age}$, as required for the planet to circularise over its lifetime. This planet is then also subject to the usual encounters in a high density environment. Thus it is still unable to circularise if $l>l_\mathrm{max}$, as depicted in Figure~\ref{fig:ea_evol}. Therefore, even HJ formation via planet-planet scattering is subject to the same suppression by tidal disruption as HJ formation via external encounters. The difference in the planet-planet scattering case is that the condition $l<l_\mathrm{max}$ may be reached earlier than it would be relying on encounters alone. This may reduce the frequency of ionisation for such a planetary system. To first order, we can ignore ionisation and estimate the suppression of HJ formation via planet-planet scattering using the ratio of the yellow-green line to the sum of the red and yellow-green lines in Figure~\ref{fig:Poc_a0}.} If such HJs originate from initial semi-major axes $a_0\gtrsim 1$~au, then the majority are unable to migrate.}

\subsubsection{Outer companions}
\label{sec:outer_comps}
{Some fraction of HEM may be induced via Kozai-Lidov oscillations within binary/multiple planet systems \citep{Kozai62, Lidov62, Naoz16, Hamers17b}. Indeed, a number of studies have suggested that massive planets/HJs preferentially have outer companions \citep{Ngo16, Fontanive19, Belokurov20}. Unlike in the planet-planet scattering case discussed above, outer companions may dominate the eccentricity evolution of planets during circularisation rather than the influence of more distant star-star encounters. We may therefore naturally ask whether we expect Kozai-Lidov oscillations to produce HJs similarly in very high density environments such as 47 Tuc.}

{Our findings indicate that if the outer companion which would in isolation be giving rise to Kozai-Lidov oscillations is at a separation $> 10$~au, then the eccentricity oscillations would be interrupted by tidal disruption or ionisation (Figure~\ref{fig:Poc_a0}). From Figure~\ref{fig:ptime_}, we see that this disruption can occur on time-scales that are $\lesssim 100$~Myr, possibly shorter than those required to produce a circularised HJ. We conclude that Kozai-Lidov may be suppressed in 47 Tuc-like environments, although this process requires further exploration with future numerical experiments.  }

\subsubsection{Low eccentricity migration}
\label{sec:LEM}

Throughout this work, we have implicitly assumed that LEM through the primordial protoplanetary disc \citep{Lin96} is not the origin of HJs. If HJs do in fact originate from migration within a disc, then there is no reason to assume this mechanism should be suppressed by dynamical perturbations. In dense environments, the massive outer planets that undergo eccentricity fluctuations could in principle result in tidal destruction, which would also presumably destroy the inner HJ. However, such massive outer systems would frequently be ionised rather than tidally destroyed.  For outer planets with $a_0\lesssim 30$~au, this would not destroy the majority of HJs that form via LEM. {Therefore, non-detection of HJs in future transit surveys of globular clusters would imply that LEM can only be the origin of the field HJs if the low metallicity environment suppresses their formation and/or migration. Thus, constraints on $f_\mathrm{HJ}$ from future surveys would supply strict conditions on planet formation across a wide dynamical range in semi-major axis.}

\subsection{Caveats for outcome probabilities}
\label{sec:caveats}

We have presented an analytic approach to computing the rates of circularisation, destruction and ionisation in dense stellar environments. Our prescription is useful for quick computation and application to dynamical models. However, a number of considerations may alter the true outcome frequencies with respect to predictions from our analytic approach. Some factors that may alter the rates of HJ formation include, but are not limited to: 
\begin{itemize}
    \item \textit{Systems of planets}: Our models apply to single planet systems. However, we have not directly considered how systems of planets may (mutually) alter eccentricity evolution in such systems. This may increase or decrease the frequencies of the various outcomes we have considered in this work. However, we would generally expect the same suppression of HJ formation due to dynamical encounters, as we discuss in Section~\ref{sec:planet_planet}.
    \item \textit{Binary fraction}: We have assumed an initial binary fraction based on that adopted by \citet{Gie11}, which was necessarily low to reproduce the observed present-day density profile. We do not here attempt to constrain this fraction. If the initial binary fraction was high, then this might result in a larger typical interaction cross section for encounters \citep[e.g.][]{Li20}, and interplay with Kozai-Lidov oscillations as discussed in Section~\ref{sec:outer_comps}. 
    \item \textit{Limits of the analytic treatment}: We have discussed that we generally assume that hyperbolic encounters dominate the eccentricity evolution of planets. This is true only for sufficiently large velocity dispersion $\sigma_v$ and/or semi-major axis $a$ such that typical encounter velocities are $v_\infty \gtrsim v_\mathrm{orb}$, the orbital velocity. In addition, in the the extreme eccentricity limits ($e\rightarrow 0,1$), the perturbation cross section may be dominated by higher order (octopole) terms, which we do not treat in this work. For very extreme eccentricities, external perturbations may be more frequent than suggested by our equations.
    \item \textit{Dynamical model uncertainties}: {As discussed in \citetalias{PaperI}, a number of possible physical mechanisms -- such as tidal shocks \citep{Gnedin99} or the formation black hole subsystems \citep{Breen13, Giersz19} -- may influence the dynamical evolution of 47 Tuc. Such mechanisms could dynamically heat the cluster, and therefore the initial core density (in which we are are most interested) may have been larger than we have assumed. If the velocity dispersion was initially larger than in our model, we would expect fewer ionisations relative to circularisation and tidal destruction. Meanwhile, higher densities favour greater numbers of planets undergoing tidal destruction with respect to circularisation. }
\end{itemize}

\section{Conclusions}
\label{sec:conclusions}

In this work, we explore the apparent absence of close-in sub-stellar companions in the globular cluster 47 Tuc from a theoretical perspective. {In particular, we are interested in how HEM of (potential) HJs is influenced by dynamical encounters in dense stellar environments. This process requires numerous close passages with the host star, over which the tides raised in the planet lead to orbital circularisation. Planets in dense environments experience eccentricity perturbations during the time they are undergoing circularisation. Eccentricity perturbations can either excite large eccentricities which promote tidal interactions with the host star, or reduce the eccentricity to curtail such interactions. The net effect of this behaviour in a dense environment can be determined statistically, such that the overall outcome of HEM in globular clusters is predictable given an initial planet population. }

{To quantify the efficiency of HJ production via HEM in dense stellar environments, we develop a theoretical prescription for the evolution of the orbital eccentricity of a planet subject to stellar encounters in dense environments. In order for a planet to circularise, the rate of this eccentricity perturbation must be slower than the rate of tidal circularisation.} Using this principle, we derive a maximum semi-latus rectum $l_\mathrm{max}$ along which a planet can circularise at a given density (equation~\ref{eq:lmax}). {Planets can only circularise when $l<l_\mathrm{max}$, while in isolation they may circularise if $l<l_\mathrm{age}$ (the semi-latus rectum for which a planet circularises during the age of the star). This situation is depicted in Figure~\ref{fig:ea_evol}, which shows the outcome of numerical experiments tracking the evolution of semi-major axis and eccentricity for planets experiencing both dynamical perturbations and circularising tides. For sufficiently small $l_\mathrm{max} < l_\mathrm{age}$, planets that may have circularised in isolation instead undergo such close passages with their host star as to experience tidal disruption rather than circularisation. Thus, above some critical density approximated by equation~\ref{eq:n_crit}, HJ formation in dense environments becomes inefficient. We quantify the fraction of would-be HJs that are tidally disrupted, coupling this with the rate at which eccentricities sufficient to circularise are excited.} Combining our prescription with previously derived cross sections for ionisation, we derive analytic expressions for the fractions of planets that undergo ionisation, tidal disruption and circularisation in dense stellar environments. We find good agreement between these expressions and the numerical experiments of \citet{Hamers17}. {These expressions demonstrate that HJ formation is inefficient at extremely high densities, and could explain the apparent discrepancy between the occurrence rates of HJs in M67 \citep{Brucalassi16} and 47 Tuc \citep{Gil00, Weldrake05}.   }

Having validated our analytic expressions, we apply them to a Monte Carlo dynamical model of 47 Tuc using the \textsc{Mocca} code \citep{Gie98, Gie01}. We demonstrate that the efficiency of HJ formation in dense stellar environments is a strong function of the initial semi-major axis. The absence of HJs found in 47 Tuc therefore has consequences for the planet formation rates in general. {We find that, assuming the same initial occurrence rate of massive planets in 47 Tuc as for field stars as a function of semi-major axis, the expected HJ occurrence rate is $f_\mathrm{HJ} \approx 2.2 \times 10^{-3}$, which remains consistent with current constraints ($f_\mathrm{HJ,max} \approx 3.6 \times 10^{-3}$). {The HJ occurrence rate is sensitive to the initial planet occurrence rate at semi-major axis $a_0 \sim 1{-}30\,\rm{au}$. Thus, applying our theoretical framework, future transit surveys have the capacity to robustly determine the efficiency of planet formation in globular clusters.} In order to rule out a occurrence rates of planets in globular clusters similar to the field, a transit survey sample size of $\gtrsim 10^5$ stars is required at a similar sensitivity as that of \citet{Gil00}. Alternatively, higher sensitivity surveys may probe the prevalence of lower mass planets. }

{In this work and in \citetalias{PaperI} we have made the case for future efforts in searching for short period sub-stellar companions in globular clusters. Present constraints on their occurrence tells us little about the physics of star and planet formation in such environments. However, with a sufficiently large sample size, searches have the potential to constrain: 
\begin{enumerate}
    \item the environmental dependence of the sub-stellar IMF, due to the expected fraction of tidal brown dwarf captures \citepalias{PaperI};
    \item the occurrence rate of massive planets with respect to the solar neighbourhood, due to the role of  encounter-induced migration in generating HJs from planets in initially wider orbits (this work).
\end{enumerate}We thus conclude that there remains much to learn from future searches for planets in globular clusters by applying the mapping of the initial sub-stellar populations to the short-period companion fractions we have presented in this work. }

\section*{Acknowledgements}

{We thank the anonymous referee for their careful reading that improved the clarity of this manuscript.} AJW acknowledges funding from an Alexander von Humboldt Stiftung Postdoctoral Research Fellowship. CJC acknowledge support from the STFC consolidated grant ST/S000623/1. This work has also been supported by the European Union’s Horizon 2020 research and innovation programme under the Marie Sklodowska-Curie grant agreement No 823823 (DUSTBUSTERS). GR acknowledges support from the Netherlands Organisation for Scientific Research (NWO, program number 016.Veni.192.233) and from an STFC Ernest Rutherford Fellowship (grant number ST/T003855/1). This project has received funding from the European Research Council (ERC) under the European Union’s Horizon 2020 research and innovation programme (grant agreement No 681601) and been supported by the DISCSIM project, grant agreement 341137 funded by the ERC under ERC-2013-ADG. 




\bibliographystyle{mnras}
\bibliography{bdbib} 


\onecolumn
\appendix
\section{Hyperbolic perturbation cross-sections}
\label{app:numeric_de}

\subsection{Scaling of eccentricity perturbation}

{We are concerned with defining a cross-section for dynamical perturbations to an initial binary (star-planet system in this case) that result in a change of orbital eccentricity $\epsilon$ greater than some threshold $\epsilon_\mathrm{thr}$. \citet{Heggie96} derived a general expression for $\epsilon$ under the influence of a hyperbolic encounter where the perturber with impact parameter $b_\mathrm{pert}$ has eccentricity:
\begin{equation}
\label{eq:epert}
  e_\mathrm{pert} = \sqrt{1+ \frac{b_\mathrm{pert}^2 v_\infty^4}{G^2 m_\mathrm{tot}^2}}.
\end{equation}Here the total mass of the three components is $m_\mathrm{tot} = m_*(1+q)+m_\mathrm{pert}$ and $v_\infty$ is the relative speed of the perturber far from the barycentre. For the secondary (of mass $qm_*$) with position $\bm{r}$ with respect to the primary (of mass $m_*$), the eccentricity can be written:
\begin{equation}
    \bm{e} = \frac{1}{G m_*(1+q)} \dot{\bm{r}} \times (\bm{r}\times \dot{\bm{r}}) - \frac{\bm{r}}{r}
\end{equation} The force of the perturbing star, separation $\bm{R}$ from the primary, on the initial binary can be written in spherical harmonic form:
\begin{equation}
   \bm{F} =  \frac{G m_{\rm{pert}}}{R} \sum_{n=0}^\infty \frac{
m_*^{n-1} -(- qm_*)^{n-1}}{[m_*(1+q)]^{n-1}} \nabla_r \left [ \left(\frac{r}{R}\right)^n P_n \left(\frac{\bm{r}\cdot \bm{R}}{rR}\right)\right],
\end{equation}where $P_n$ is the $n^{\rm{th}}$ Legendre polynomial. The resulting acceleration of the secondary with respect to the primary is:
\begin{equation}
    \ddot{\bm{r}} = -\frac{Gm_*(1+q)\bm{r}}{r^3}  + \bm{F},
\end{equation} and the eccentricity changes as:
\begin{equation}
\label{eq:dot_e}
    \dot{\bm{e}} = \frac{2(\bm{F} \cdot \dot{\bm{r}})\bm{r} - (\bm{r} \cdot \dot{\bm{r}})\bm{F} - (\bm{F}\cdot \bm{r})\dot{\bm{r}}}{Gm_*(1+q)}.
\end{equation} In principle, equation~\ref{eq:dot_e} can now be integrated over time to give the change in eccentricity for a given perturber trajectory. In practice, this requires taking the lowest terms that do not vanish ($n=2$, quadrupole) and making the assumptions that the encounter is tidal and slow (see discussion in Section~\ref{sec:analytic_caveats}). The approach for this is given by \citet{Heggie75} and again in Appendix A of \citet{Heggie96}, with a sign correction. }

The resultant perturbation to eccentricity is necessarily dependent on all three of the usual orbital angles defining the orientation of the perturbers orbit with respect to the eccentric planet orbit. The line of nodes is the line of intersection of the orbital planes of the pertuber and the binary. The ascending node is then the point along this line where the perturber crosses the plane of the binary. The longitude of this ascending node $\Omega$ is defined in the sense of the binary angular momentum vector. The inclination between the two orbital planes is $i$, and $\omega$ be the longitude of pericentre of the third body, measured in its plane of motion from the ascending node, in the sense of its motion around the binary. With these definitions, for a binary with initial eccentricity $e_0$ the firsr order change in eccentricity is:
\begin{equation}
\label{eq:delta_e_general}
    \epsilon \approx \alpha \, y \, a_0^{3/2} r_\mathrm{p}^{-3/2} \left\{\Theta_1(\Omega, i) \chi +  \left[\Theta_2 (\Omega, i, \omega)+ \Theta_3(\Omega, i, \omega) \right] \psi \right\},
\end{equation} where we have defined:
\begin{equation}
    y \equiv e_0 \sqrt{1-e_0^2} \frac{m_\mathrm{pert}}{\sqrt{(1+q)m_* m_\mathrm{tot}}}
\end{equation} and
\begin{equation}
    \alpha = -\frac{15}{4}  {(1+e_\mathrm{pert})^{-3/2}} , \qquad \chi = \underbrace{\arccos\left(\frac{-1}{e_\mathrm{pert}} \right)}_{\chi_1} + \underbrace{\sqrt{e_\mathrm{pert}^2-1}}_{\chi_2}, \qquad \psi  = \frac 1 3\frac{(e_\mathrm{pert}^2- 1)^{3/2}}{e_\mathrm{pert}^2},
\end{equation}and $r_\mathrm{p}$ is the closest approach distance. We have absorbed all of the dependence on orientation into the $\Theta_k$:
\begin{equation}
\label{eq:def_theta}
    \Theta_1 = \sin^2 i \sin 2\Omega, \qquad \Theta_2 = (1+\cos^2 i)\cos 2 \omega \sin 2 \Omega, \qquad \Theta_3 = 2\cos i \sin 2 \omega \cos 2\Omega. 
\end{equation} This notation is convenient since each component $\Theta_k$ may vanish under particular assumptions about the orbit (e.g. $\Theta_{2,3}$ for a parabolic perturber orbit, $\Theta_1$ for $i=0^\circ$, $\Theta_3$ for $i=90^\circ$). 

\label{Encounter cross section}

Converting equation~\ref{eq:delta_e_general} into a cross section can be approached in two ways. The approach of \citet{Heggie96} is to find the domain $\mathcal{E}$ for which $\epsilon > 0$ (or equivalently $<0$), then intergrate out to the maximum impact parameter $b_\mathrm{max}$ for  which $|\epsilon| > \epsilon_\mathrm{pert}$. The case where the encounter is parabolic ($e_\mathrm{pert} \rightarrow 1$) is far simpler because $\psi=0$ and the dependence on $\Theta_2$ and $\Theta_3$ (and therefore $\omega$) is removed. The domain $\mathcal{E}$ in which $\epsilon$ is positive or negative is therefore trivially dependent on the sign of $\Theta_1$, and the corresponding domain of $\Omega$ is simply $[\pi/2, \pi]$ and $[3\pi/2, 2\pi]$ for positive $\epsilon$ (and the complement for negative). In this case, the perturbation cross section can be easily written as equation~\ref{eq:sig_pert1}, as established by \citet{Heggie96}. 

However, this approach does not work if $e_\mathrm{pert}\neq 1$. In this case we must take a slightly different (numerical) approach. If we do not care about the sign of $\epsilon$, then we are free to integrate over the full range of $\Omega$, $i$, and $\omega$; at each angle there exists some impact parameter $b_\mathrm{max}$ such that $|\epsilon|>\epsilon_\mathrm{thr}$ for $b_\mathrm{pert}< b_\mathrm{max}$ and $\epsilon_\mathrm{thr}>0$. The cross section is then:
\begin{equation}
\label{eq:sigma_epert_int}
    \sigma_\mathrm{pert} = \frac{1}{4\pi^2} \oint \mathrm{d}\mathrm{\Omega}\! \oint \! \mathrm{d}\omega \oint \! \mathrm{d}i\, \, \frac 1 2 \sin i \int_0^{b_\mathrm{max}(\Omega, i,\omega)} \!\!\!\mathrm{d}b \, 2\pi b = \frac{1}{4\pi} \int_0^{2\pi} \! \mathrm{d}\mathrm{\Omega} \! \int_0^{2\pi} \mathrm{d}\omega \! \int_0^{\pi} \! \mathrm{d}i\,  \,  \sin i \cdot {b_\mathrm{max}^2(\Omega, i,\omega)} .
\end{equation} We then need to solve equation~\ref{eq:delta_e_general} for $b_\mathrm{max}$ more generally. Noting that 
\begin{equation}
\label{eq:bmax}
   b_\mathrm{max}^2 = r_\mathrm{p,max}^2\left(   1+ \frac{2 G m_\mathrm{tot}}{r_\mathrm{p,max} v_\infty^2}\right)
\end{equation}for corresponding periastron distance $r_\mathrm{p,max}$, we write:
\begin{equation}
    r_\mathrm{p,max}^{3/2} = a_0^{3/2}\epsilon_\mathrm{thr}^{-1}  y \cdot \left| {\alpha} \left[\Theta_1 \chi +  \left(\Theta_2 + \Theta_3 \right) \psi \right]\right|.
\label{eq:rperi}
\end{equation}The right hand side of equation~\ref{eq:rperi} is valid because we allow positive or negative $\epsilon$ (i.e. $|\epsilon| > \epsilon_\mathrm{thr}>0$). Here $\alpha$, $\chi$ and $\psi$ are dependent on $e_\mathrm{pert}$, which is in turn dependent on $r_\mathrm{p, max}$ (via equations~\ref{eq:epert} and~\ref{eq:bmax}). No analytic solution is forthcoming for $r_\mathrm{p,max}$, hence we solve equation~\ref{eq:rperi} numerically across a range of $v_\infty$ to give $r_\mathrm{p,max}$ across a grid in $\Omega$, $i$, $\omega$. We then integrate equation~\ref{eq:sigma_epert_int} numerically substituting in the upper limit $b_\mathrm{max}$ from equations~\ref{eq:bmax} and~\ref{eq:rperi}. 

\begin{figure}
    \centering
    \includegraphics[width=0.5\textwidth]{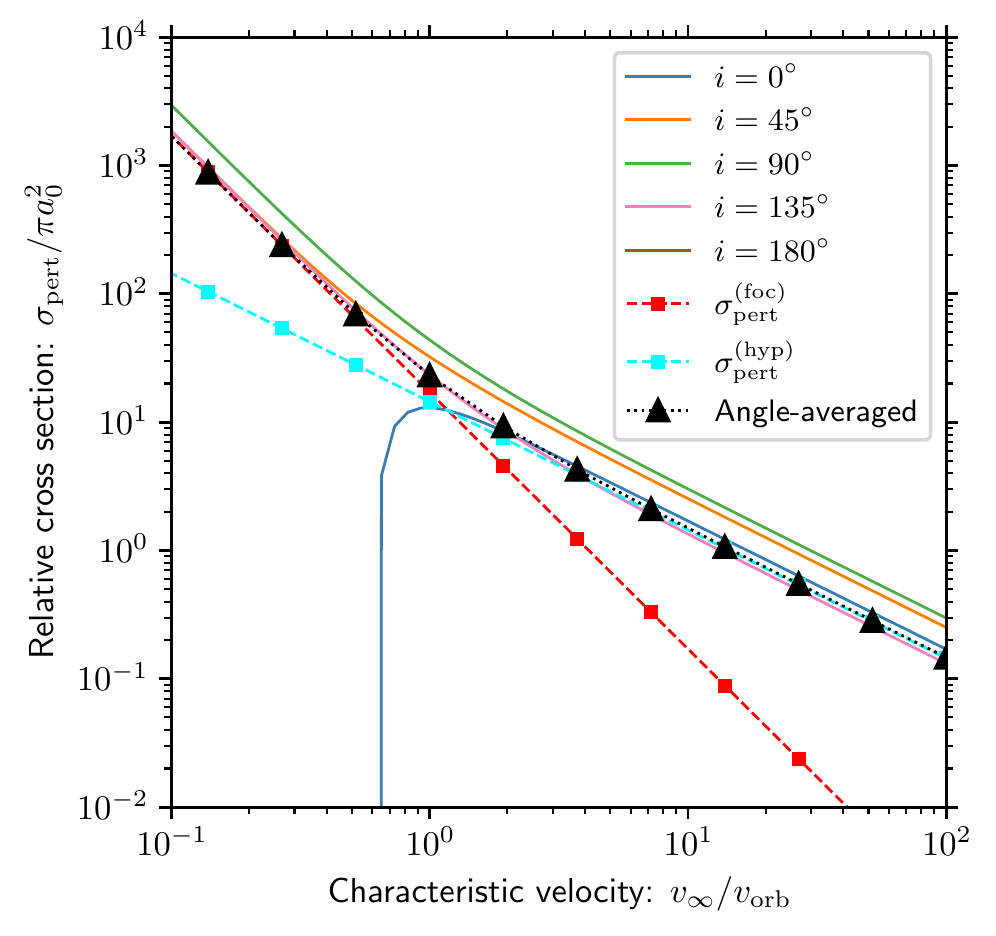}
    \caption{The cross section for perturbation, defined to be an interaction with $|\epsilon|>\epsilon_\mathrm{thr} = 0.05$, as a function of velocity of the perturber at infinity $v_\infty$ normalised by the characteristic orbital speed $v_\mathrm{orb}$ of the planet. In all cases, the initial eccentricity is $e_0=0.9$ and the perturber mass $m_\mathrm{pert}=0.5\,M_\odot = m_*$, the host star mass, while $q=0$. We show results of numerical evaluations for specific orientations with fixed $\omega=\Omega=15^\circ$ and varying inclination $i$ (the cross section vanishes for $i=180^\circ$). The angle averaged results calculated by numerically integrating equation~\ref{eq:sigma_epert_int} are shown by black triangles. The red squares show the corresponding cross sections in the gravitationally focused limit, as calculated by \citet{Heggie96}. The squares show the hyperbolic limit that we derive, with normalisation constant fitted to the angle averaged results.  }
    \label{fig:pert_cross}
\end{figure}
\begin{figure}
    \centering
   \subfloat[\label{subfig:pc_mp}]{ \includegraphics[width=0.45\textwidth]{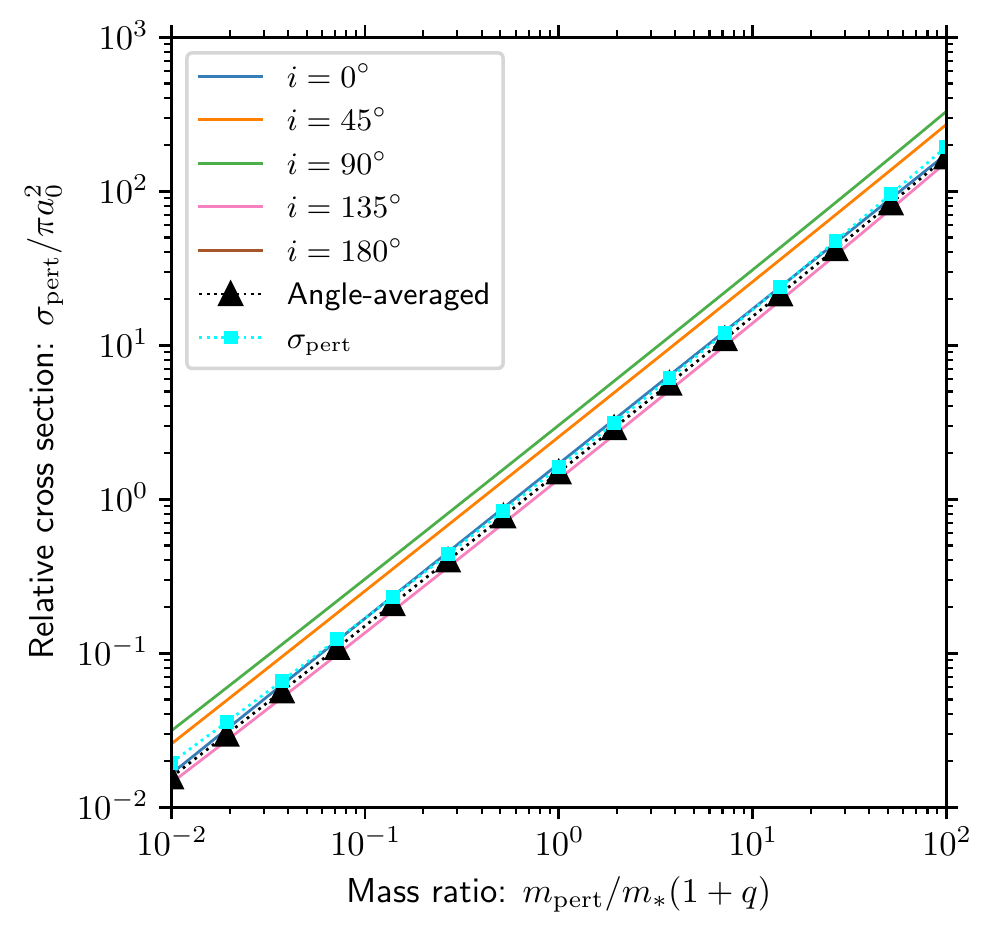}}
   \subfloat[\label{subfig:pc_e0}]{ \includegraphics[width=0.45\textwidth]{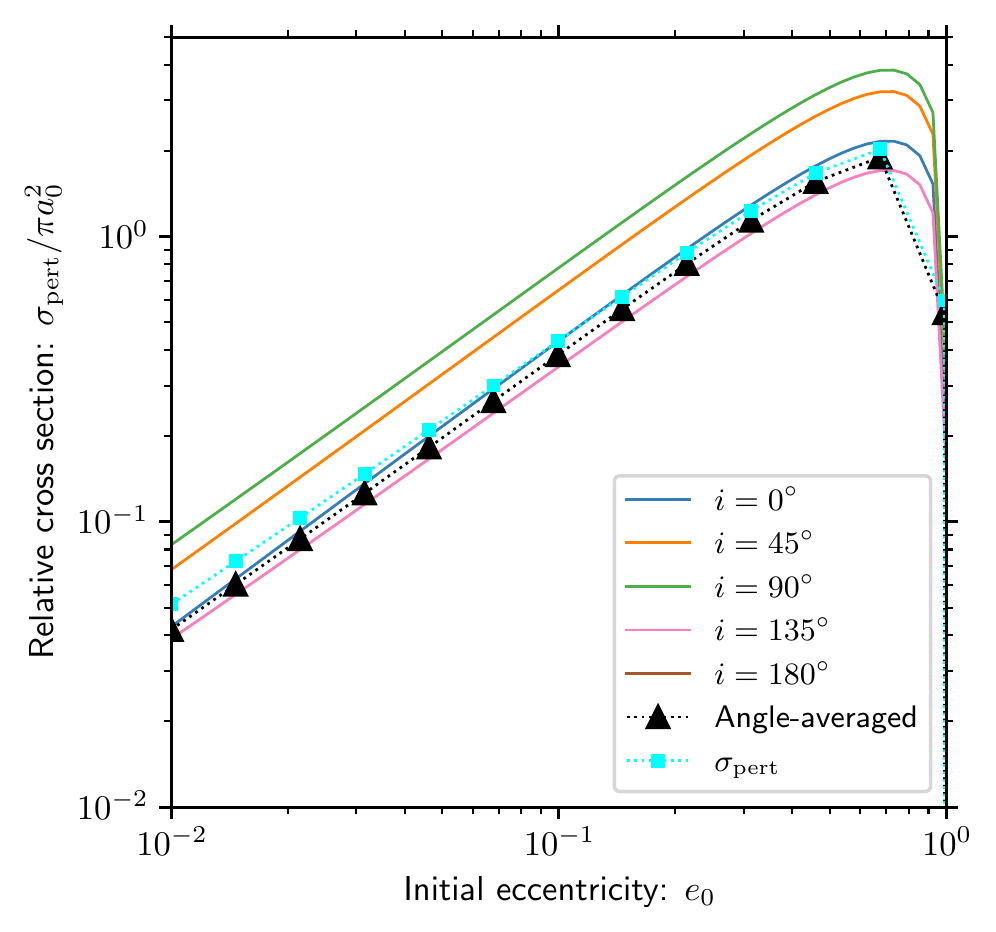}}
    \caption{As in Figure~\ref{fig:pert_cross} except for fixed $v_\infty/v_{\mathrm{orb}} = 10$ and varying $m_\mathrm{pert}$ (Figure~\ref{subfig:pc_mp}) and $e_0$ (Figure~\ref{subfig:pc_e0}). Because $v_\infty/v_{\mathrm{orb}} \gg 1 $ we are always in the hyperbolic limit, such that $\sigma_\mathrm{pert} \approx \sigma_\mathrm{pert} ^{\mathrm{(hyp)}}$. }
    \label{fig:pert_cross_emp}
\end{figure}

The results of this exercise are shown in Figure~\ref{fig:pert_cross} for our fiducial parameters ($a_0=5$~au, $e_0=0.9$, $m_\mathrm{p} = m_* = 0.5\,M_\odot$, $\epsilon_\mathrm{thr}=0.05$). We have also adopted $q=0$, the test particle limit, although this assumption would not significantly alter the results while $q\ll 1$. A number of features are require highlighting. Firstly, we consider the $i=0^\circ$ results (for $\omega=\Omega=15^\circ$), for which only the hyperbolic component contributes. In this case, we have $\Theta_1=0$ but $\Theta_{2,3}\neq 0$. This is in contrast to the general $e_\mathrm{pert}=1$ case where $\psi=0$ and hence $\Theta_2$ and $\Theta_3$ both effectively vanish. When $\psi\neq 0$, $\epsilon \propto  a_0^{3/2} y r_\mathrm{p}^{-3/2}e_\mathrm{pert}^{-1/2} \propto  a_0^{3/2} y r_\mathrm{p}^{-2} v_\infty^{-1} m_\mathrm{tot}^{1/2}$. Substituting these expressions back into equation~\ref{eq:sigma_epert_int}, we have:
\begin{equation}
\label{eq:sig_hyp_prop}
\sigma_\mathrm{pert}^\mathrm{(hyp)}  \propto b_\mathrm{max}^2 \propto r_\mathrm{p,max}^2 \propto  a_0^{3/2} y v_\infty^{-1}\epsilon_\mathrm{thr}^{-1}  m_\mathrm{tot}^{1/2} \propto a_0^2 \cdot  y \epsilon_\mathrm{thr}^{-1}   \left(\frac{v_\infty}{v_\mathrm{orb}} \right)^{-1} \sqrt{1+ \frac{m_\mathrm{pert}}{m_*(1+q)}},
\end{equation} which is shallower in $v_\infty$ than the focused version given by equation~\ref{eq:sig_pert1}. Meanwhile, taking the $\psi$ component for $e_\mathrm{pert}\rightarrow 1$ we have $\epsilon \propto y r_\mathrm{p}^{-3/2} (e_\mathrm{pert}-1)$. Hence $r_\mathrm{p, max}$ vanishes for fixed $\epsilon_\mathrm{thr}$ as $v_\infty \rightarrow v_\mathrm{orb}$ for decreasing $v_\infty$. In this case, it is possible that $|\epsilon|$ would in fact be dominated by the second order terms that apply to initially circular binaries \citep[see][]{Heggie96}. Since the contribution for $i=0^\circ$ is a special case that does not strongly influence our angle averaged result, we do not consider the second order terms here. Another special case where the first order terms vanish is $i=180^\circ$: $\Theta_1=0$ and $\Theta_2 = -\Theta_3$ when $\omega=\Omega$. Again, in this case second order terms must be computed, although this would not influence the angle averaged results.

More generally, if $\Theta_1 \neq 0$ then as $e_\mathrm{pert} \rightarrow \infty$ we have $\chi \propto  \psi$ such that $\epsilon \propto a_0^{3/2} y r_\mathrm{p}^{-2} v_{\infty}^{-1}$ as before. However, in the limit $e\rightarrow 1$ we have $\psi\rightarrow 0$ and $\chi_2 \rightarrow 0$, but in this case $\chi \rightarrow \chi_1 \rightarrow \pi$. We then have $\epsilon \propto a_0^{3/2} y r_\mathrm{p}^{-3/2}$ only -- i.e. independent of $v_\infty$. Then 
\begin{equation}
    \sigma_\mathrm{pert}^\mathrm{(foc)}  \propto b_\mathrm{max}^2  \propto  r_\mathrm{p, max} m_\mathrm{tot} v_\infty^{-2} \propto a_0 \epsilon^{-2/3} y^{2/3} m_\mathrm{tot} v_\infty^{-2} \propto a_0^2 \cdot  \epsilon_\mathrm{thr}^{-2/3} y^{2/3} \left(\frac{v_\infty}{v_\mathrm{orb}}\right)^{-2} \left[1 +\frac{m_\mathrm{pert}}{m_*(1+q)} \right],
\end{equation} as in equation~\ref{eq:sig_pert1}. We can therefore write the ratio of the hyperbolic to focused cross section:
\begin{equation}
    \Delta_\mathrm{hf} = \sigma_\mathrm{pert}^\mathrm{(hyp)} /\sigma_\mathrm{pert}^\mathrm{(foc)} = C_\mathrm{hf} \Delta'_\mathrm{hf} \propto a_0^{1/2}\epsilon_\mathrm{thr}^{-1/3} y^{1/3} m_\mathrm{tot}^{-1/2} v_\infty \propto \epsilon_\mathrm{thr}^{-1/3} y^{1/3}  \frac{v_\infty}{v_\mathrm{orb}} \left[1 +\frac{m_\mathrm{pert}}{m_*(1+q)} \right]^{-1/2},
\end{equation}where $C_\mathrm{hf}$ is a constant which can be obtained by numerically computing the angle averaged cross section in the hyperbolic case. We therefore have a general encounter cross section:
\begin{equation}
\label{eq:app_spert}
   \sigma_\mathrm{pert} = \sigma_\mathrm{pert}^{\mathrm{(foc)}} \left[ 1+\Delta_\mathrm{hf} \frac{v_\infty}{v_{\mathrm{orb}}}  \right] = \sigma_\mathrm{pert}^{\mathrm{(foc)}} \left\{ 1+C_\mathrm{hf}  {y}^{1/3}{\epsilon_\mathrm{thr}}^{-1/3}  \left[1 +\frac{m_\mathrm{pert}}{m_*(1+q)} \right]^{-1/2} \frac{v_\infty}{v_{\mathrm{orb}}}  \right\},
\end{equation}where $C_\mathrm{hf} \approx 0.67$ is a factor numerically computed from the angle averaged results (Figure~\ref{fig:pert_cross}) for our fiducial parameters: $e_0=0.9$, $a_0=5$~au, $m_\mathrm{pert}=m_* = 0.5 \, M_\odot$, $q=0$ and $\epsilon_\mathrm{thr} = 0.05$. 

As a sanity check, we compare the equation~\ref{eq:app_spert} with the fitted constant $C_\mathrm{hf}$ to the numerical integrations with variable $m_\mathrm{pert}$ and $e_0$ in Figure~\ref{fig:pert_cross_emp} for the hyperbolic limit ($v_\infty/v_\mathrm{orb}=10$). In both cases, we find good agreement between the analytic scaling and the numerical calculation. For variable $m_\mathrm{pert}$ (Figure~\ref{subfig:pc_mp}), we obtain the linear scaling with $m_\mathrm{pert}$ we expect because:
\begin{equation}
    \sigma_\mathrm{pert}^{\mathrm{(hyp)}} \propto y \sqrt{1+q_\mathrm{pert}} = e_0 \sqrt{1-e_0^2} q_\mathrm{pert},
\end{equation}where $q_\mathrm{pert}\equiv {m_\mathrm{pert}}/{(1+q) m_*}$. In the varying $e_0$ case, Figure~
\ref{subfig:pc_e0}, the perturbation peaks at $e_0 = \sqrt{2}/2$. We conclude that equation~\ref{eq:app_spert} is a valid approximation for the effective perturbation cross section of an eccentric binary (or star-planet system). 

\subsection{Applicability of the analytic cross section}
\label{sec:analytic_caveats}
{The two primary assumptions made by \citet[][see also \citealt{Heggie75}]{Heggie96} are that encounters are \textit{tidal} and \textit{slow}. The tidal condition is that the closest approach distance $r_\mathrm{p}$ of the perturber considerably exceeds the semi-major axis $a_0$. In the hyperbolic limit, we have $\sigma_\mathrm{pert} \approx \pi r_\mathrm{p}^2$ and therefore from equation~\ref{eq:sigma_hyp} we require:
\begin{equation}
\label{eq:tidal_cond}
 y\sqrt{1+ \frac{m_\mathrm{pert}}{m_*(1+q)}} \left(\frac{v_\infty}{v_\mathrm{orb}} \right)^{-1} \gg C_\mathrm{hyp}^{-1} \epsilon_\mathrm{thr} ,
\end{equation}where $C_\mathrm{hyp}$ is the constant factor for the RHS of equation~\ref{eq:sig_hyp_prop} that turns the expression into an equality. The second requirement, that encounters are slow, is used to average over the binary (star-planet) orbit to obtain the change of eccentricity. This requires that the angular velocity of the perturber at closest approach is slower than the angular velocity of the binary. In the hyperbolic limit, this is equivalent to the condition:
\begin{equation}
\label{eq:slow_cond}
 \sqrt{\frac{m_\mathrm{pert}}{m_\mathrm{tot}} \left(1+\frac{m_\mathrm{pert}}{m_*(1+q)} \right)} \frac{r_\mathrm{p}}{a_0}  \gg \frac{v_\infty}{v_\mathrm{orb}} ,
\end{equation} or
\begin{equation}
   y {\frac{m_\mathrm{pert}}{m_\mathrm{tot}} \left(1+\frac{m_\mathrm{pert}}{m_*(1+q)} \right)^{3/2}}   \left(\frac{v_\infty}{v_\mathrm{orb}} \right)^{-3} \gg  C_\mathrm{hyp}^{-1} \epsilon_\mathrm{thr}.
\end{equation} The second of these expressions is more restrictive in our case where $v_\infty\gtrsim v_\mathrm{orb}$ and both are satisfied for sufficiently small $\epsilon_\mathrm{thr}$. In general we are interested in the many encounters that result in small changes in the eccentricity, such that we expect the approximations to hold. In either case, we are also never clearly in the fast encounter regime where the opposite of equation~\ref{eq:slow_cond} applies. It is therefore not helpful to rework the cross sections in this limit \citep[although see Section 3.2 and Appendix A4 of][]{Heggie96}. In this work, we always adopt the slow encounter expressions. }


\section{Statistical eccentricity evolution}
\label{app:stat_e_evol}
In Appendix~\ref{app:numeric_de} we derived the general perturbation cross section for an encounter with a binary that yields a change of eccentricity of magnitude $\epsilon_\mathrm{thr}$. In the limit of small $\epsilon_\mathrm{thr} \rightarrow 0$, the time-scale for encounters $\tau_\mathrm{pert} \rightarrow 0$ and we are in the continuum limit of many distant encounters. In this case, we can model the evolution of the planet eccentricity $e$ as a random walk. To do this, we first assume that for a small change $\tau$ from time $t$, the probability of a change eccentricity $e_0$ of magnitude greater than $\epsilon_0$ is $\psi(e_0, \epsilon_0)$. For initial eccentricity within $\epsilon_0/2$ of $e_0$, the chance of having a new eccentricity $e>e_0$ where $e \in [e_1 - \epsilon_{\rm{l}}, e_1+\epsilon_{\rm{r}}]$ where $\epsilon_{\rm{l}} = \epsilon_{\rm{r}} = \epsilon_1/2$ after time $\tau$ is therefore:
\begin{equation}
\label{eq:p_plus}
  \epsilon_1  p_{+}(e_0; e_1, t+\tau)= \frac{\epsilon_0 }{2}p(e_0, t) \left[\psi(e_0, e_1-e_0-\epsilon_{\rm{l}}) -  \psi(e_0, e_1-e_0+\epsilon_{\rm{r}}) \right] ,
  \end{equation}where the factor $1/2$ comes from the positive $\epsilon_\mathrm{thr}$ part of the cross section computed in Appendix~\ref{app:numeric_de}. In the limit of small $\epsilon_\mathrm{l}$ this becomes:
  \begin{equation}
     \epsilon_1  p_{+}(e_0; e_1, t+\tau)=     -\frac{\epsilon_0 \epsilon_1}{2} p(e_0, t) \partial_{(e_1-e_0)} \psi =  \frac{\epsilon_0 \epsilon_1}{2}  p(e_0, t) |\partial_{e_0} \psi|.
\end{equation}Considering also the probability density $p_-$ of a planet being scattered away from the neighbourhood of $e_1$ and the probability density $p_0$ of it already occupying the neighbourhood without being scattered out, we have:
\begin{equation}
   \epsilon_1 p(e_1, t+\tau) = \epsilon_1 p_0(\epsilon_1, t) - \epsilon_1 p_-(e_1, t) +  \epsilon_1 \int_{\varepsilon} p_+(\tilde{e}; e_1, t) \,\mathrm{d}\tilde e,
\end{equation}where $\varepsilon$ is the complement of the local eccentricity space $\varepsilon = \overline{[e_1-\epsilon_{\rm{l}}, e_1+\epsilon_{\rm{r}}]}$. The probability of evacuating the enclosed region is:
\begin{equation}
   \epsilon_1 p_-(e_1,t) =  \epsilon_{\rm{l}} p(e_1, t)\psi(e_1,\epsilon_\mathrm{l} ) +  \epsilon_{\rm{r}} p(e_1, t)\psi(e_1,\epsilon_\mathrm{r} ) .
\end{equation}
The probability of having eccentricity within a small range $\epsilon$ of $e$ is therefore:
\begin{equation}
    \epsilon p(e, t+\tau)  = \epsilon p(e,t)  \left[1- \frac{1}{2}\psi(e,\epsilon_{\rm{l}}) - \frac{1}{2} \psi(e, \epsilon_{\rm{r}})  \right]+ \frac{\epsilon_\mathrm{l}}{2}\int_0^{e-\epsilon_{\rm{l}}} p(\tilde e, t) |\partial_{\tilde e}  \psi(\tilde e, e- \tilde e)| \, \mathrm{d} \tilde e +\frac{\epsilon_\mathrm{r}}{2} \int_{e+\epsilon_{\rm{r}}}^1 p(\tilde e, t) | \partial_{\tilde e} \psi(\tilde e,  \tilde e-e)| \, \mathrm{d} \tilde e.
\end{equation}We can then integrate the last terms by parts, noting that the sign of the derivative $\partial_{e_1-e_0} \psi$ changes sign for $e_0<e_1$ and $e_0>e_1$, as does the sign of the differential distance from $e$ in the two integrals. However, this is not necessarily true for the derivative $\partial_{e_1-e_0} p$, hence the sign of the integral is changed above and below $e$. Finally, we must approximate the functional form of the probability $\psi$, which comes from the instantaneous rate $\Gamma(e,\epsilon)$ of pertubations to $e$ greater than $\epsilon$:
\begin{equation}
    \psi(e,\epsilon) = 1 -\exp\left(-\int_0^\tau \, \Gamma(e,\epsilon) \mathrm{d}t \right) \approx \Gamma \tau = \gamma e\sqrt{1-e^2} \frac{\tau}{\epsilon},
\end{equation}where we have used the hyperbolic perturbation rate computed in Section~\ref{sec:pert_rate}. In this case:
\begin{equation}
\Gamma \equiv \frac{\epsilon}{\tau} = \gamma e\sqrt{1-e^2} \epsilon^{-1},
\end{equation} such that equation~\ref{eq:partial_de} be written in terms of the drift diffusion equation with variable diffusivity:
\begin{equation}
\label{eq:PDE}
    \partial_t p(e, t) =\frac{\gamma}{2} \partial_{e} \left[ e \sqrt{1-e^2} \partial_e p(e,t) \right] -  \lim_{\epsilon \rightarrow 0} \left\{\Delta_- -\Delta_+\right\}.
\end{equation}The last two terms are:
\begin{equation}
    \Delta_- =  \frac \gamma 2 \int_0^{e-
    \epsilon} \partial_{\tilde e} p(\tilde e, t) \frac{ \tilde e \sqrt{1-\tilde e^2}}{e - \tilde e}  \, \mathrm{d} \tilde e \qquad    \Delta_+ =  \frac \gamma 2 \int_{e+
    \epsilon}^1 \partial_{\tilde e} p(\tilde e, t) \frac{ \tilde e \sqrt{1-\tilde e^2}}{ \tilde e-e}  \, \mathrm{d} \tilde e ,
\end{equation} which are the drift terms. Both of these expressions appear to diverge as $\epsilon \rightarrow 0$. However, the two contributions actually cancel close to $e$. To show this, let us assume that $\Delta_{\pm}$ are both dominated by the contribution of the integrand close to $e$. Then as $\epsilon \rightarrow 0$, we consider a small region of size $\epsilon' \gg \epsilon$ around $e$ over which we estimate the value of the integral by the midpoint approximation:
\begin{equation}
  \Delta_- \approx   \frac \gamma 2 \epsilon' \frac{\partial_e p (e-\epsilon'/2, t)\cdot (e-\epsilon'/2)\sqrt{1-(e-\epsilon'/2)^2}}{\epsilon'/2} \qquad   \Delta_+  \approx  \frac \gamma 2 \epsilon' \frac{\partial_e p (e+\epsilon'/2, t)\cdot (e+\epsilon'/2)\sqrt{1-(e+\epsilon'/2)^2}}{\epsilon'/2}.
\end{equation}Thus, if $p(e,t)$ is twice continuously differentiable at $e$ we have:
\begin{equation}
    \lim_{\epsilon' \rightarrow 0} \left\{\Delta_+ -\Delta_- \right\} \approx \gamma \epsilon' \partial_{e} \left[ e\sqrt{1-e^2} \partial_e p(e, t) \right] \longrightarrow 0.
\end{equation}We have therefore shown that the contribution of $\Delta_- - \Delta_+$ is finite as $\epsilon \rightarrow 0$. The PDE described by equation~\ref{eq:PDE} can therefore be computed numerically. 

\section{Random walk circularisation experiments}
\label{app:rw_exp}

\subsection{Circularisation radii in the rapid encounter limit}

\begin{figure}
    \centering
    \includegraphics{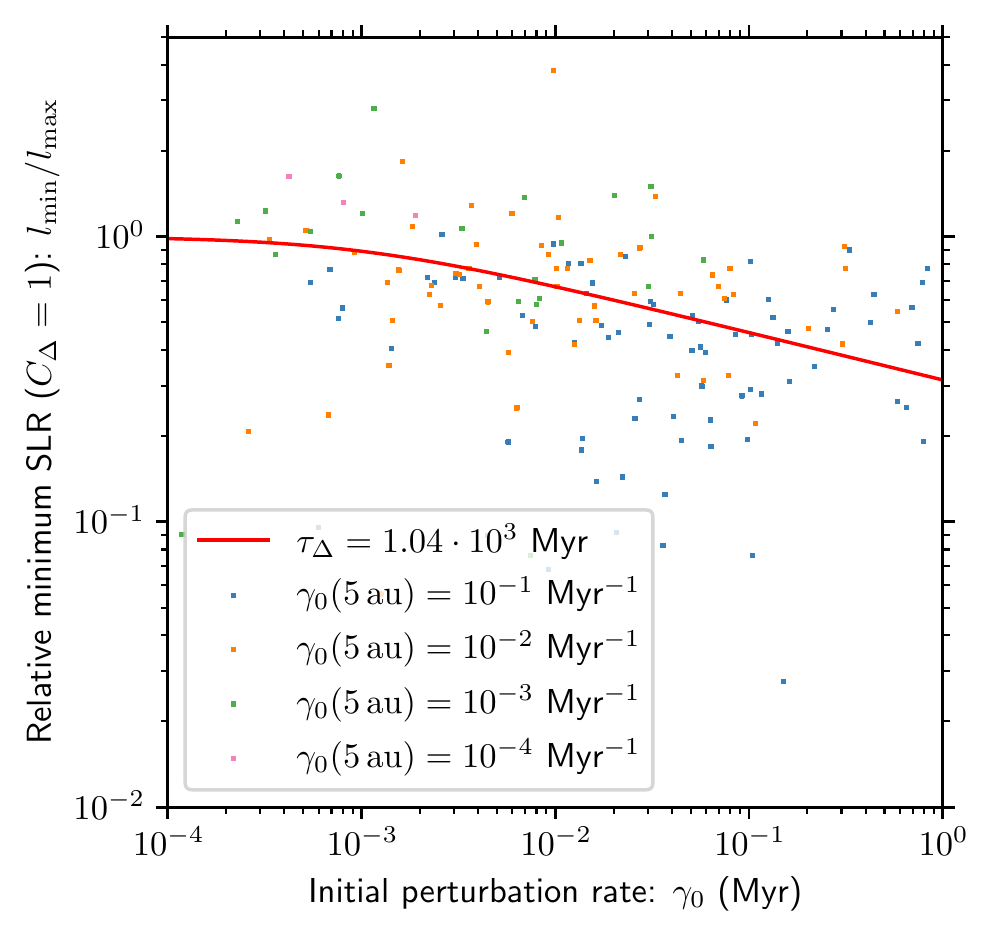}
    \caption{The distribution of the minimum semi-latus rectum (SLR, $l_\mathrm{min}$) distribution for planets that circularise during our fiducial random walk experiments ($e_0=1$) as a function of the initial perturbation rate $\gamma_0$. The ratio of the minimum SLR to the maximum value $l_\mathrm{max}$ predicted with the pre-factor $\mathcal{C}_\Delta=1$ is adopted. The red line shows the best fit assuming the functional form described by equation~\ref{eq:C_delta}, with the fitting parameter $\tau_\mathrm{\Delta}=1.04\cdot 10^3$~Myr.    }
    \label{fig:C_Delta_fit}
\end{figure}

During the analytic derivation of the maximum semi-latus rectum (SLR) along which a planet can circularise, we first assumed in Section~\ref{sec:tidal_acc} that we are able to ignore the non-local terms in the PDE ($\Delta$ in equation~\ref{eq:partial_de}) that describes the statistical evolution of the planet eccentricity. However, this may not always be the case. In particular, for extreme eccentricities $e\rightarrow 1$, the diffusion coefficient that scales with $e\sqrt{1-e^2}$ becomes small. Thus if the minimum eccentricity $e_\mathrm{min}$ required for a planet to circularise (equation~\ref{eq:emin}) is sufficiently large, then individual encounters that result in comparatively large changes in $e$  -- i.e. $|\epsilon | \gg 1-e_\mathrm{min}$ -- can dominate for circularising planets over many weak encounters.  In this appendix, we will refer to encounters with a change of eccentricity $|\epsilon| \gg 1-e_\mathrm{min}$ as `strong' encounters, while $|\epsilon| \lesssim 1-e_\mathrm{min}$ are `weak' encounters. In the strong encounter regime, we must apply a correction factor for the contribution of these encounters in producing circularising planets. 

    Our approach for quantifying this correction factor is semi-empirical. We reason that the factor is $\mathcal{C}_\Delta=1$ for the weak encounter regime. In this case, strong encounters rarely yield circularisation outcomes because $e_\mathrm{min}\sqrt{1-e_\mathrm{min}^2}$ remains large. By contrast, when encounter rates are frequent, the required $e_\mathrm{min}\sqrt{1-e_\mathrm{min}^2}$ for circularisation becomes small. In the latter case, strong encounters may not be followed by sufficient numbers of weak encounters to influence the final circularisation radius. Thus the maximumum SLR (SLR, $l_\mathrm{max}$) scales more steeply with the initial encounter rate $\gamma_0$ than suggested by equation~\ref{eq:lmax}. From the governing PDE, equation~\ref{eq:partial_de}, when the diffusion coefficient $D \propto \gamma e\sqrt{1-e^2}$ (equation~\ref{eq:diffusion_coeff}) becomes small then the additional term $\Delta \cdot \gamma/2$ additionally contributes to the rate of eccentricity evolution. This non-local term becomes important when:
    \begin{equation}
         \gamma\frac{e_\mathrm{min}\sqrt{1-e_\mathrm{min}^2}}{1-e_\mathrm{min}}  \gtrsim \tau_\mathrm{circ}^{-1},
    \end{equation}where $ \tau_\mathrm{circ}$ is defined at semi-latus rectum $l_\mathrm{max}$ and corresponding eccentricity $e_\mathrm{min}$. In general, we have $ l_\mathrm{max} \propto 1-e_\mathrm{min}^2 \approx 2(1-e_\mathrm{min}) $, which is thus only weakly dependent on the local encounter rate (equation~\ref{eq:lmax}). Hence the relative importance of the non-local term is $\propto \tau_\mathrm{circ} \gamma$ in the rapid encounter rate limit. We thus estimate the correction factor:
\begin{equation}
\label{eq:C_delta}
    \mathcal{C}_\Delta \approx \left[1+  \tau_\mathrm{\Delta} \gamma_0 \right]^{-1} ,
\end{equation}where $\tau_\mathrm{\Delta}$ represents a constant time-scale that is an empirical fitting parameter. The second term on the RHS of equation~\ref{eq:C_delta} scales with $\gamma_0$ as $e_\mathrm{min}\rightarrow 1$ and the non-local terms dominate the encounter rate, while it remains of order unity for moderate $e_\mathrm{min}$. 

In Figure~\ref{fig:C_Delta_fit} we show the outcome of the circularisation experiments we present in Section~\ref{sec:circ_radii}. Specifically, we show the minimum SLR $l_\mathrm{min}$ achieved by each planet undergoing a random eccentricity walk and subject to tidal forces. We normalise each $l_\mathrm{min}$ by the maximum SLR $l_\mathrm{max}$ predicted by equation~\ref{eq:lmax} with $\mathcal{C}_{\Delta}=1$. We then consider this ratio as a function of $\gamma_0$, and fit an appropriate value for $\tau_\Delta$ in equation~\ref{eq:C_delta} using the \textsc{Scipy} \citep{Virtanen20} package \texttt{optimize.minimize}. We obtain $\tau_\Delta = 1.04\cdot 10^3$~Myr, which appears to reproduce the suppression in $l_\mathrm{max}$ at extreme $\gamma_0$ values (Figure~\ref{fig:C_Delta_fit}). We therefore adopt the corresponding definition of $\mathcal{C}_\Delta$.

\subsection{Circularisation time-scale and eccentricity dependence}

\begin{figure*}
    \centering
    \subfloat[\label{subfig:tcirc_e01}$e_0=0.1$]{\includegraphics[width=0.5\textwidth]{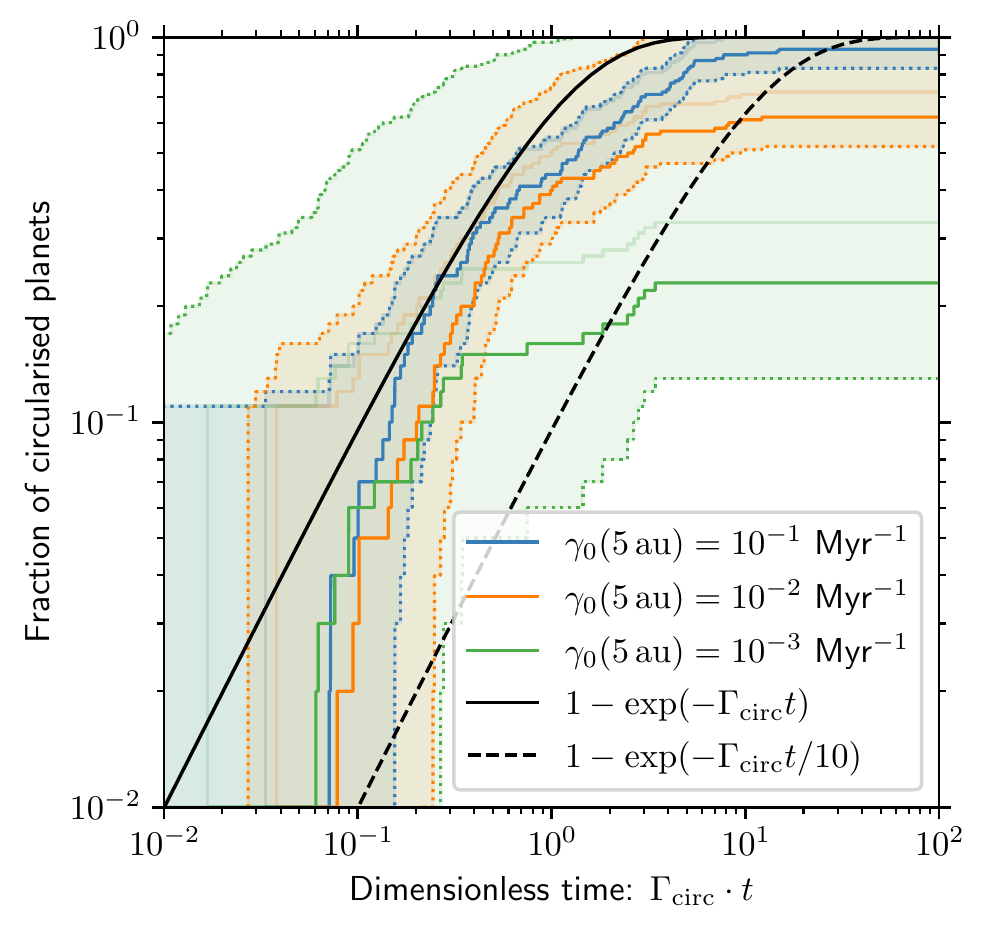}}
    \subfloat[\label{subfig:tcirc_e06} $e_0=0.6$~au]{\includegraphics[width=0.5\textwidth]{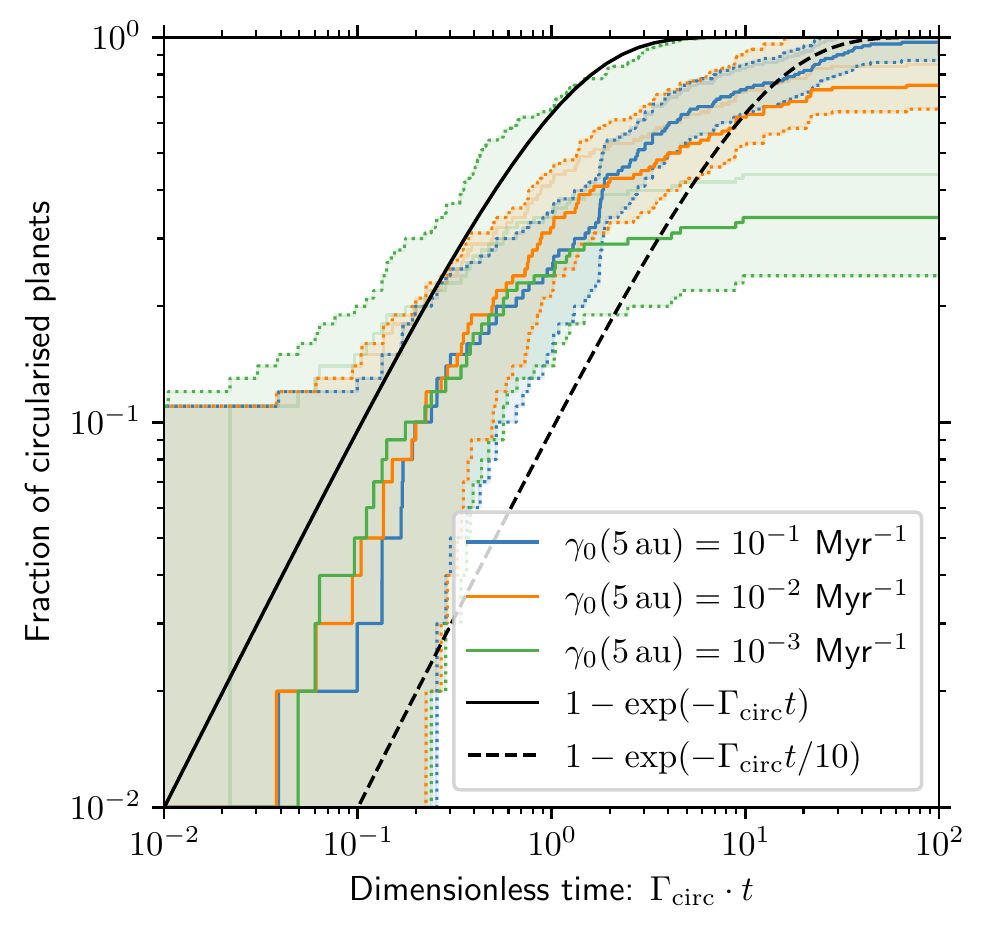}}
    \caption{Distribution of the circularisation times obtained from random walk experiments with initial eccentricities $e_0=0.1$ (Figure~\ref{subfig:tcirc_e01}) and $e_0=0.6$ (Figure~\ref{subfig:tcirc_e06}). The time coordinate is mutliplied by $\Gamma_\mathrm{circ}$ as approximated by equation~\ref{eq:Gamma_circ}, which is the rate at which circularisation outcomes occur due to a single encounter. The solid black line shows the expected circularisation fraction if the true rate is $\Gamma_\mathrm{circ}$, while the dashed line shows the equivalent fraction if the rate is $\Gamma_\mathrm{circ}/10$. Circularisation over long time-scales is limited by the the integration time of our simulation, which is $10$~Gyr. This contributes to the statistical uncertainties in the fraction of circularised planets, which are shown as shaded regions. }
    \label{fig:tcirc}
\end{figure*}

We are interested in understanding how the typical time-scale required for circularisation depends on the initial eccentricity of the planet $e_0$. To do this, we first make an analytic estimate of the expected rate at which planet circularise due to dynamical perturbations. To first order, we adopt the single-encounter approximation:
\begin{equation}
\label{eq:Gamma_circ}
    \Gamma_\mathrm{circ} \approx \frac{\gamma_0 e_0 \sqrt{1-e_0^2}}{2\, (e_\mathrm{min}-e_0)},
\end{equation}which is equivalent to the $\Gamma_\mathrm{pert}^{\rm{(hyp)}}/2$ with $\epsilon_\mathrm{thr}=e_\mathrm{min}-e_0$ from equation~\ref{eq:Gamma_pert_hyp}. This is not an exact rate at which circularisation is instigated for a perturbed planetary systems because we assume that a single large encounter produces the required change in eccentricity. We thus ignore the many smaller encounters that result in a random walk in eccentricity that may increase or reduce the circularisation time-scale.

In Figure~\ref{subfig:tcirc_e01} we show the fraction of circularised planets from our numerical experiments presented in Section~\ref{sec:circ_radii}, with initial eccentricity $e_0=0.1$. In addition, we show the same experiment with an initial eccentricity $e_0=0.6$ in Figure~\ref{subfig:tcirc_e06}. We normalise the time coordinate by multiplying by $\Gamma_\mathrm{circ}$. We can then show the expected fraction of circularised planets:
\begin{equation}
\label{eq:Pcirc}
    P_\mathrm{circ} = 1- \exp(-\Gamma_\mathrm{circ} t)
\end{equation}
as a solid black line in Figure~\ref{fig:tcirc}. We find that the distribution of normalised circularisation time-scales is similar for both $e_0=0.1$ and $e_0=0.6$. For planets that circularise in time $\Gamma_\mathrm{circ}t \sim 0.1{-}1$, both distributions are well-described by equation~\ref{eq:Pcirc}. However, the distribution deviates from this expectation at the extreme ends of the distribution. For large $\Gamma_\mathrm{circ}t$, we are limited by the integration time-scale ($10$~Gyr). The fraction of planets that circularise at early times is limited by the initial time-step ($0.1$~Myr). In general, the true value of $\Gamma_\mathrm{circ}$ is well approximated by equation~\ref{eq:Gamma_circ} within the uncertainties in our random walk experiments.   

\section{Analytic estimate vs. \textsc{Fewbody}}
\label{app:analytic_vs_MC}

 \begin{figure}
 \centering
 \includegraphics[width=0.5\columnwidth]{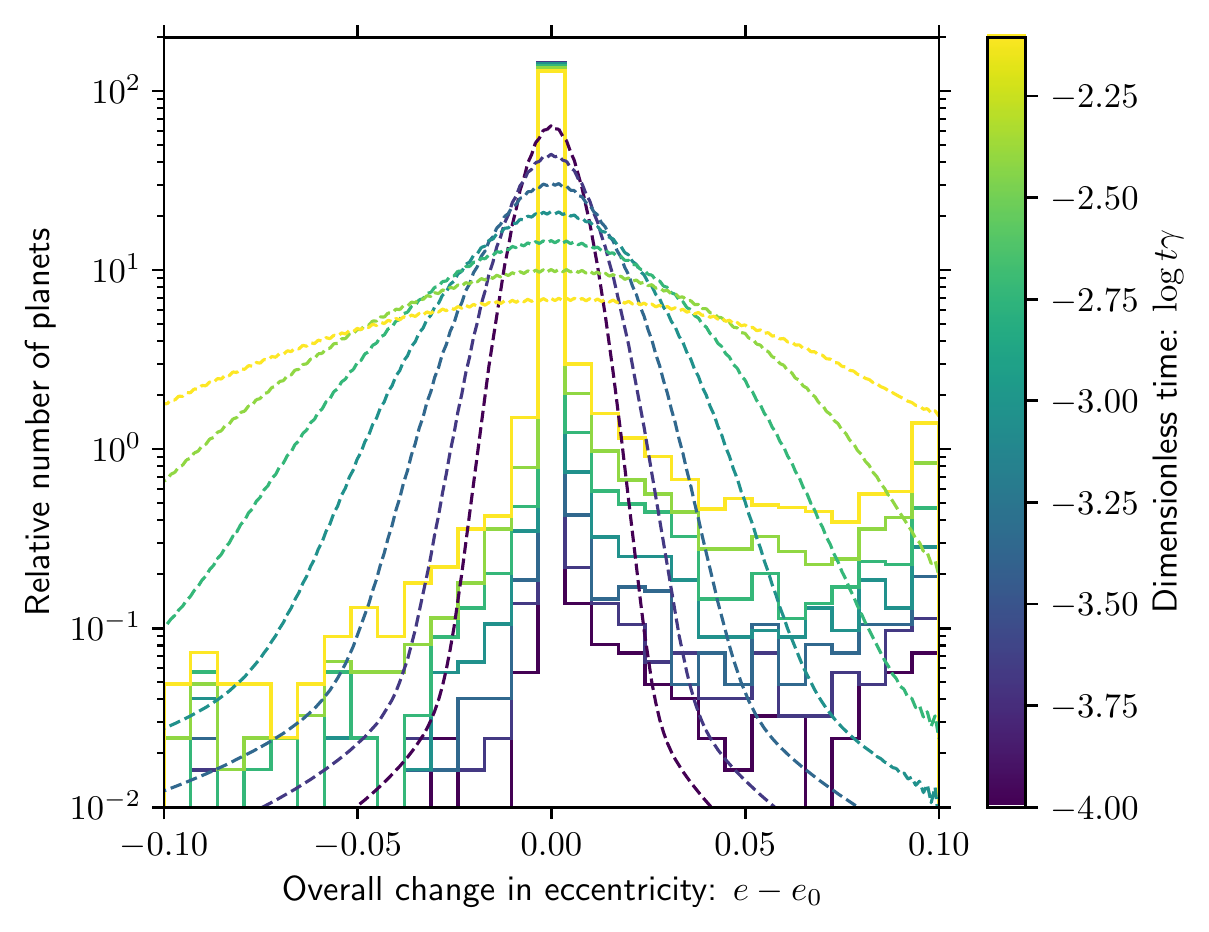}
    \caption{Statistical distribution of the orbital eccentricities for planets evolving in our dynamical model of 47 Tuc. The solid histograms are obtained directly from the results of \textsc{Fewbody}, implemented within the \textsc{Mocca} code. The dashed lines are the solution of the PDE described by equation~\ref{eq:PDE}. The colour bar shows the time evolution normalised by the encounter rate $\gamma$. } 
    \label{fig:eccdiff_comp}
\end{figure}

  \begin{figure}
    \centering
     \includegraphics[width=0.5\columnwidth]{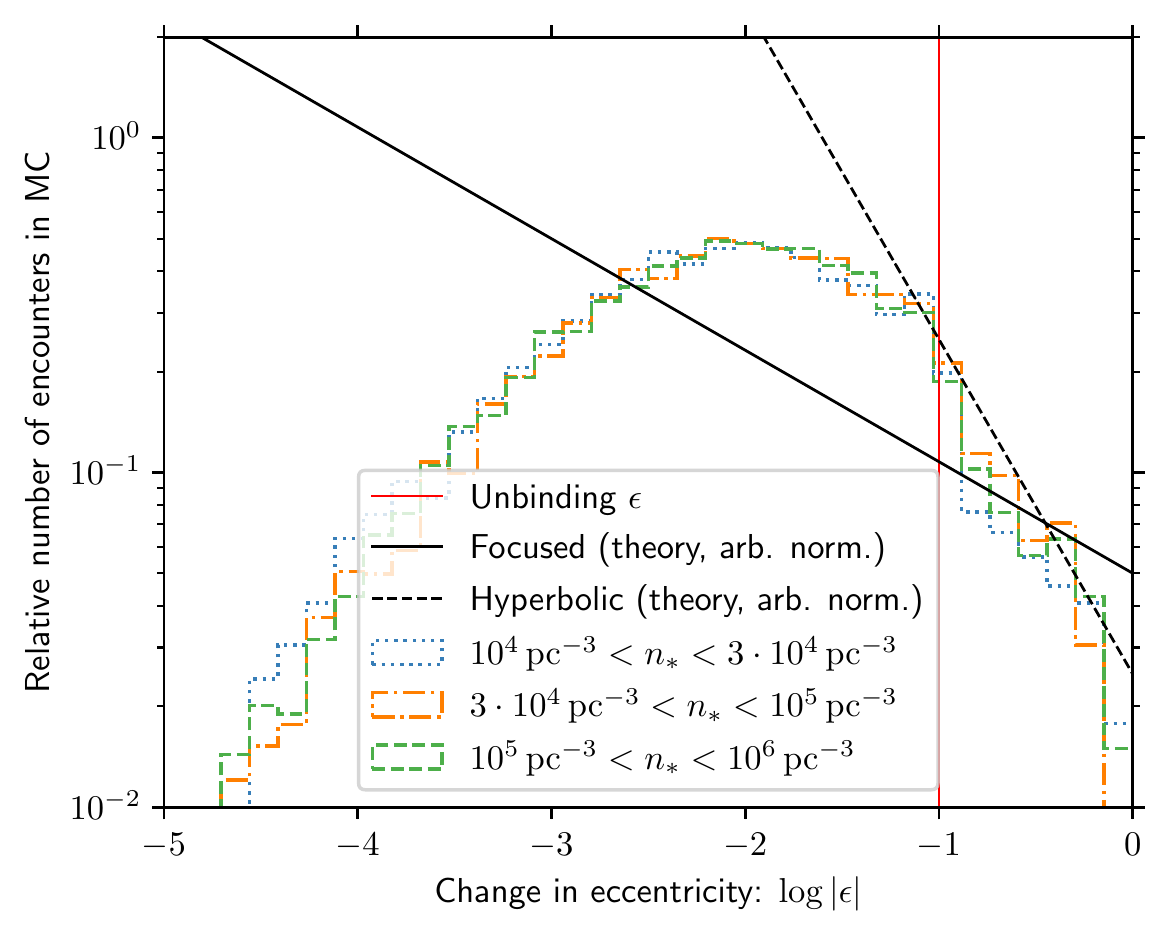}
     \caption{Histogram of the relative number of encounters that change a planets eccentricity by $|\epsilon|$ during our \textsc{Mocca} simulation, binned by initial local stellar density. We include only encounters that do not unbind the planet. For changes in eccentricity $|\epsilon|>0.1$ (vertical red line) we apply a weighting of $2$ because only encounters with a negative $\epsilon$ are possible. The two black lines show the theoretical distribution for focused encounters (solid line) and hyperbolic encounters (dashed line).   }
    \label{fig:rel_encs}
\end{figure}

Here we compare the rate at which eccentric planets are perturbed in the Monte Carlo model using the \textsc{Fewbody} code with the theoretical eccentricity evolution. To achieve this we normalise the time coordinate in both cases by the characteristic time-scale $\gamma^{-1}$, where $\gamma$ is defined in equation~\ref{eq:gamma}. In the case of the Monte Carlo simulation, $\gamma$ is tracked at $100$~Myr time intervals, such that the time-coordinate is in fact the estimated integral sum of $\gamma \mathrm{\Delta} t$. We then compute the distribution of eccentricities expected following equation~\ref{eq:PDE}, for an initial eccentricity dispersion $\sigma_{e,0} =10^{-3}$ around $e_0=0.9$. 

The two distributions are compared in Figure~\ref{fig:eccdiff_comp}. We generally see poor agreement between the two prescriptions. This is expected, and is a consequence of the algorithm used to compute encounters in \textsc{Mocca}. In the first instance, the prescription is designed to capture the physics of energy transfer, which influences the dynamical evolution of the cluster, rather than the evolution of the orbital eccentricity. Because energy transfer drops exponentially with periastron distance $r_\mathrm{p}$, the maximum closest approach $R_\mathrm{enc}$ can be limited to consider only encounters within a comparatively small radius $R_\mathrm{enc} = X\cdot a$, where $X>1$ is some factor and $a$ is the semi-major axis of the binary (star-planet system in this case). In producing Figure~\ref{fig:eccdiff_comp} we have adopted $X=2$, which results in many encounters that yield eccentricity changes $|\epsilon|\lesssim 0.05$ to be ignored. This can be seen in the evolution of the distribution of the orbital eccentricities, where at early times the fraction of systems which have $e-e_0\gtrsim 0.05$ are much better produced than those with smaller changes in the Monte Carlo model. 

Unfortunately, increasing the value of $X$ is not a solution to this problem. When $X$ becomes large, the number of encounters that occur on a single time-step also becomes large. In the \textsc{Mocca} framework, the large number of small encounters is replaced with a single close encounter, drawn from the appropriate distribution in relative velocity and closest approach distance. This is acceptable for energy transfer, but not for computing the orbital eccentricity evolution. In principle one could decrease the time-step, however this would ultimately undermine the purpose of the Monte Carlo prescription and quickly become computationally impracticable. In the limit of large $X$, this would also replicate something similar to the experiment by \citet{Hamers17}, with no benefit in terms of the parameter space exploration. 

We can however extract some quantitative comparison between the theoretical prediction and Monte Carlo results. This comparison is the relative number of encounters that result in a change of eccentricity of size $|\epsilon|$. This distribution (in log space) is shown in Figure~\ref{fig:rel_encs}. For hyperbolic encounters we expect the relative number of encounters to scale with $|\epsilon|^{-2}$. This is what we find for sufficiently large eccentricity ($|\epsilon|\gtrsim 0.05$) within the Monte Carlo model, independently of the local stellar density. While we are unable to directly compare the Monte Carlo and theoretical predictions for the eccentricity evolution, this exercise somewhat justifies our prescription. A further benchmarking exercise is performed in Section~\ref{sec:comp_numerical}, where we compare to the more accurate numerical experiments by \citet{Hamers17}.

\bsp	
\label{lastpage}
\end{document}